\documentclass[
 reprint,
 floatfix,
 amsmath,amssymb,
 aps,
 superscriptaddress,
]{revtex4-2}

\usepackage{graphicx}
\usepackage{dcolumn}
\usepackage{bm}
\usepackage{hyperref}
\usepackage{placeins}
\usepackage{dblfloatfix}
\usepackage{capt-of}
\graphicspath{{./}{../}}
\newcommand{\BibDatabase}{paper}
\IfFileExists{\BibDatabase.bib}{}{%
  \renewcommand{\BibDatabase}{main/paper}
}

\begin{document}
\raggedbottom

\title{Bayesian inference constraints on jet quenching across centrality, beam energy, and observable classes in LHC heavy-ion collisions}
\author{Dongguk Kim}
\affiliation{Department of Physics, Sungkyunkwan University, Suwon City, Republic of Korea}

\author{D.~J.~Kim}
\affiliation{University of Jyv\"askyl\"a, Department of Physics, P.O.\ Box 35, FI-40014 University of Jyv\"askyl\"a, Finland}
\affiliation{Helsinki Institute of Physics, P.O.\ Box 64, FI-00014 University of Helsinki, Finland}

\author{Jeongsu Bok}
\affiliation{Department of Physics, Pusan National University, Busan, Republic of Korea}

\author{Beomkyu Kim}
\affiliation{Department of Physics, Sungkyunkwan University, Suwon City, Republic of Korea}
\date{\today}

\begin{abstract}
Jet quenching in heavy-ion collisions probes parton energy loss in the quark--gluon plasma (QGP), but the extracted transport properties may not be universally constrained across centrality, beam energy, and observable class. In this work, we perform an analysis of the compatibility and predictive transferability of Bayesian constraints obtained from a six-parameter JETSCAPE effective energy-loss model across these subsets. The model is calibrated to charged-hadron and inclusive-jet data from ALICE, ATLAS, and CMS in PbPb collisions at $\sqrt{s_{\mathrm{NN}}}=5.02$ and $2.76$ TeV. We find that centrality-dependent posteriors are largely compatible, whereas beam-energy and observable-class splits exhibit moderate shifts within overlapping credible regions, indicating that posterior overlap alone does not guarantee predictive universality. This is further examined by propagating subset posteriors to complementary datasets without refitting, where predictive performance varies across subsets. These results indicate that different observables probe distinct aspects of jet--medium interactions and motivate leading-hadron-selected jet observables to bridge hadron-biased and jet-inclusive constraints.
\end{abstract}

\maketitle

\section{Introduction}

Ultra-relativistic heavy-ion collisions at RHIC and LHC create a deconfined Quantum Chromodynamics (QCD) medium whose collective behavior is consistent with the formation of a strongly interacting quark--gluon plasma (QGP)~\cite{Adams:2005dq,Adcox:2004mh,Arsene:2004fa,Back:2004je,Busza:2018rrf,Braun-Munzinger:2015hba,Shuryak:2008eq,Heinz:2013th}.
Hard partons produced at early times lose energy as they traverse this medium, producing jet quenching, which is observed through the suppression and modification of high-$p_T$ hadron and jet yields, angular correlations, and jet substructure~\cite{Gyulassy:1990ye,Wang:1991xy,Majumder:2010qh,Qin:2015srf,Cao:2020wlm}.
These probes sample the full evolution of medium and are commonly characterized through the jet transport coefficient $\hat{q}$, an effective measure of transverse momentum broadening and jet--medium coupling~\cite{Baier:1996sk,Burke:2013yra}.

Extracting $\hat{q}$ is difficult because the measured suppression depends simultaneously on the microscopic interaction, collision geometry, parton path length, medium evolution, and observable selection bias.
Charged-hadron and inclusive-jet nuclear modification factors, $R_{\mathrm{AA}}$, therefore give complementary constraints: high-$p_T$ charged hadrons are biased toward jets with hard leading fragments and stronger surface bias, whereas reconstructed jets remain sensitive to a broader shower-energy distribution.
Centrality and beam energy further change the geometry, temperature history, and hard-parton spectra, so subset $\hat{q}/T^3$ posteriors must be checked for mutual compatibility and predictive transferability.

Bayesian parameter estimation provides a practical framework for this problem by combining multidimensional experimental data with computationally expensive jet-quenching simulations in a statistically controlled manner.
Within JETSCAPE, Gaussian-process emulators and Markov-chain Monte Carlo sampling have enabled high-dimensional jet-sector calibrations.
The first hard-sector extraction calibrated a MATTER+LBT multistage energy-loss model to inclusive charged-hadron suppression at RHIC and the LHC, constraining the effective transport coefficient and the switching scale between high- and low-virtuality evolution~\cite{JETSCAPE:2021ehl}.
Later JETSCAPE studies extended the calibration to combined charged-hadron and inclusive-jet suppression data, refined the model--data comparison metric, and showed that added jet information tightens constraints while exposing residual dependence on observable class and kinematic selection~\cite{Fan:2023metric,Ehlers:2024bayesjet}.
These studies establish global calibration as a baseline, but a global posterior does not by itself show whether constraints inferred from one subset can predict another without refitting.
Here we use universality in this operational sense: subset-restricted calibrations should be compatible in parameter space and transferable in observable space.

In this paper, we calibrate a six-parameter effective jet-energy-loss model in the JETSCAPE multistage framework to charged-hadron and inclusive-jet $R_{\mathrm{AA}}$ measurements from ALICE, ATLAS, and CMS in PbPb collisions at $\sqrt{s_{\mathrm{NN}}}=5.02$ and $2.76$ TeV.
We first obtain the all-observables posterior, then decompose it by centrality, beam energy, and observable class.
We focus on charged-hadron and inclusive-jet $R_{\mathrm{AA}}$ because they are the only high-$p_T$ classes with sufficient coverage at both energies to support matched subset comparisons.
Subset posteriors are compared at a common reference temperature and parton energy and then propagated to target observables without refitting; a complementary sensitivity analysis identifies which parameter combinations are constrained by the current dataset.
Since our observable set, covariance treatment, and inference goal differ from Ref.~\cite{Ehlers:2024bayesjet}, we do not attempt a point-by-point numerical reproduction; instead, we focus on consistency and residual discrepancy patterns within the present setup.
The paper is organized as follows. Section~\ref{sec:physics-model} introduces the physics model, Section~\ref{sec:bayesian-inference} presents the Bayesian setup and calibration strategy, Section~\ref{sec:results} reports the posterior and subset-decomposition results, Section~\ref{sec:discussion} summarizes cross-prediction tests, sensitivity analysis, and a proposed observable to bridge hadron-biased and jet-inclusive constraints, and Section~\ref{sec:conclusion} concludes. Implementation and validation details are collected in Appendices~\ref{app:production-details}--\ref{app:mcmc-diagnostics}.

\section{PHYSICS MODEL}
\label{sec:physics-model}

\subsection{Framework overview}
This work uses the JETSCAPE multistage event framework to simulate both $pp$ and PbPb collisions in a common setup~\cite{JETSCAPE:2020mzn,Ehlers:2024bayesjet,Fan:2023metric}. For each design point, hard partons are first produced in vacuum, then propagated through an expanding QCD medium, and finally converted to hadrons for observable construction. The same baseline generator and final-state analysis definitions are applied to $pp$ and PbPb collisions to isolate medium effects in the ratio observables.

\subsection{Hard-scattering}
\label{subsec:hard-scattering}
The initial hard scattering is generated with the PYTHIA event generator within the JETSCAPE workflow~\cite{Sjostrand:2014zea}.
PYTHIA provides a general-purpose Monte Carlo description of the perturbative hard process and the subsequent parton-shower evolution in high-energy collisions.
In the present context, it supplies the vacuum partonic configuration before any medium-induced modification is applied.

Because the charged-hadron and inclusive-jet observables used in this work extend to relatively high transverse momentum, the event generation is performed in $\hat{p}_{\mathrm{T}}$-binned samples rather than in a single fully inclusive sample.
This improves the statistical coverage of the rare high-$p_T$ region while keeping the total computational cost manageable.
For each sampled interval $\ell$, PYTHIA provides the corresponding hard-scattering cross section $\hat{\sigma}_{\ell}$.
The final spectra are reconstructed by combining the bin-wise yields with these cross-section weights,
\begin{equation}
R_{\mathrm{AA}}^{X}(p_{\mathrm{T}};c)
=
\frac{\mathrm{d}\sigma_{\mathrm{AA}}^{X,c}/\mathrm{d}p_{\mathrm{T}}}
     {\mathrm{d}\sigma_{pp}^{X}/\mathrm{d}p_{\mathrm{T}}}
=
\frac{\displaystyle \sum_{\ell}
\left(
\frac{\mathrm{d}N_{\mathrm{AA},\ell}^{X,c}}{\mathrm{d}p_{\mathrm{T}}}
\, \hat{\sigma}_{\ell}(\hat{p}_{\mathrm{T}})
\right)}
{\displaystyle \sum_{\ell}
\left(
\frac{\mathrm{d}N_{pp,\ell}^{X}}{\mathrm{d}p_{\mathrm{T}}}
\, \hat{\sigma}_{\ell}(\hat{p}_{\mathrm{T}})
\right)}.
\label{eq:raa}
\end{equation}
Here $X$ labels the observable species, $c$ is the centrality class, $\ell$ labels the sampled hard-scattering $\hat{p}_{\mathrm{T}}$ intervals, $\mathrm{d}N_{\ell}/\mathrm{d}p_{\mathrm{T}}$ denotes the simulated yield in a given $p_{\mathrm{T}}$ bin for interval $\ell$, and $\hat{\sigma}_{\ell}$ is the corresponding PYTHIA cross section. This recombination preserves the physical spectrum shape while keeping the rare high-$p_T$ region statistically tractable. The technical bin-stitching and local smoothing steps used to suppress isolated Monte Carlo artifacts are summarized in Appendix~\ref{app:production-details}. The same hard-process setup is applied to both $pp$ and PbPb events, so medium effects enter only through the subsequent in-medium propagation.

\subsection{Initial-state geometry from TRENTo}
The event-by-event initial entropy-density profiles used for the medium background are generated with the TRENTo model~\cite{Moreland:2014oya,Bernhard:2016tnd}.
TRENTo provides a reduced-thickness parametrization of the initial-state geometry that captures the dominant event-by-event structure controlling the later medium evolution. We generate the TRENTo events in advance and use the same geometric model assumptions at $\sqrt{s_{\mathrm{NN}}}=5.02$ and $2.76$ TeV, varying only the overall normalization between the two systems. The numerical TRENTo settings and event counts are collected in Appendix~\ref{app:production-details}.

\subsection{Evolution of the QCD medium}
Jet partons are embedded in a time-dependent QCD medium background and interact with the local thermodynamic environment along their trajectories.
The background medium is constructed from the event-by-event TRENTo initial conditions discussed above, passed through a short freestream stage, and then evolved with boost-invariant 2+1D viscous MUSIC hydrodynamics~\cite{Song:2007ux}. The resulting temperature and flow fields define the local medium sampled by the propagating jet partons. All background parameters are held fixed to the production baseline, so the present study tests the relative consistency of the jet-sector extraction within a fixed background campaign rather than performing a simultaneous bulk-plus-jet calibration. The numerical freestream+MUSIC settings are summarized in Appendix~\ref{app:production-details}. The hydrodynamic start time $\tau_{\mathrm{hydro}}$ used in this fixed background evolution is distinct from the calibrated medium-onset parameter $\tau_0$ introduced below.

\subsection{Energy-loss parametrization}
The jet--medium interaction is described with an effective model with 6 parameters,
\[
\theta=(Q_0,\tau_0,A,B,C,\alpha_s),
\]
which is calibrated by Bayesian inference.
Here $\tau_0$ sets the onset time for medium-induced energy loss in the MATTER stage (distinct from the fixed hydrodynamic start time $\tau_{\mathrm{hydro}}$ used in the background evolution), while $Q_0$, $A$, $B$, $C$, and $\alpha_s$ determine the type-6 virtuality-dependent transport coefficient.
For a gluon probe, it is useful to separate the implemented type-6 form into an HTL baseline and a virtuality-dependent factor:
\begin{equation}
\hat{q}_g(T,E,Q^2)
=
\hat{q}_g^{\mathrm{HTL}}(T,E)\,F_6(E,Q^2),
\label{eq:qhat-type6}
\end{equation}
where the HTL baseline is
\begin{equation}
\hat{q}_g^{\mathrm{HTL}}(T,E)
=
\left(C_A\,\frac{50.4864}{\pi}\right)
\alpha_s^{\mathrm{run}}(\mu_{\mathrm{th}}^2)\,\alpha_s\,T^3
\ln\!\left(\frac{\mu_{\mathrm{th}}^2}{m_D^2}\right),
\label{eq:qhat-htl-base}
\end{equation}
with
\[
\begin{aligned}
\mu_{\mathrm{th}}^2&=\max(1,2ET),\\
m_D^2&=4\pi\alpha_s T^2\,\frac{6+n_f}{6},
\qquad n_f=3.
\end{aligned}
\]
Here $\alpha_s^{\mathrm{run}}(\mu_{\mathrm{th}}^2)$ denotes the one-loop running coupling evaluated at the thermal scale $\mu_{\mathrm{th}}^2$, while the separate factor $\alpha_s$ is the fixed calibration parameter that sets the overall HTL normalization and also enters the Debye mass.
For $\mu_{\mathrm{th}}^2<1$, the running coupling is frozen to this same value, $\alpha_s$.
Eq.~\eqref{eq:qhat-htl-base} contains no explicit virtuality dependence; all dependence on the parent-parton virtuality enters through the multiplicative factor $F_6(E,Q^2)$.
This extra factor is needed because the MATTER stage evolves a virtuality-ordered parton shower, so partons with the same energy $E$ propagating through the same medium temperature $T$ can still probe the medium at different virtuality scales.
Without such a factor, the model would assign the same interaction strength to all shower partons at fixed $(T,E)$; the type-6 prefactor instead provides an effective scale dependence while being normalized to recover the HTL baseline at the reference scale $Q^2=Q_0^2$.
The factor is written as
\begin{equation}
F_6(E,Q^2)=
\begin{cases}
1, & Q^2 \le Q_0^2, \\[0.6em]
R(Q^2)\,
\dfrac{e^{C(1-x_B)}-1}{e^{C(1-x_{B0})}-1},
& C>0,\ x_B<0.99, \\[1.0em]
R(Q^2)\,\dfrac{1-x_B}{1-x_{B0}},
& C=0,\ x_B<0.99, \\[1.0em]
0, & \text{otherwise},
\end{cases}
\label{eq:f6-type6}
\end{equation}
with
\[
\begin{aligned}
x_B&=\frac{Q^2}{2E},
\qquad
x_{B0}=\frac{Q_0^2}{2E},\\
L(z)&=\ln\!\left(\frac{z}{Q_{\mathrm{ref}}^2}\right),
\qquad
N(z)=1+A\,L(z)+B\,L^2(z),\\
R(z)&=\frac{N(Q_0^2)}{N(z)},\\
Q_{\mathrm{ref}}^2&=0.04~\mathrm{GeV}^2.
\end{aligned}
\]
The quark transport coefficient is obtained from
$\hat{q}_q=(C_F/C_A)\hat{q}_g$.
In this parametrization, $Q_0$ fixes the reference virtuality, $A$ and $B$ control the logarithmic virtuality dependence in the denominator, and $C$ controls the large-$x_B$ suppression in the numerator.
This parametrization is used consistently for emulator training, posterior inference, and the derived $\hat{q}/T^3$ results.

\subsection{Hadronization and observable reconstruction}
After in-medium propagation, partons are hadronized and analyzed with the same observable definitions used for the experimental comparisons.
In the present setup, the hadronization step is performed with the PYTHIA colorless hadronization module provided within the JETSCAPE framework.
Charged-hadron and inclusive-jet spectra are computed in each centrality class, and the final $R_{\mathrm{AA}}$ observables are formed from matched PbPb and $pp$ references.
For inclusive jets, we reconstruct jets with the FastJet anti-$k_T$ algorithm using the same jet-radius selections listed in Table~\ref{tab:inclusive-jet-raa}.
In the postprocessing analysis, we also evaluate the hole-subtracted jet transverse momentum, denoted here as $p_{\mathrm{T}}^{\mathrm{sub,hole}}$, using the JETSCAPE hole-subtraction prescription.
Here the hole hadrons denote hadronized thermal partons sampled from the medium during the jet--medium interaction; in our event record, they are tagged by $p_{\mathrm{stat}}=-1$.
Explicitly, the hole-subtracted jet four-momentum is defined as
\begin{equation}
p_{\mathrm{jet}}^{\mu,\mathrm{sub,hole}}
=
p_{\mathrm{jet}}^\mu
-\sum_{i \in \{p_{\mathrm{stat}}=-1\}}
p_i^\mu\,
\Theta\!\left(R-\Delta R_{i,\mathrm{jet}}\right),
\end{equation}
where $\Delta R_{i,\mathrm{jet}}=\sqrt{(\Delta \eta)^2+(\Delta \phi)^2}$ and $R$ is the jet radius.
The corresponding transverse momentum is then evaluated as
\begin{equation}
p_{\mathrm{T}}^{\mathrm{sub,hole}}
=
\sqrt{
\left(p_{x,\mathrm{jet}}-\sum_i p_{x,i}\right)^2
+
\left(p_{y,\mathrm{jet}}-\sum_i p_{y,i}\right)^2
}.
\end{equation}
This hole-subtraction procedure is applied event by event and provides the jet observable used for the corresponding hole-subtracted distributions.

\section{BAYESIAN INFERENCE}
\label{sec:bayesian-inference}

We now summarize the Bayesian inference framework used throughout the analysis. We infer the calibration parameter vector
$\theta=(Q_0,\tau_0,A,B,C,\alpha_s)$
from the experimental observable vector
$y^{\mathrm{exp}}$
using Bayes' theorem:
\begin{equation}
p(\theta \mid y^{\mathrm{exp}})
\propto
\mathcal{L}(y^{\mathrm{exp}} \mid \theta)\,p(\theta),
\label{eq:posterior}
\end{equation}
where $p(\theta)$ is the prior and $\mathcal{L}$ is the likelihood. Direct evaluation of the full JETSCAPE event-generation chain at every trial point in parameter space is computationally prohibitive, so we replace the full simulation inside the inference loop with a Gaussian-process (GP) emulator trained on a finite set of design points. For any new parameter point $\theta$, the emulator returns both a mean prediction $\mu(\theta)$ for the observable vector and an associated predictive covariance, which are then propagated into the likelihood.

For an observalbe vector of dimension $m$, we assume a multivariate normal likelihood, Equation, where
\begin{equation}
\mathcal{L}(y^{\mathrm{exp}} \mid \theta)
=
\frac{\exp\!\left[-\frac{1}{2}\,\Delta y(\theta)^T
\Sigma_{\mathrm{tot}}^{-1}(\theta)\,\Delta y(\theta)\right]}
{\sqrt{(2\pi)^m\,\left|\Sigma_{\mathrm{tot}}(\theta)\right|}},
\label{eq:likelihood}
\end{equation}
with $\mu(\theta)$ the emulator prediction and $\Sigma_{\mathrm{tot}}(\theta)$ the total covariance matrix. In the implementation used here, the likelihood is evaluated block by block and the corresponding log-likelihoods are summed. For each block,
\[
\Sigma_{\mathrm{tot}}(\theta)
=
\Sigma_{\mathrm{exp}}
+ \Sigma_{\mathrm{emu}}(\theta),
\]
where $\Delta y(\theta)=y^{\mathrm{exp}}-\mu(\theta)$ and the two covariance terms denote the experimental covariance and the emulator covariance, respectively. In the implementation used here, $\Sigma_{\mathrm{exp}}$ includes the experimental statistical covariance and the available systematic covariance contributions. The inferred posteriors, therefore, quantify consistency with the stated experimental and emulator uncertainties, but they do not marginalize over unmodeled theory systematics such as background-medium or fragmentation uncertainties.

In the present analysis, no separate $\Sigma_{\mathrm{model}}(\theta)$ term is added to the likelihood. Finite-statistics fluctuations of the model calculations are instead absorbed into $\Sigma_{\mathrm{emu}}(\theta)$ through the GP noise treatment. The emulator itself is constructed in a PCA-reduced output space with separate hadron and jet blocks that are reassembled in a common observable ordering; the predictive covariance combines the GP uncertainty of the retained components with the residual variance from the truncated ones. The detailed standardization, weighting, kernel, hyperparameter, and MCMC settings are summarized in Appendix~\ref{app:inference-details}, while the chain-quality criteria are given in Appendix~\ref{app:mcmc-diagnostics}. This Bayesian-GP-MCMC setup defines the statistical framework used in the simulation and calibration steps described below.

\subsection{Simulation setup}

With the statistical framework specified, we now summarize the training design and parameter ranges. The parameters are briefly described as follows:
\begin{itemize}
\item $Q_0$: reference virtuality scale in the type-6 transport-coefficient model.
\item $\tau_0$: onset time for medium-induced energy loss in the MATTER stage (not the hydrodynamic start time $\tau_{\mathrm{hydro}}$).
\item $A$: coefficient of the linear logarithmic term in the type-6 virtuality denominator.
\item $B$: coefficient of the quadratic logarithmic term in the type-6 virtuality denominator.
\item $C$: coefficient controlling the $x_B$-dependent numerator in the type-6 prefactor.
\item $\alpha_s$: fixed calibration parameter entering the HTL normalization and Debye mass, and setting the infrared-frozen value of $\alpha_s^{\mathrm{run}}$.
\end{itemize}
\begin{table}[htbp]
\caption{Calibration parameter ranges.}
\label{tab:calibration-parameters}
\begin{ruledtabular}
\scriptsize
\begin{tabular*}{\columnwidth}{@{\extracolsep{\fill}}lcccccc}
Parameter & $Q_0$ & $\tau_0$ & $A$ & $B$ & $C$ & $\alpha_s$ \\
Range & 0.5--4.0 & 0.1--1.5 & 5.0--15.0 & 50.0--150.0 & 0.1--0.3 & 0.1--0.5 \\
\end{tabular*}
\end{ruledtabular}
\end{table}
The bounds listed in Table~\ref{tab:calibration-parameters} are selected based on Refs.~\cite{Ehlers:2024bayesjet,Fan:2023metric}.

A design point denotes a unique parameter vector ($Q_0$, $\tau_0$, $A$, $B$, $C$, $\alpha_s$) at which a full event simulation is performed. We use 50 Latin-hypercube-sampled design points drawn from independent uniform priors and evaluate each point with the same production chain described in Sec.~\ref{sec:physics-model}. This finite design is a pragmatic compromise between parameter-space coverage and computational cost, and its adequacy is assessed empirically through the emulator-validation and closure tests presented below. Detailed simulation statistics and the LHS coverage diagnostics are collected in Appendix~\ref{app:inference-details}.
\subsection{Experimental data setup}
\label{sec:experimental-data-setup}
With the simulation design fixed, we now specify the experimental constraints. In this work, we use PbPb data at $\sqrt{s_{\mathrm{NN}}}=5.02$ and $2.76$~TeV from ALICE, ATLAS, and CMS~\cite{ALICE:2018chg,ATLAS:2015cps,CMS:2017cnmf,CMS:2012hpts,ALICE:2020ijs,ATLAS:2019nmfjet,ATLAS:2015jetr276,CMS:2021lajets,CMS:2017ijc}.
Table~\ref{tab:charged-hadron-raa} and Table~\ref{tab:inclusive-jet-raa} summarize the experimental datasets used for charged hadron and inclusive jet $R_{\mathrm{AA}}$, respectively.
The observables are the nuclear modification factors $R_{\mathrm{AA}}$ for charged hadrons and inclusive jets, defined in Eq.~\eqref{eq:raa}, with $X\in\{\mathrm{ch},\mathrm{jet}\}$.
The centrality intervals used for each experiment and collision energy are listed in Tables~\ref{tab:charged-hadron-raa} and \ref{tab:inclusive-jet-raa}; they comprise central selections ($0$--$5\%$ and/or $0$--$10\%$) and mid-central selections ($30$--$50\%$ and/or $40$--$50\%$, as available).

This dataset composition is chosen to test how the preferred $\hat{q}$ depends on medium scale and in-medium path length.
Within the observables used in this calibration, high-$p_{\mathrm{T}}$ hadron and jet $R_{\mathrm{AA}}$ provide strong sensitivity to $\hat{q}$ because their suppression is primarily controlled by parton energy loss accumulated along the in-medium path.
In addition, the ratio structure of $R_{\mathrm{AA}}$ reduces several normalization and production-rate systematics, making changes induced by transport coefficients more clearly identifiable.

Using two collision energies ($2.76$ and $5.02$~TeV) provides systems with different initial energy densities and lifetimes, i.e., different effective medium sizes, which gives leverage on possible medium-size dependence of the inferred $\hat{q}$.
For centrality, we treat $0$--$5\%$ and $0$--$10\%$ as a central-collision group and the mid-central bins in Tables~\ref{tab:charged-hadron-raa} and \ref{tab:inclusive-jet-raa} as a mid-central group.
Comparing these two groups probes geometry-related differences, including differences in the typical in-medium path length, because central events have, on average, longer in-medium path lengths than mid-central events.
Taken together, these choices strengthen the physics motivation for the selected datasets and support the validity of the calibration constraints.

\begin{table}[htbp]
\caption{Charged hadron $R_{\mathrm{AA}}$ dataset summary.}
\label{tab:charged-hadron-raa}
\begin{ruledtabular}
\begin{tabular}{cccc}
Experiment & $\sqrt{s_{\mathrm{NN}}}$ (TeV) & Centrality (\%) & Acceptance \\
\hline
ALICE~\cite{ALICE:2018chg} & 5.02 & 0-5, 40-50 & $|\eta| < 0.8$ \\
ALICE~\cite{ALICE:2018chg} & 2.76 & 0-5, 40-50 & $|\eta| < 0.8$ \\
ATLAS~\cite{ATLAS:2015cps} & 2.76 & 0-5, 40-50 & $|\eta| < 2.0$ \\
CMS~\cite{CMS:2017cnmf} & 5.02 & 0-5, 30-50 & $|\eta| < 1.0$ \\
CMS~\cite{CMS:2012hpts} & 2.76 & 0-5 & $|\eta| < 1.0$ \\
\end{tabular}
\end{ruledtabular}
\end{table}

\begin{table}[htbp]
\caption{Inclusive jet $R_{\mathrm{AA}}$ dataset summary.}
\label{tab:inclusive-jet-raa}
\begin{ruledtabular}
\begin{tabular}{ccccc}
Experiment & $\sqrt{s_{\mathrm{NN}}}$ (TeV) & Centrality & Jet radius & Acceptance \\
\hline
ALICE~\cite{ALICE:2020ijs} & 5.02 & 0-10 & $R = 0.4$ & $|\eta_{\mathrm{jet}}| < 0.5$ \\
ATLAS~\cite{ATLAS:2019nmfjet} & 5.02 & 0-10, 40-50 & $R = 0.4$ & $|\eta_{\mathrm{jet}}| < 2.8$ \\
ATLAS~\cite{ATLAS:2015jetr276} & 2.76 & 40-50 & $R = 0.4$ & $|\eta_{\mathrm{jet}}| < 2.1$ \\
CMS~\cite{CMS:2021lajets} & 5.02 & 0-10, 30-50 & $R = 0.4$ & $|\eta_{\mathrm{jet}}| < 2.0$ \\
CMS~\cite{CMS:2017ijc} & 2.76 & 0-5 & $R = 0.4$ & $|\eta_{\mathrm{jet}}| < 2.0$ \\

\end{tabular}
\end{ruledtabular}
\end{table}

\subsection{Subset strategy and evidentiary logic}
\label{subsec:subset-strategy}

With the dataset now specified, the central physics question is not simply whether a broad global posterior can accommodate the full measurements, but whether the inferred effective transport coefficient remains stable when the calibration is restricted to physically distinct subsets.
We therefore evaluate universality at two levels.
First, posterior decomposition tests parameter-space consistency: subset-restricted calibrations should give compatible $\hat{q}/T^3$ when evaluated at a common reference temperature and parton energy if a single effective description is adequate.
Second, cross-prediction tests observable-space consistency: a posterior calibrated on one subset should describe a different subset without refitting.
Posterior overlap is thus necessary but not sufficient for universality.

The subset definitions follow the structure of the available measurements.
We group the $0$--$5\%$ and $0$--$10\%$ selections into a central class and the $30$--$50\%$ and $40$--$50\%$ selections into a mid-central class.
Because the exact centrality binning and detector acceptances are not identical across experiments, these are tests between central-class and mid-central-class calibrations rather than literal one-to-one comparisons of identical geometry slices.
Likewise, the beam-energy split compares all available $5.02$ and $2.76$ TeV observables, and the observable-class split compares all charged-hadron and inclusive-jet observables in hadron-only and jet-only calibrations.

This construction also explains the restricted observable basis.
Charged-hadron and inclusive-jet $R_{\mathrm{AA}}$ are the two high-$p_T$ channels that are available across both beam energies and multiple experiments with enough coverage to support matched subset decompositions and cross-predictions.
More differential observables would be valuable, but including them here would make the subset comparisons much less uniform across systems.
The present analysis, therefore, prioritizes a controlled universality test on a common observable basis.

\subsection{Prior distributions}
\begin{figure}[!tbp]
\centering
\begin{minipage}{0.48\columnwidth}
\centering
\includegraphics[width=\linewidth]{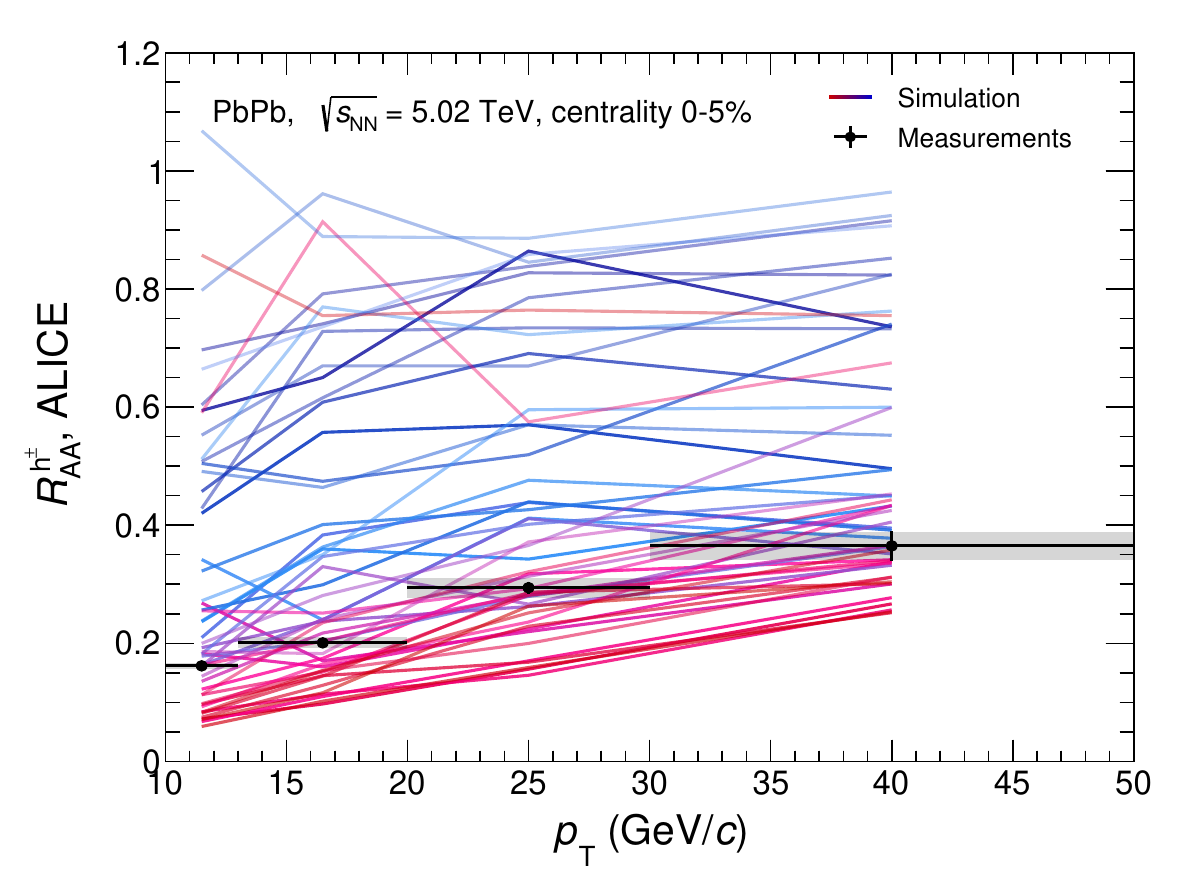}
\end{minipage}
\hfill
\begin{minipage}{0.48\columnwidth}
\centering
\includegraphics[width=\linewidth]{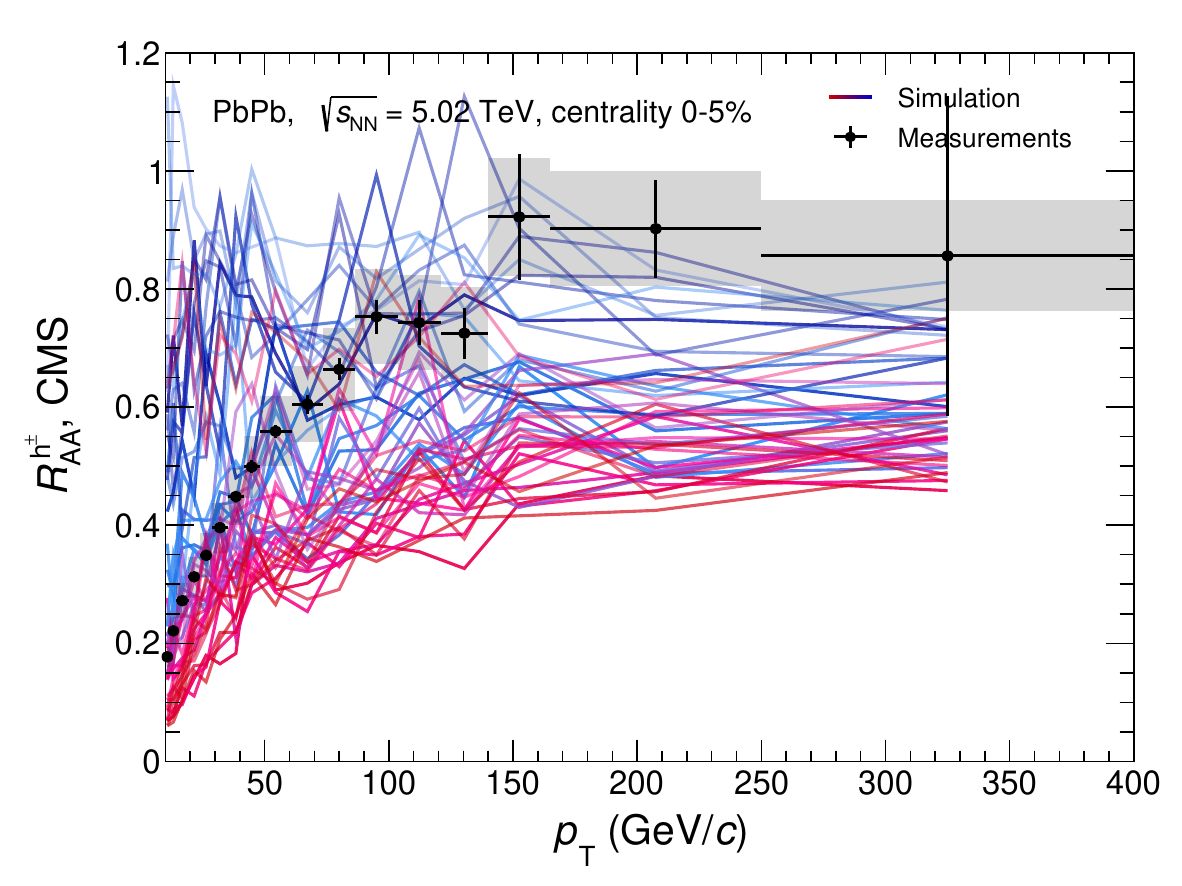}
\end{minipage}

\vspace{0.5em}

\begin{minipage}{0.48\columnwidth}
\centering
\includegraphics[width=\linewidth]{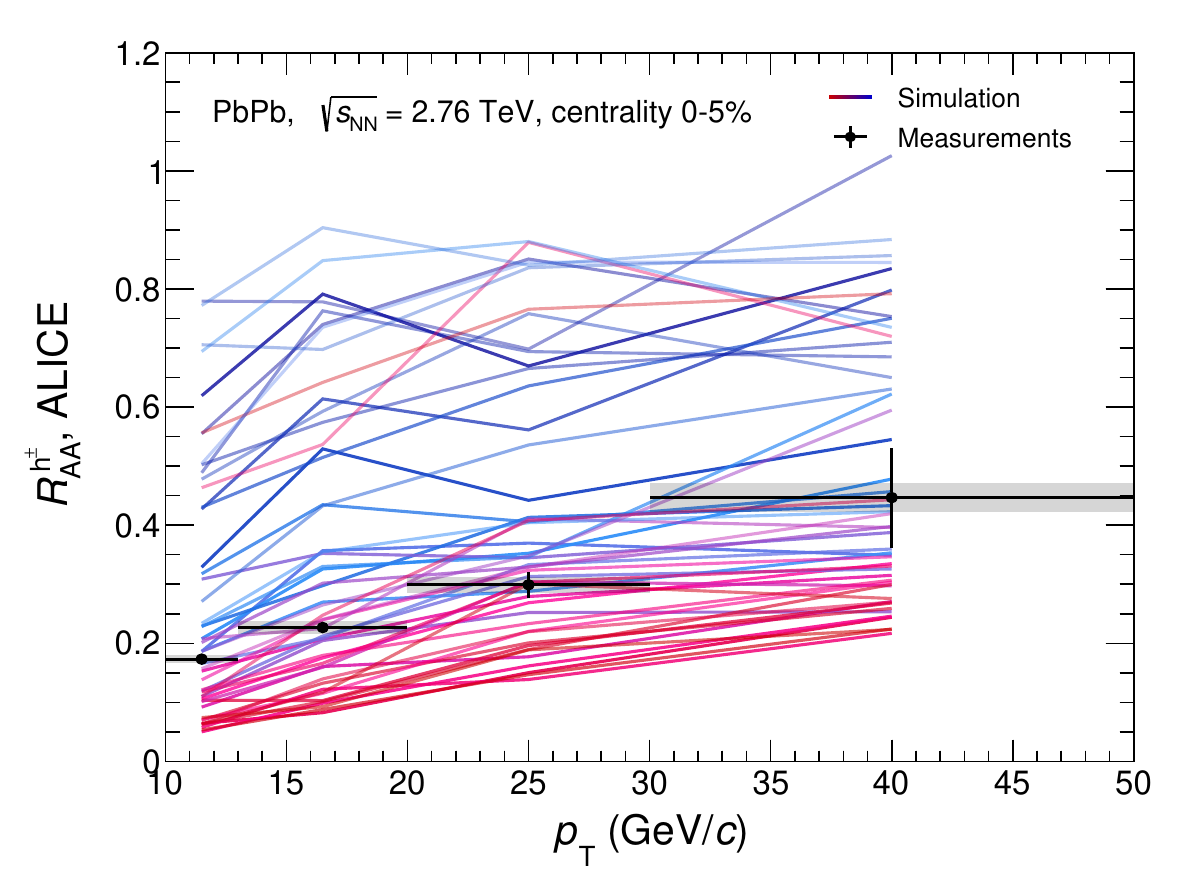}
\end{minipage}
\hfill
\begin{minipage}{0.48\columnwidth}
\centering
\includegraphics[width=\linewidth]{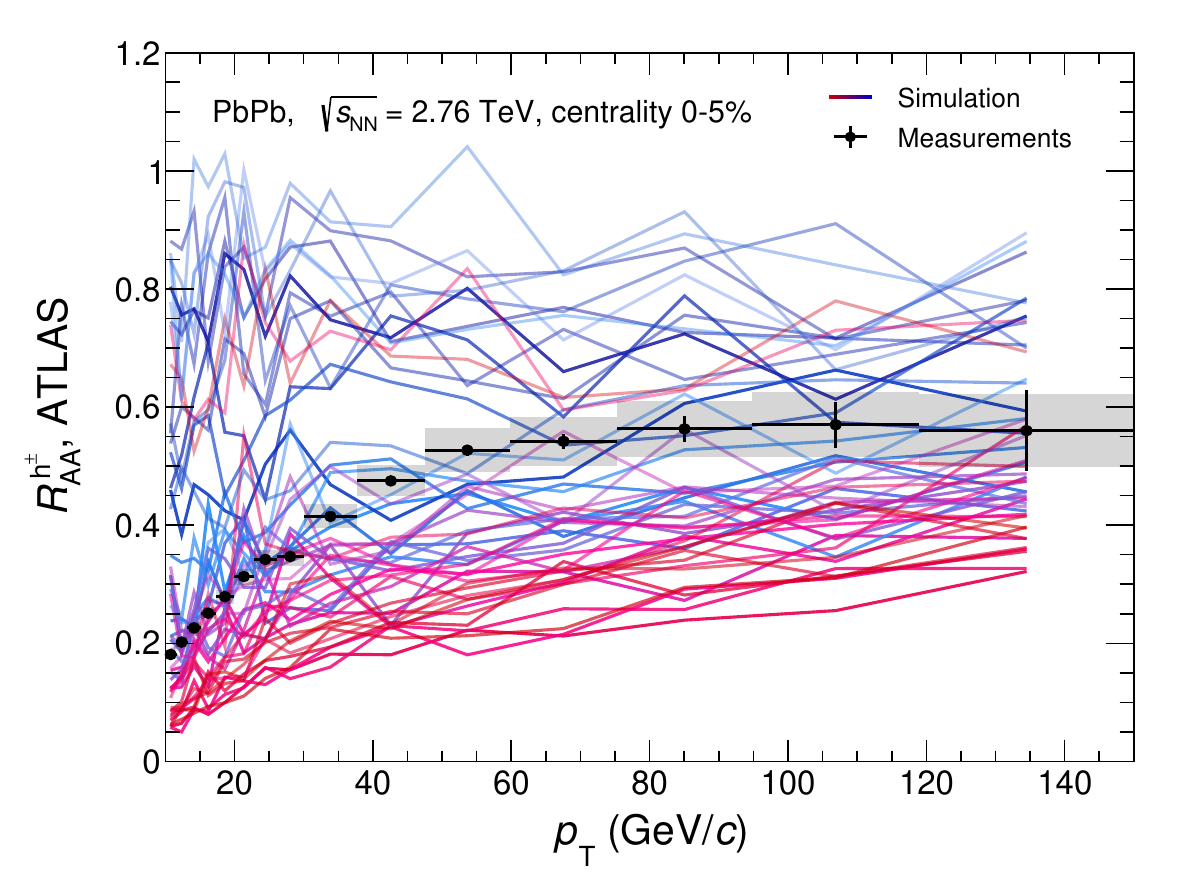}
\end{minipage}

\vspace{0.5em}

\begin{minipage}{0.48\columnwidth}
\centering
\includegraphics[width=\linewidth]{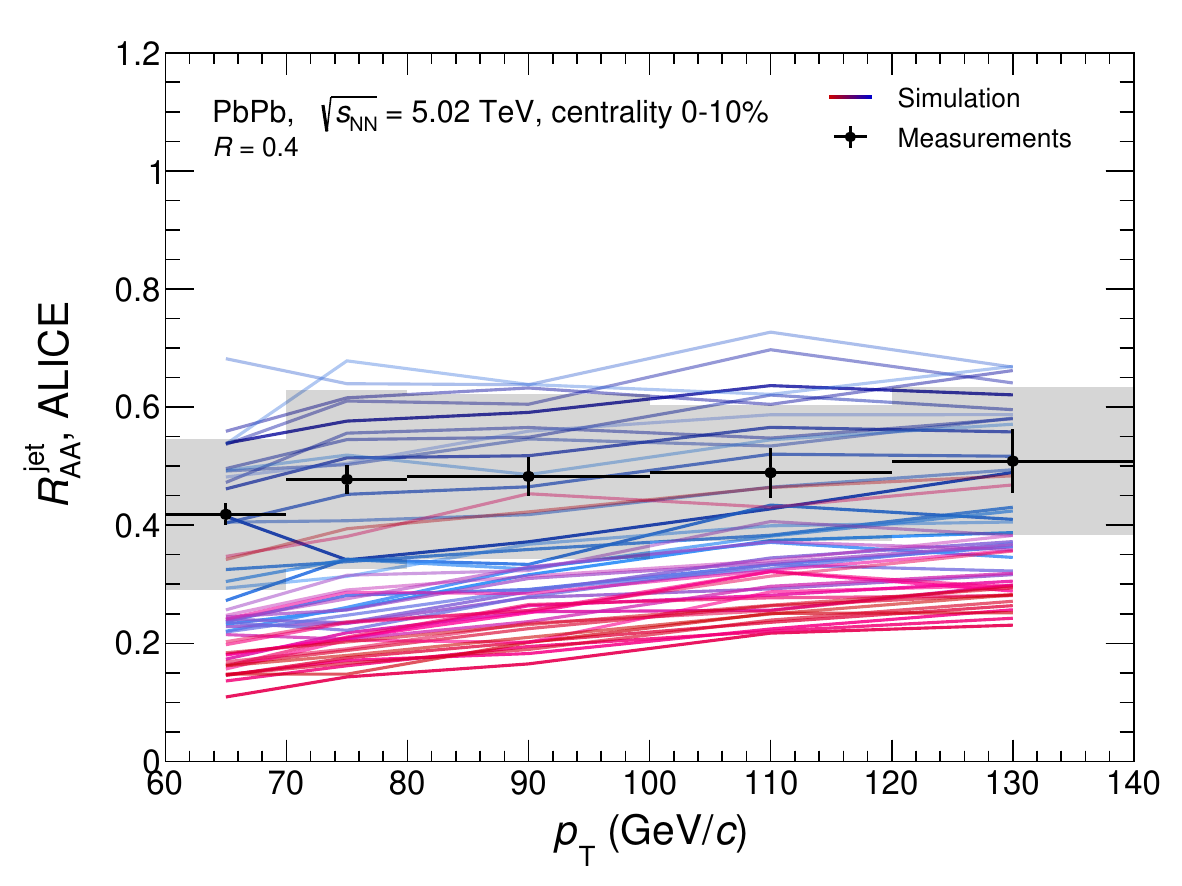}
\end{minipage}
\hfill
\begin{minipage}{0.48\columnwidth}
\centering
\includegraphics[width=\linewidth]{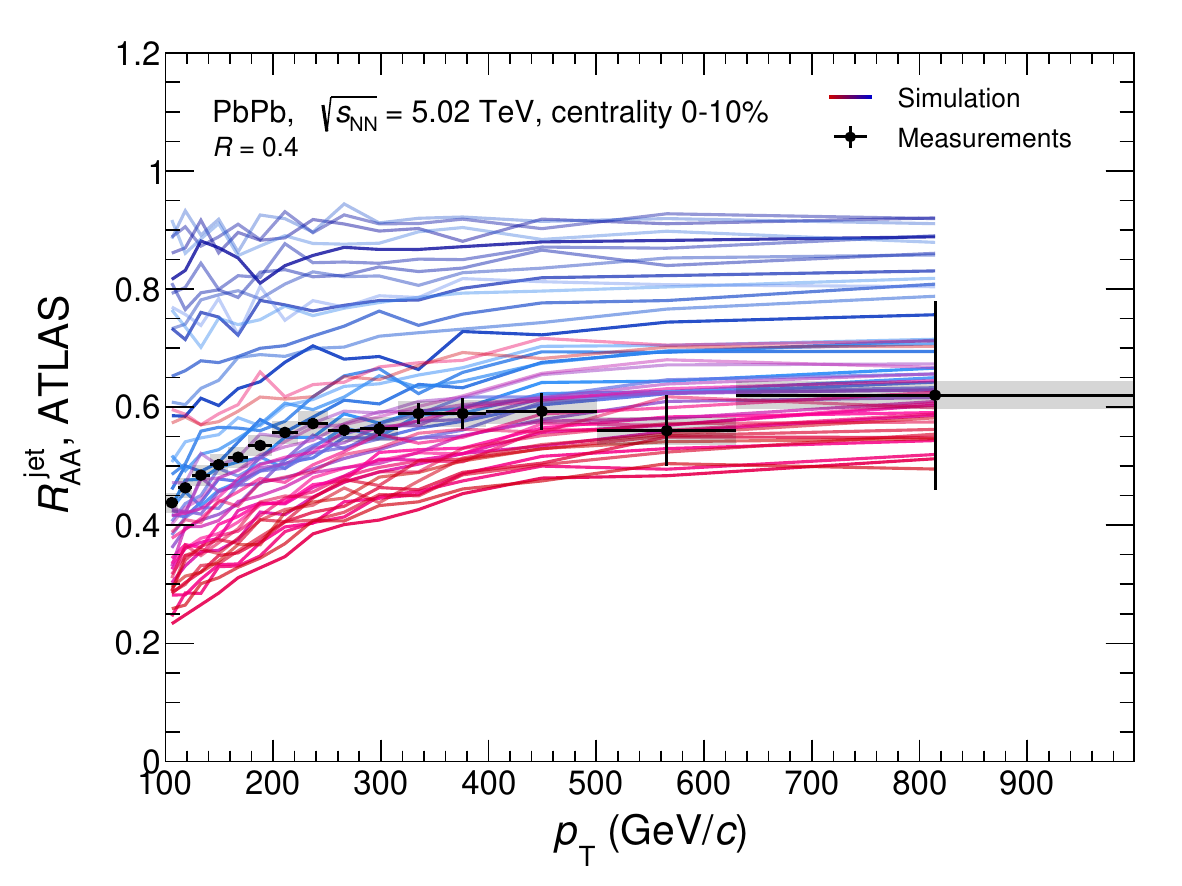}
\end{minipage}
\caption{Representative prior distributions for the fitted observables. The first row shows charged-hadron $R_{\mathrm{AA}}$ at $\sqrt{s_{\mathrm{NN}}}=5.02$ TeV for ALICE and CMS in the $0$--$5\%$ centrality class. The second row shows charged-hadron $R_{\mathrm{AA}}$ at $\sqrt{s_{\mathrm{NN}}}=2.76$ TeV for ALICE and ATLAS in the $0$--$5\%$ centrality class. The third row shows inclusive-jet $R_{\mathrm{AA}}$ at $\sqrt{s_{\mathrm{NN}}}=5.02$ TeV for ALICE and ATLAS in the $0$--$10\%$ centrality class.}
\label{fig:prior-panels-main}
\end{figure}

Before turning to emulator performance, we first check that the prior-prediction range covers the data. The prior-prediction distributions are organized by observable.
For the charged-hadron $R_{\mathrm{AA}}$, only data points with $p_T > 10~\mathrm{GeV}/c$ are included in order to restrict the analysis to the perturbative regime.
The same lower cut is imposed when assembling the charged-hadron emulator-training outputs, whereas the inclusive-jet training set is kept at the experimental jet binning without an additional emulator-specific $p_T$ cut.
Before training the GP emulators, we verify that the dynamic range of the model calculations sufficiently covers that of the experimental data across all considered centrality intervals, ensuring that the data-constraining region lies within the sampled design space.
This coverage is illustrated in Fig.~\ref{fig:prior-panels-main} for representative charged-hadron and inclusive-jet observables; the remaining prior panels are collected in Appendix~\ref{app:prior-panels}.
The color scale represents the value of $\alpha_s$: red indicates larger $\alpha_s$, while progressively bluer tones correspond to smaller values.

\subsection{Emulator validation}
\begin{figure}[!tbp]
\centering
\begin{minipage}{0.48\columnwidth}
\centering
\includegraphics[width=\linewidth]{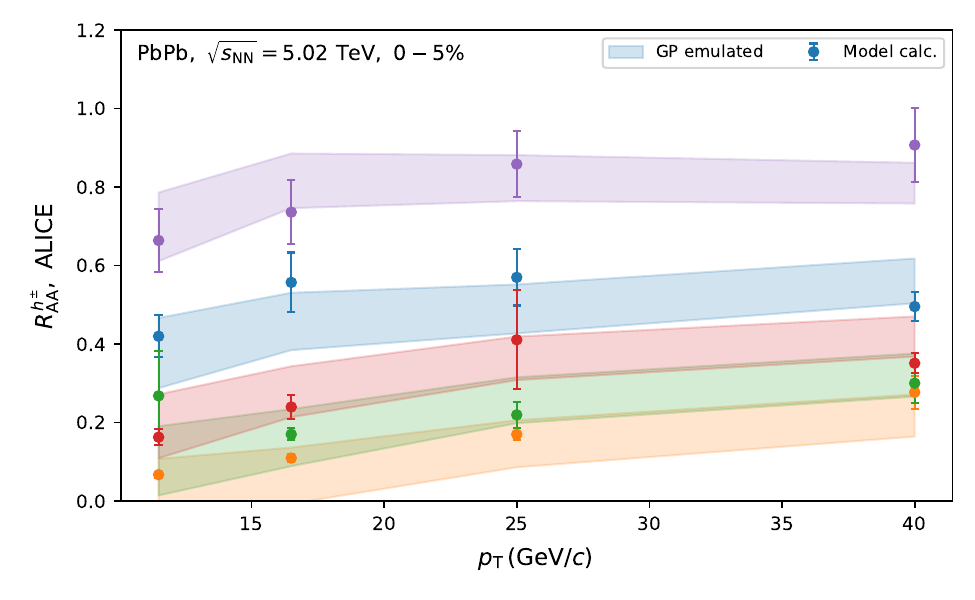}
\end{minipage}
\hfill
\begin{minipage}{0.48\columnwidth}
\centering
\includegraphics[width=\linewidth]{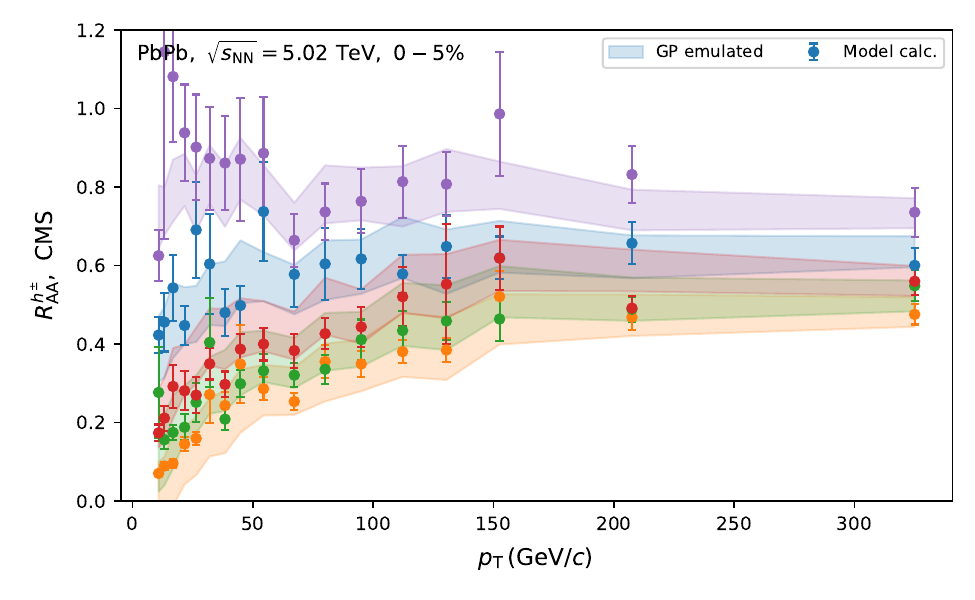}
\end{minipage}

\vspace{0.5em}

\begin{minipage}{0.48\columnwidth}
\centering
\includegraphics[width=\linewidth]{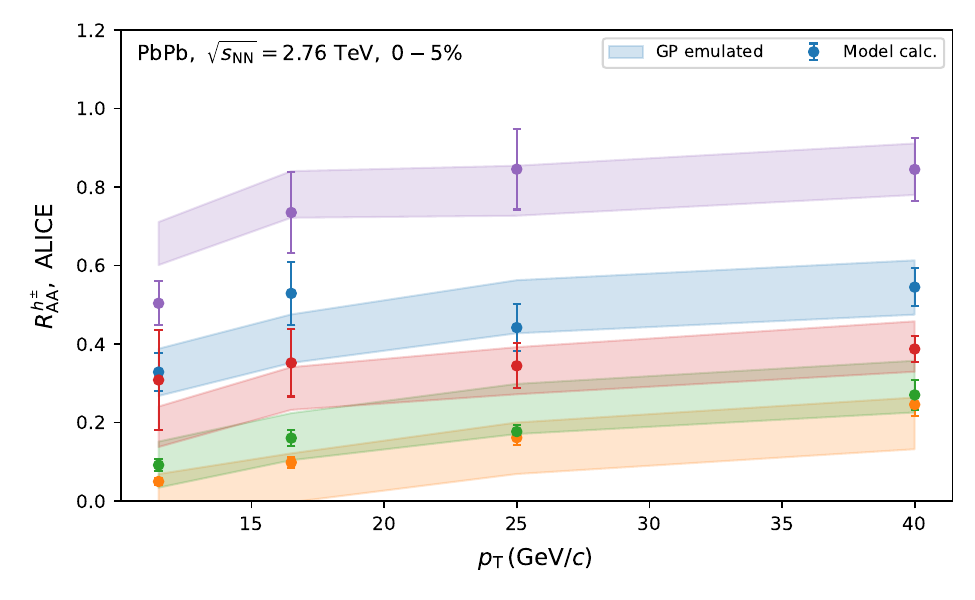}
\end{minipage}
\hfill
\begin{minipage}{0.48\columnwidth}
\centering
\includegraphics[width=\linewidth]{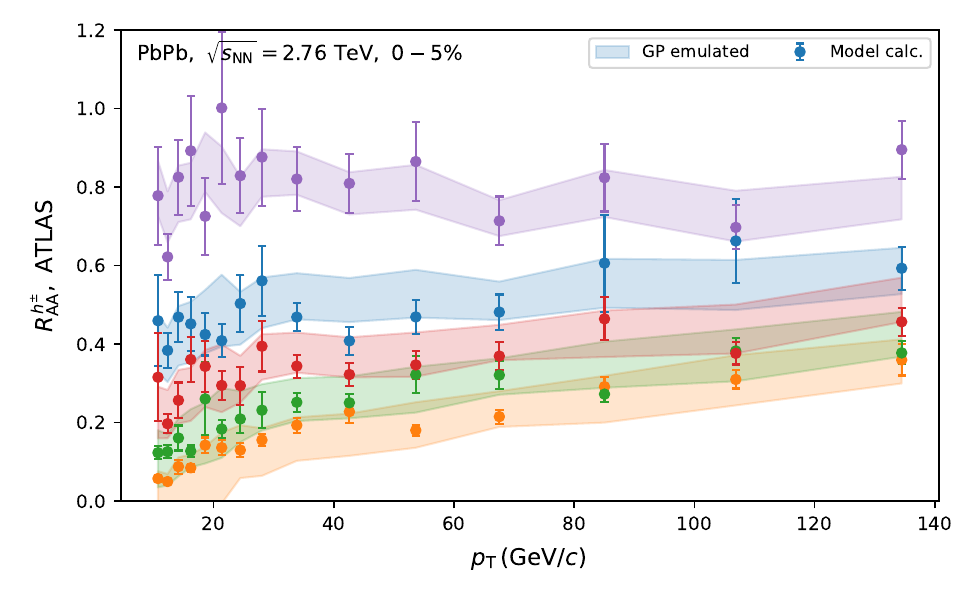}
\end{minipage}

\vspace{0.5em}

\begin{minipage}{0.48\columnwidth}
\centering
\includegraphics[width=\linewidth]{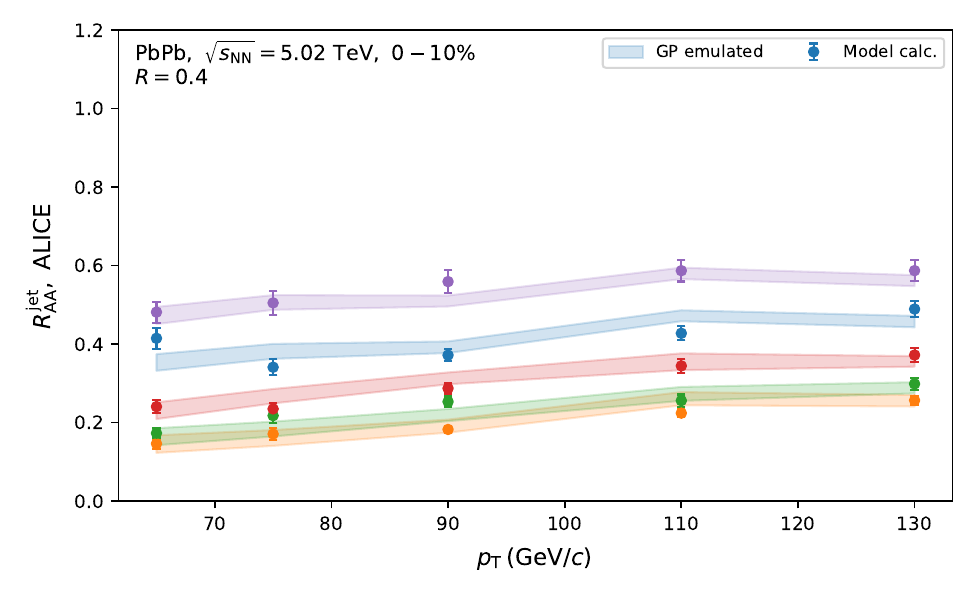}
\end{minipage}
\hfill
\begin{minipage}{0.48\columnwidth}
\centering
\includegraphics[width=\linewidth]{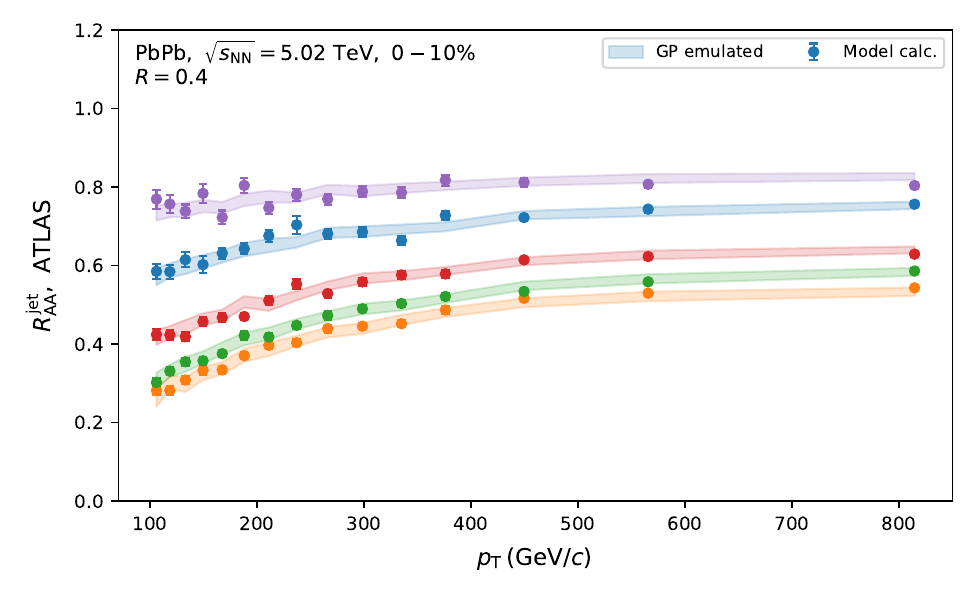}
\end{minipage}
\caption{Emulator validation (Method 1): partial leave-one-out test for the representative observables. The first row shows ALICE and CMS charged-hadron $R_{\mathrm{AA}}$ at $\sqrt{s_{\mathrm{NN}}}=5.02$ TeV in the $0$--$5\%$ centrality class, the second row shows ALICE and ATLAS charged-hadron $R_{\mathrm{AA}}$ at $\sqrt{s_{\mathrm{NN}}}=2.76$ TeV in the $0$--$5\%$ centrality class, and the third row shows ALICE and ATLAS inclusive-jet $R_{\mathrm{AA}}$ at $\sqrt{s_{\mathrm{NN}}}=5.02$ TeV in the $0$--$10\%$ centrality class.}
\label{fig:closure-method1-main}
\end{figure}

\begin{figure}[!tbp]
\centering
\begin{minipage}{0.48\columnwidth}
\centering
\includegraphics[width=\linewidth]{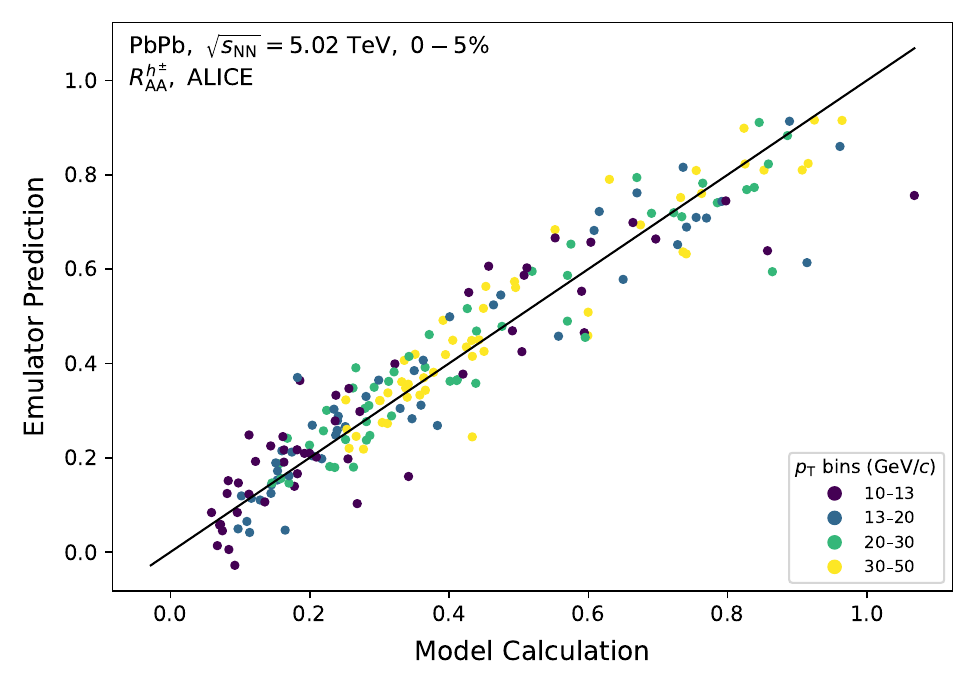}
\end{minipage}
\hfill
\begin{minipage}{0.48\columnwidth}
\centering
\includegraphics[width=\linewidth]{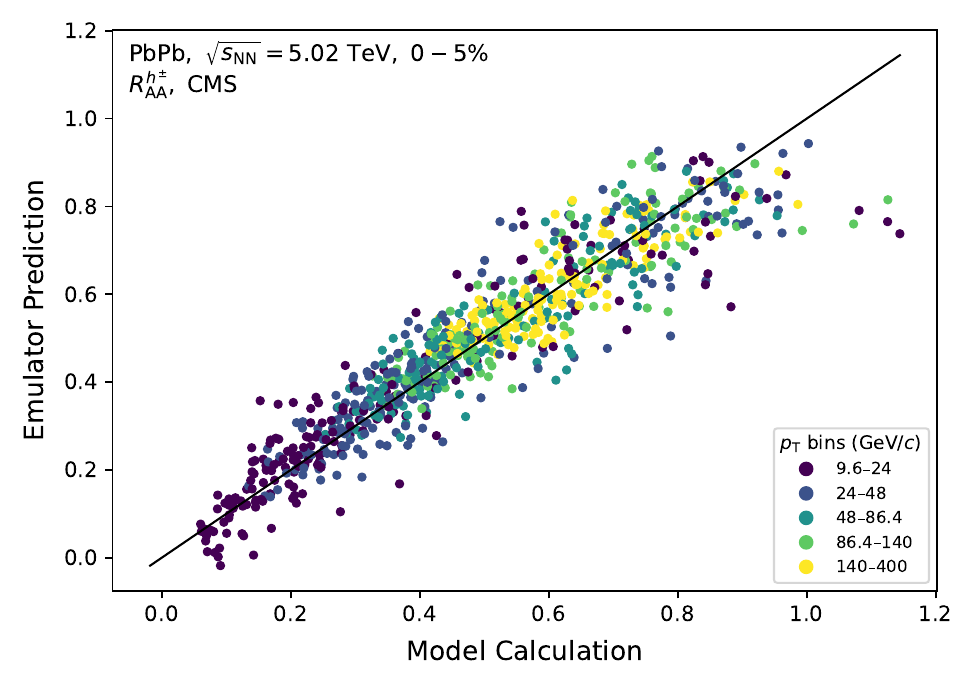}
\end{minipage}

\vspace{0.5em}

\begin{minipage}{0.48\columnwidth}
\centering
\includegraphics[width=\linewidth]{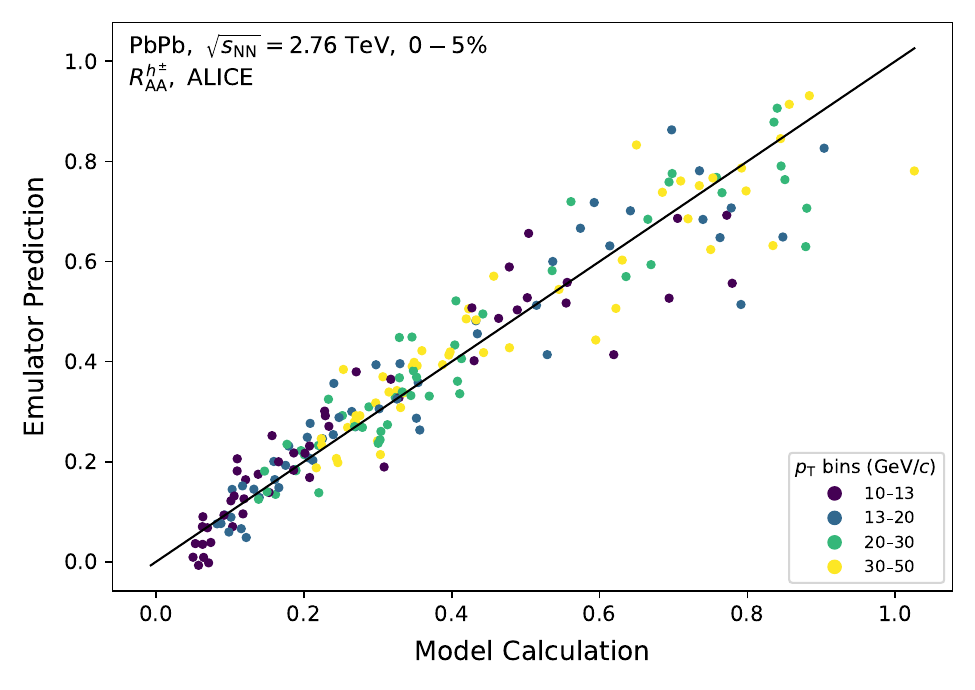}
\end{minipage}
\hfill
\begin{minipage}{0.48\columnwidth}
\centering
\includegraphics[width=\linewidth]{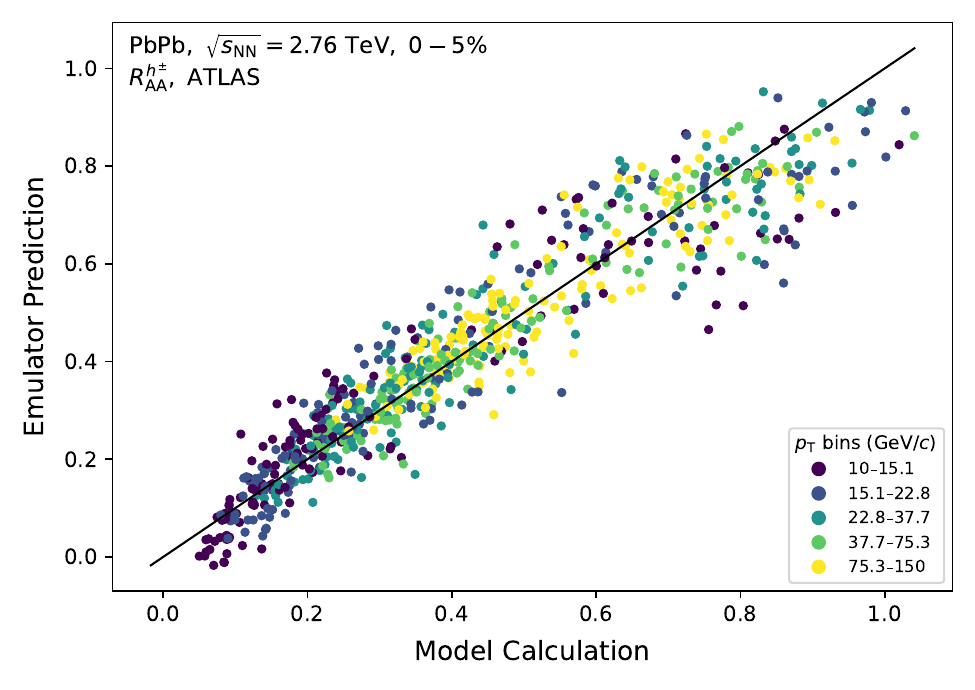}
\end{minipage}

\vspace{0.5em}

\begin{minipage}{0.48\columnwidth}
\centering
\includegraphics[width=\linewidth]{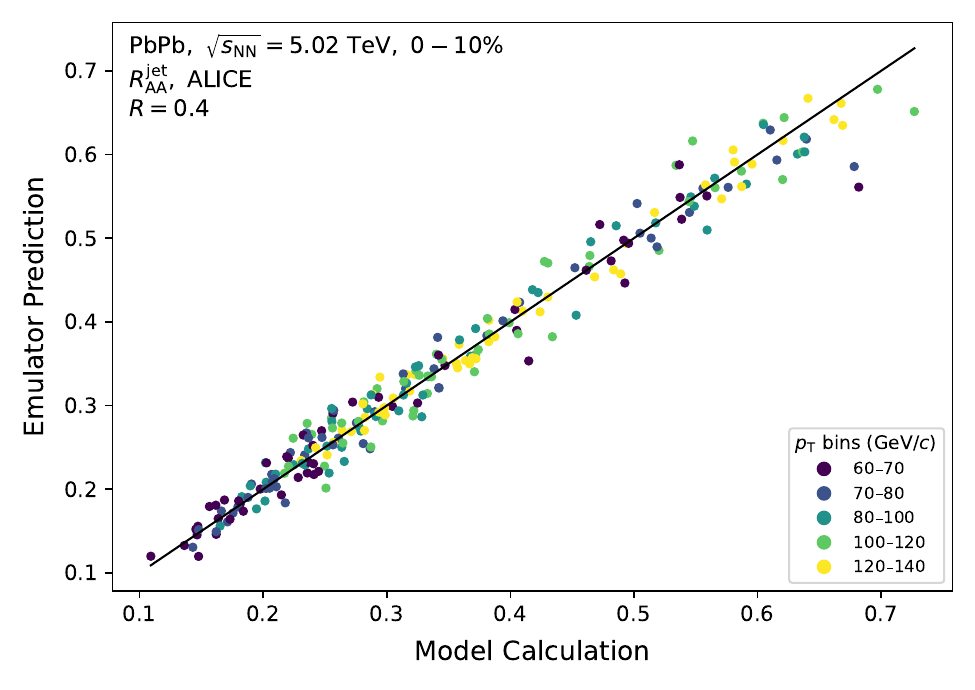}
\end{minipage}
\hfill
\begin{minipage}{0.48\columnwidth}
\centering
\includegraphics[width=\linewidth]{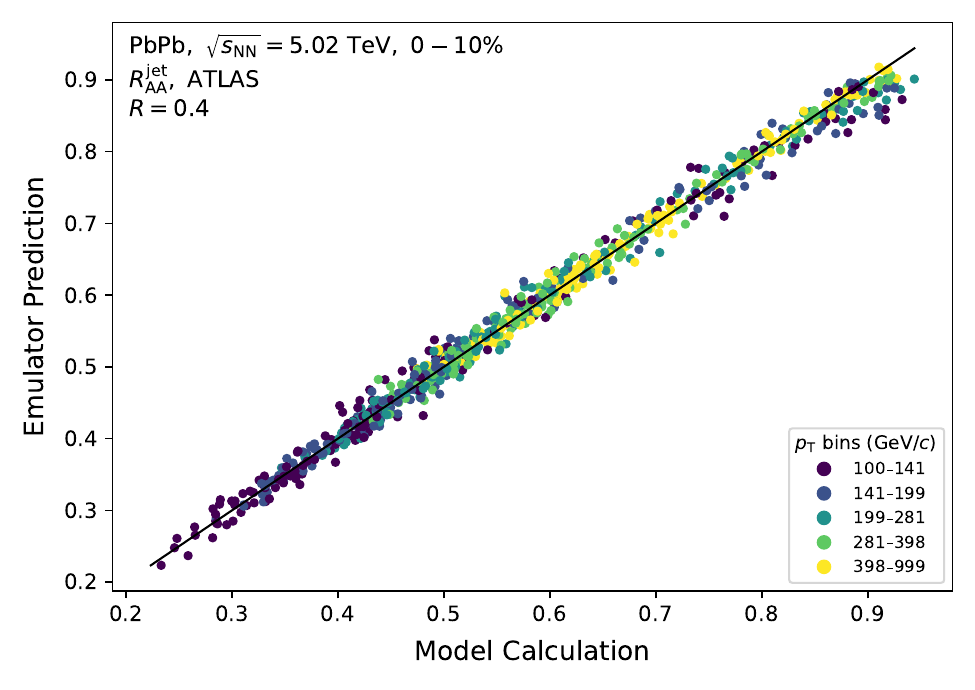}
\end{minipage}

\vspace{0.5em}

\makebox[\columnwidth][c]{%
\begin{minipage}{0.48\columnwidth}
\centering
\includegraphics[width=\linewidth]{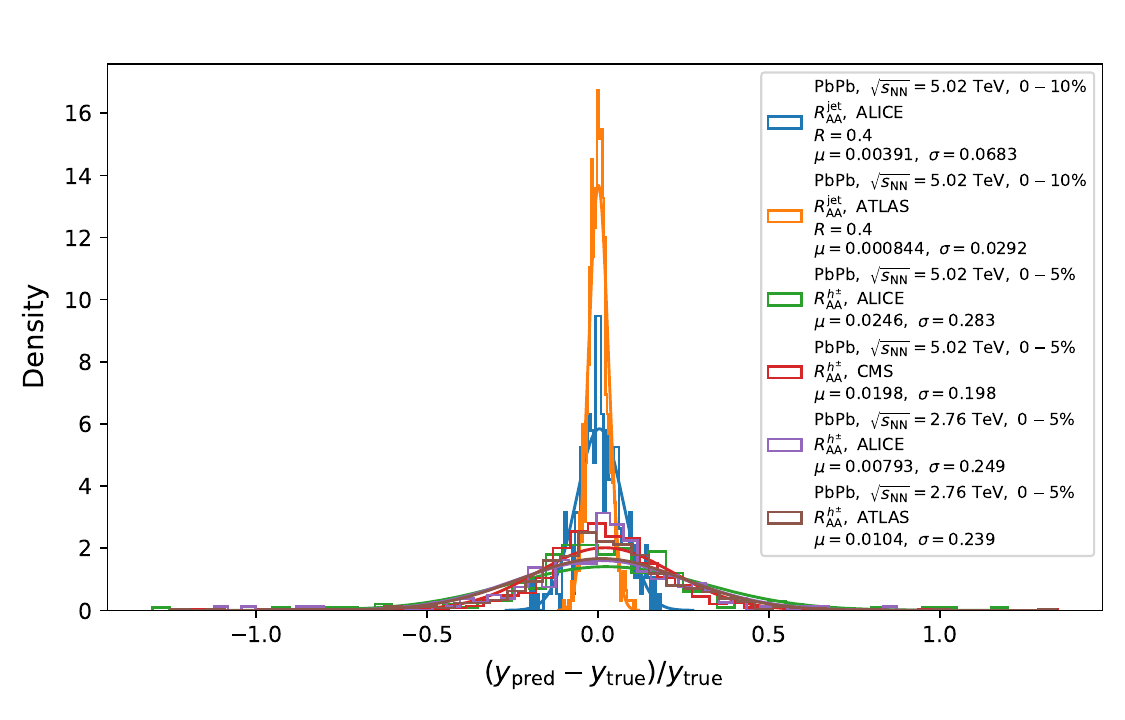}
\end{minipage}%
}
\caption{Emulator validation (Method 2): full-performance check for the representative observables. The first row shows ALICE and CMS charged-hadron $R_{\mathrm{AA}}$ at $\sqrt{s_{\mathrm{NN}}}=5.02$ TeV in the $0$--$5\%$ centrality class, the second row shows ALICE and ATLAS charged-hadron $R_{\mathrm{AA}}$ at $\sqrt{s_{\mathrm{NN}}}=2.76$ TeV in the $0$--$5\%$ centrality class, the third row shows ALICE and ATLAS inclusive-jet $R_{\mathrm{AA}}$ at $\sqrt{s_{\mathrm{NN}}}=5.02$ TeV in the $0$--$10\%$ centrality class, and the last panel shows the fit-overlay residual distributions for these six observables.}
\label{fig:closure-method2-main}
\end{figure}

We validate the GP emulator with complementary leave-one-out and full-design checks. Across the representative observables shown in Figs.~\ref{fig:closure-method1-main} and \ref{fig:closure-method2-main}, the emulator reproduces the underlying model calculations well for both charged-hadron and inclusive-jet $R_{\mathrm{AA}}$, and the fit-overlay residuals show no localized breakdown. Detailed validation procedures are summarized in Appendix~\ref{app:validation-details}, and the full panel set is collected in Appendices~\ref{app:method1-panels} and \ref{app:method2-panels}. Taken together, these diagnostics indicate that the present 50-point design is adequate for stable emulator-based inference at the precision level targeted here, although it does not provide exact local parameter resolution throughout the prior volume.

\subsection{Closure test}
\begin{figure}[!tbp]
\centering
\includegraphics[width=\columnwidth]{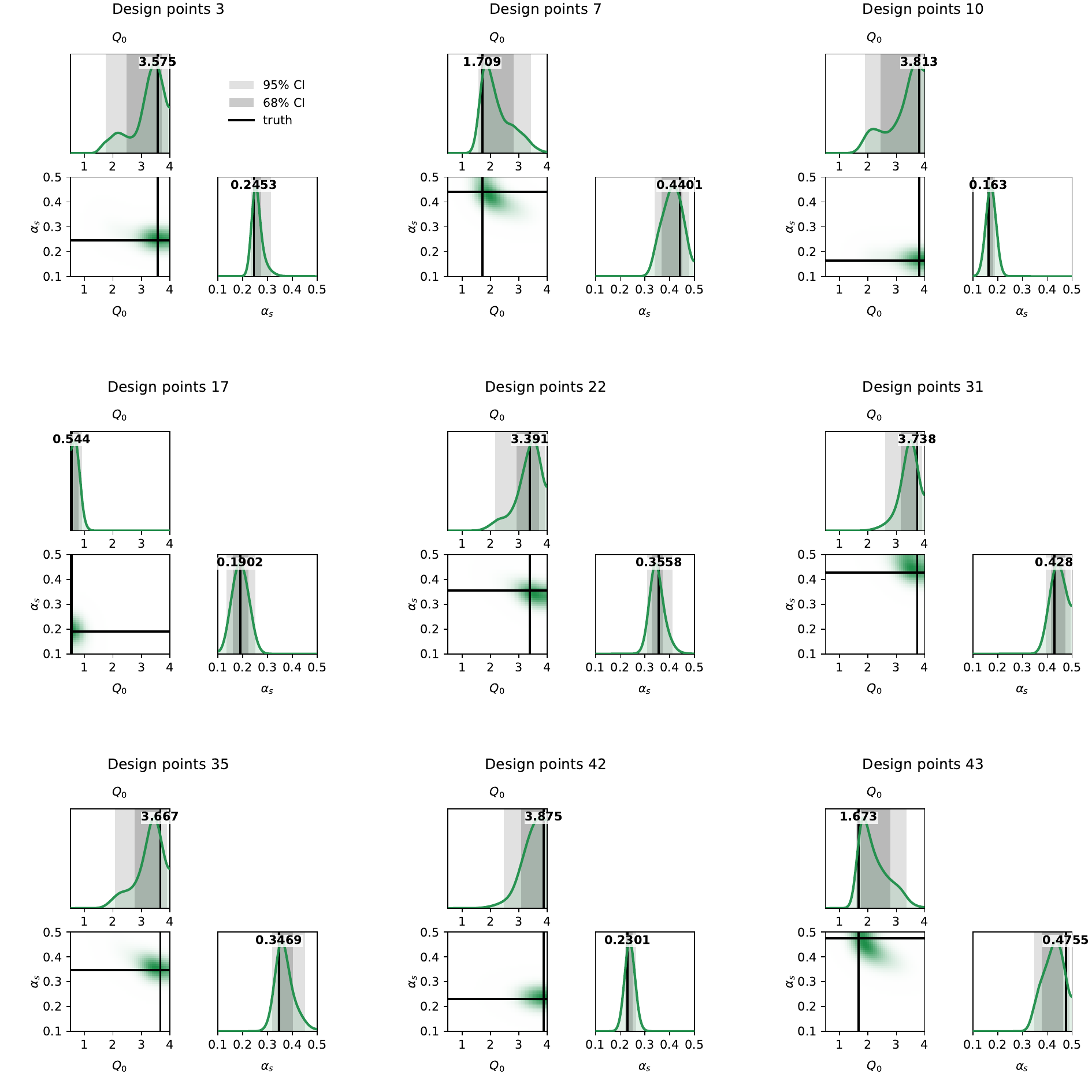}
\caption{Closure-test posterior panel for 9 randomly selected held-out design points in inclusive-jet observables at $0$--$10\%$ centrality. Only $Q_0$ and $\alpha_s$ are displayed; the other parameters remain weakly constrained and close to uniform over the prior ranges.}
\label{fig:closure-posterior-panel}
\end{figure}

Even if the GP emulator predicts observables accurately, this alone does not guarantee that the inferred posterior is well constrained. We therefore perform leave-one-design-point-out closure tests to verify that the emulator-plus-sampling workflow can qualitatively recover held-out pseudo-truth parameters. Fig.~\ref{fig:closure-posterior-panel} summarizes nine representative runs for the $0$--$10\%$ centrality class. The truth values are generally covered by the posterior support for the parameters to which the present dataset is most sensitive, while the weaker parameters remain broadly unconstrained. This is the behavior expected from a design that resolves the dominant quenching scale but not all shape directions. The detailed closure procedure is described in Appendix~\ref{app:validation-details}, and the displayed runs satisfy the chain-quality criteria summarized in Appendix~\ref{app:mcmc-diagnostics}.

With the inference workflow validated, we now turn to the calibration results and their subset-resolved comparison.

\section{RESULTS}
\label{sec:results}

\subsection{All-observables calibration}
\begin{figure}[!htbp]
\centering
\includegraphics[width=\columnwidth]{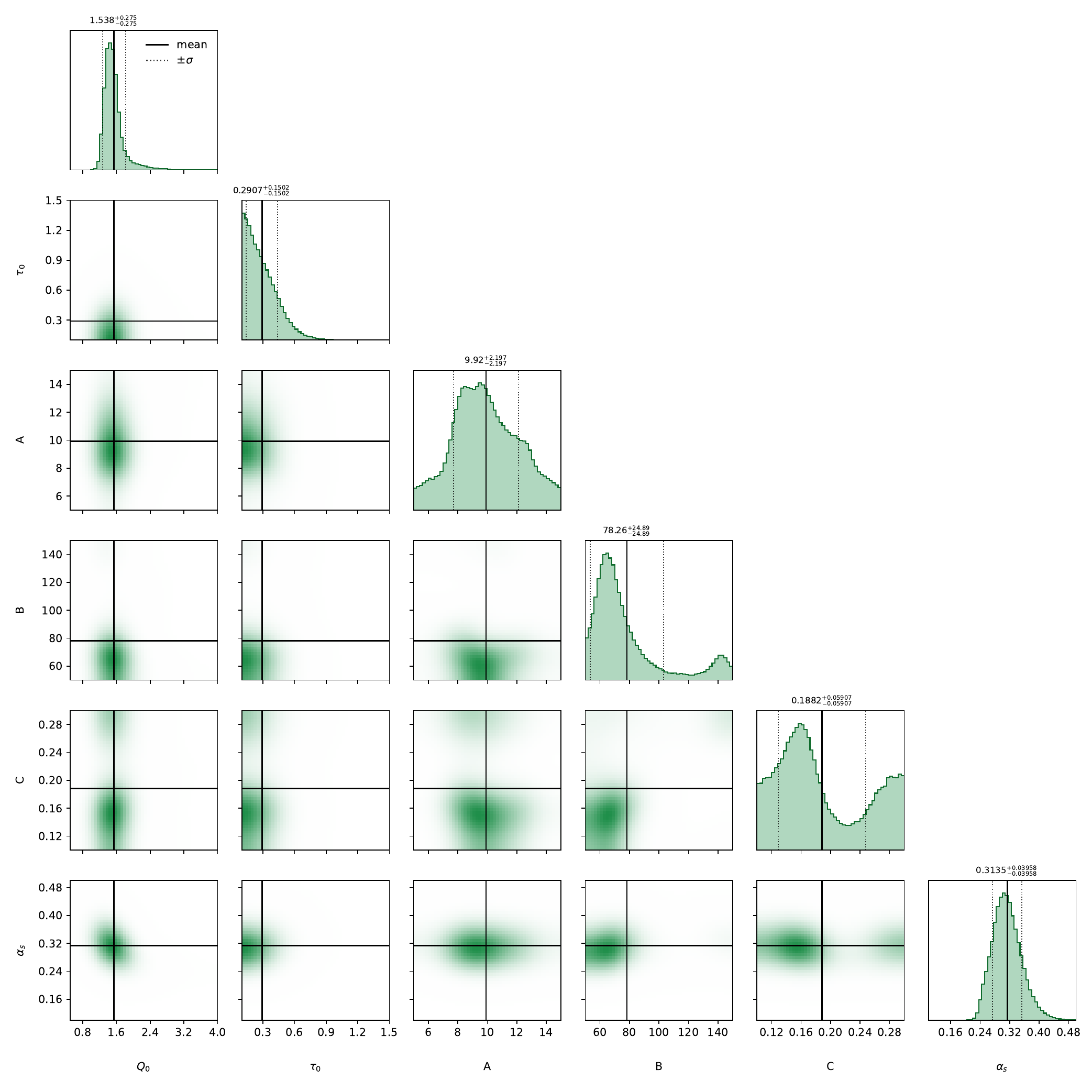}
\caption{Posterior distributions of calibration parameters from Bayesian inference using the full observable set.}
\label{fig:experiment-params-posterior}
\end{figure}

The full observable most clearly set constrains the calibration along the overall quenching-scale directions, as seen in the parameter posterior distributions in Fig.~\ref{fig:experiment-params-posterior}.
Compared with the prior ranges in Table~\ref{tab:calibration-parameters}, the posterior contraction is most visible for $Q_0$ and $\alpha_s$, whose marginalized means and widths are approximately $Q_0=1.54\pm0.27$ and $\alpha_s=0.314\pm0.040$.
These two parameters are therefore the best constrained by the current dataset and exhibit the clearest pairwise structure in the corner plot, with a pronounced anticorrelation.
By contrast, $\tau_0$, $A$, $B$, and $C$ remain broader and more non-Gaussian.
In particular, $\tau_0$ is pulled toward the lower prior boundary, $A$ retains a broad but still visible preference around $A\sim 10$, $B$ shows a dominant low-$B$ mode with a weaker secondary high-$B$ branch, and $C$ is concentrated mainly near $C\sim 0.15$ while retaining additional support toward the upper edge of its prior range.
This weaker constraint is expected because the present calibration remains most sensitive to the global suppression scale, while the current charged-hadron and inclusive-jet datasets have less leverage on the subleading model structure.
These trends are qualitatively consistent with previous Bayesian jet-quenching studies in the JETSCAPE framework~\cite{Fan:2023metric,Ehlers:2024bayesjet}.

\begin{figure}[!htbp]
\centering
\begin{minipage}{\columnwidth}
\centering
\includegraphics[width=\linewidth]{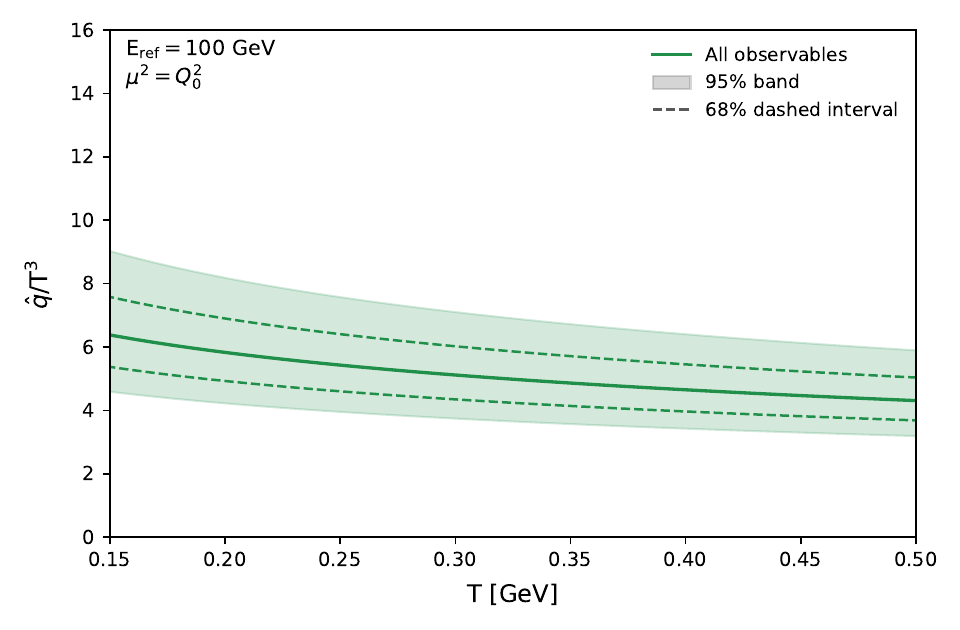}
\end{minipage}
\par\smallskip
\begin{minipage}{\columnwidth}
\centering
\includegraphics[width=\linewidth]{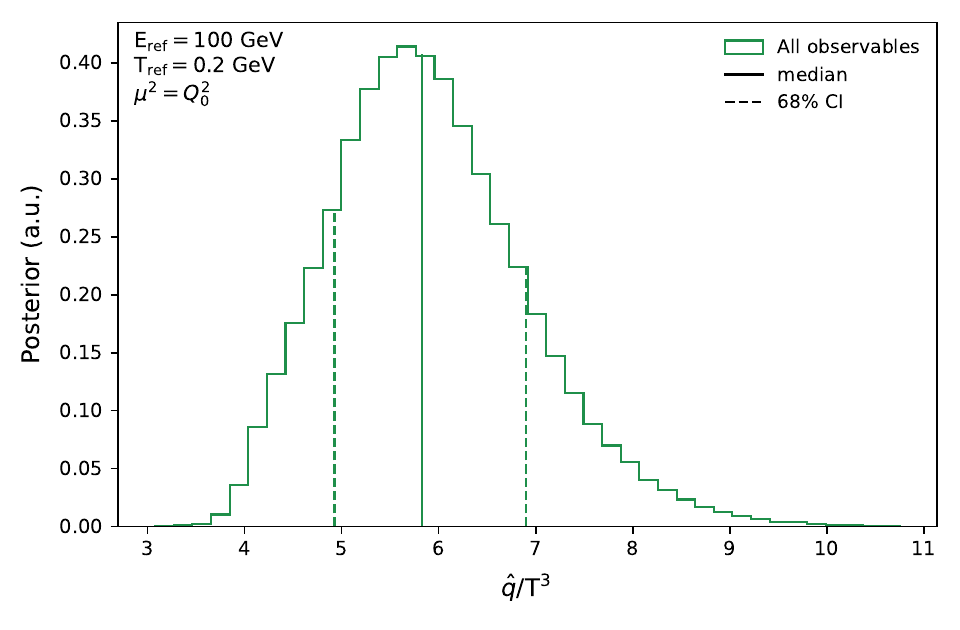}
\end{minipage}
\caption{Baseline posterior estimates of $\hat{q}/\mathrm{T}^3$ from the all-observables calibration. Top: temperature-dependent posterior for the quark case at fixed $\mathrm{E}_{\mathrm{ref}}=100~\mathrm{GeV}$. Bottom: posterior at the common reference point, $\mathrm{E}_{\mathrm{ref}}=100~\mathrm{GeV}$ and $\mathrm{T}_{\mathrm{ref}}=0.2~\mathrm{GeV}$.}
\label{fig:qhat-over-t3-all-observables}
\end{figure}

The same all-observables calibration defines the baseline transport-coefficient scale used for the subset comparisons below, as summarized in Fig.~\ref{fig:qhat-over-t3-all-observables}. At the reference virtuality scale, $Q^2=Q_0^2$, the type-6 prefactor is normalized to the HTL baseline, so the top panel isolates the temperature dependence of the calibrated baseline for the quark case at fixed $\mathrm{E}_{\mathrm{ref}}=100~\mathrm{GeV}$. The mild decrease of $\hat{q}/T^3$ over the displayed temperature range should not be interpreted as a decrease of the absolute transport coefficient in hotter matter; after the explicit $T^3$ scaling is divided out, the remaining trend reflects the running-coupling and Debye-logarithm structure of Eq.~\eqref{eq:qhat-htl-base}. The bottom panel gives the corresponding posterior at the common reference point, $\mathrm{E}_{\mathrm{ref}}=100~\mathrm{GeV}$ and $\mathrm{T}_{\mathrm{ref}}=0.2~\mathrm{GeV}$. This point is used below only as a common coordinate for comparing subset shifts.

In observable space, the same common calibration narrows the prior predictive spread and tracks the measured $R_{\mathrm{AA}}$ trends across the fitted $p_{\mathrm{T}}$ range for representative charged-hadron and inclusive-jet measurements, as shown in Fig.~\ref{fig:experiment-observables-posterior-main}; the remaining posterior-predictive panels are collected in Appendix~\ref{app:posterior-panels}.

At the same time, the residual trends are not identical for hadrons and jets. For the charged-hadron $R_{\mathrm{AA}}$ panels, both the $\sqrt{s_{\mathrm{NN}}}=5.02$ and $2.76$ TeV results show a systematic momentum-dependent preference: at lower $p_{\mathrm{T}}$ the posterior tends to favor slightly less quenching than indicated by the data, while toward higher $p_{\mathrm{T}}$ it prefers somewhat stronger quenching. The inclusive-jet panels exhibit the opposite tendency, with the posterior favoring stronger quenching in the lower-$p_{\mathrm{T}}$ region and becoming progressively closer to the data toward the highest jet $p_{\mathrm{T}}$.

These opposite residual trends do not by themselves establish a separate inconsistency, but they motivate the observable-class tests below. High-$p_{\mathrm{T}}$ charged hadrons preferentially tag jets with a hard leading fragment, thereby emphasizing hard-fragmenting, more quark-dominated, and more surface-biased jet populations, whereas inclusive jets remain sensitive to the broader shower and to energy transported to large angles, including partial recovery inside the jet cone from recoil and medium response~\cite{CasalderreySolana:2019hadjet,Du:2022jettomography,He:2019inclusivejet}. The remaining slope difference in Fig.~\ref{fig:experiment-observables-posterior-main} is therefore a sign that the present effective parametrization captures much of the overall suppression scale while leaving a possible momentum-scale mismatch between the two observable classes~\cite{Fan:2023metric,Ehlers:2024bayesjet}.

Taken together, Figs.~\ref{fig:experiment-params-posterior}, \ref{fig:qhat-over-t3-all-observables}, and \ref{fig:experiment-observables-posterior-main} define the common-calibration baseline preferred by the full dataset within the present model. The subset-restricted tests below ask whether this baseline is genuinely stable across physically distinct parts of the data or instead partly averages over compensating pulls.

\begin{figure}[!htbp]
\centering
\begin{minipage}{0.48\columnwidth}
\centering
\includegraphics[width=\linewidth]{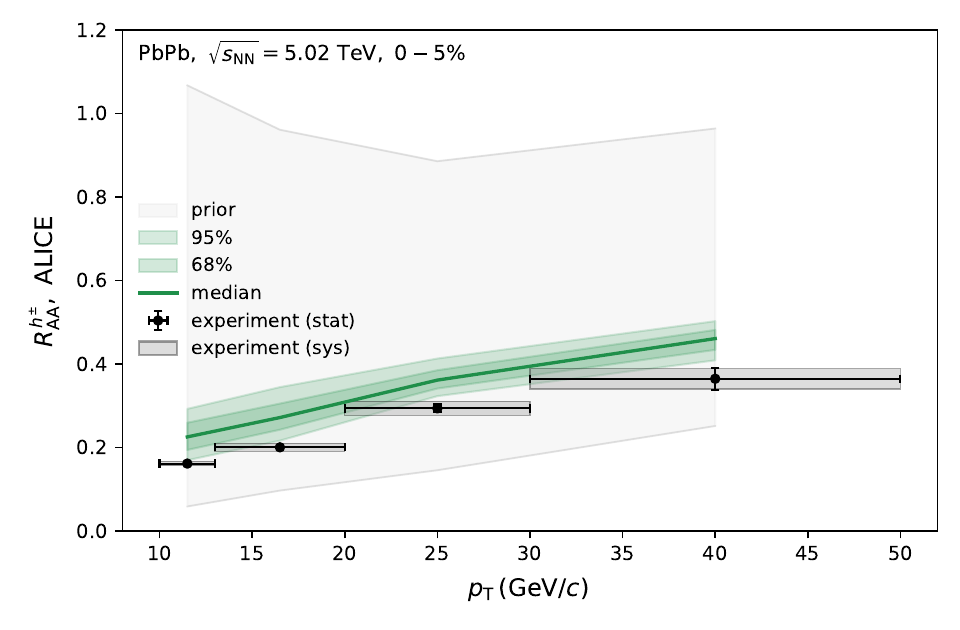}
\end{minipage}
\hfill
\begin{minipage}{0.48\columnwidth}
\centering
\includegraphics[width=\linewidth]{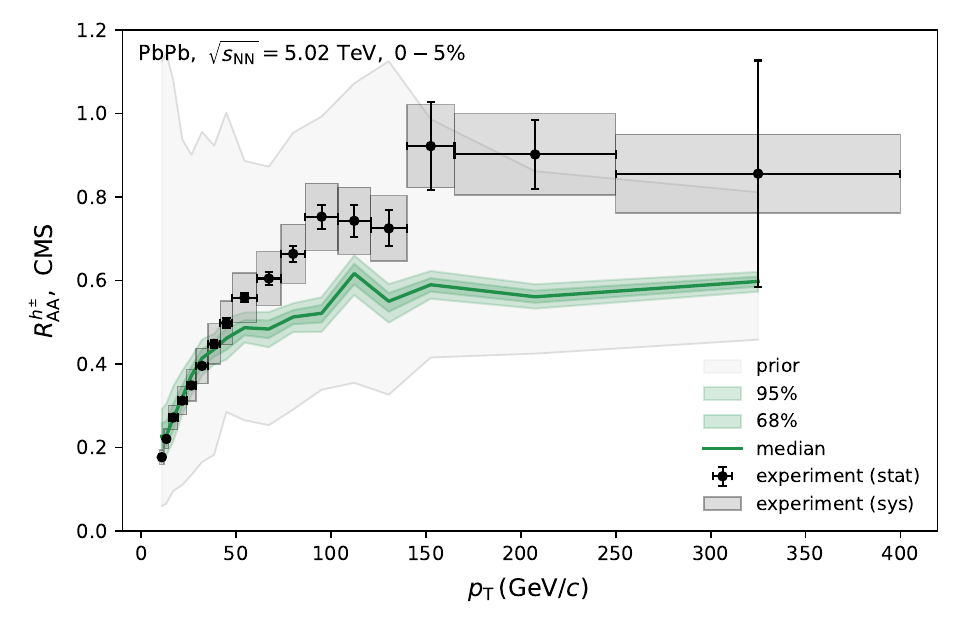}
\end{minipage}

\vspace{0.5em}

\begin{minipage}{0.48\columnwidth}
\centering
\includegraphics[width=\linewidth]{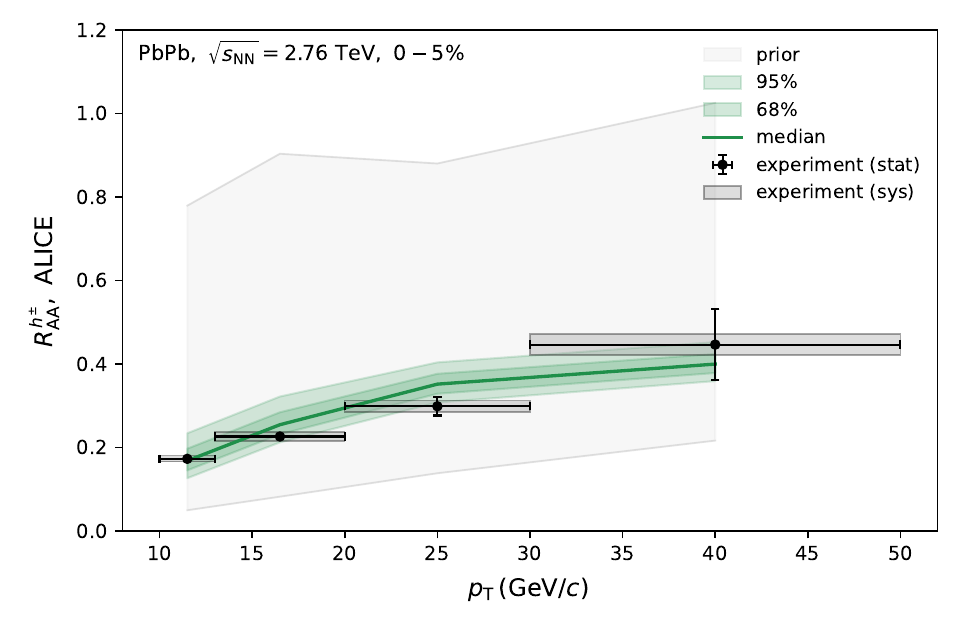}
\end{minipage}
\hfill
\begin{minipage}{0.48\columnwidth}
\centering
\includegraphics[width=\linewidth]{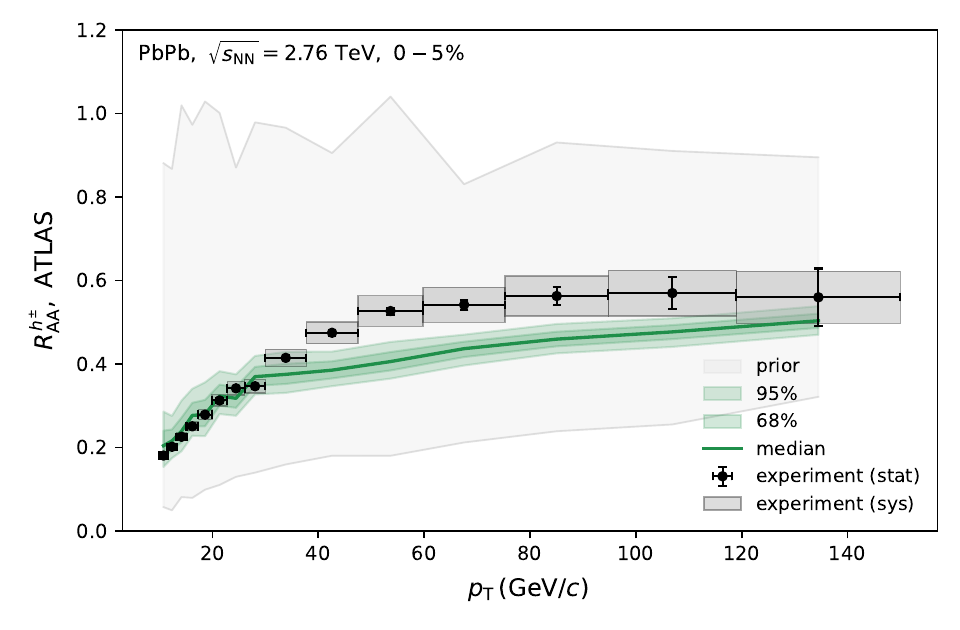}
\end{minipage}

\vspace{0.5em}

\begin{minipage}{0.48\columnwidth}
\centering
\includegraphics[width=\linewidth]{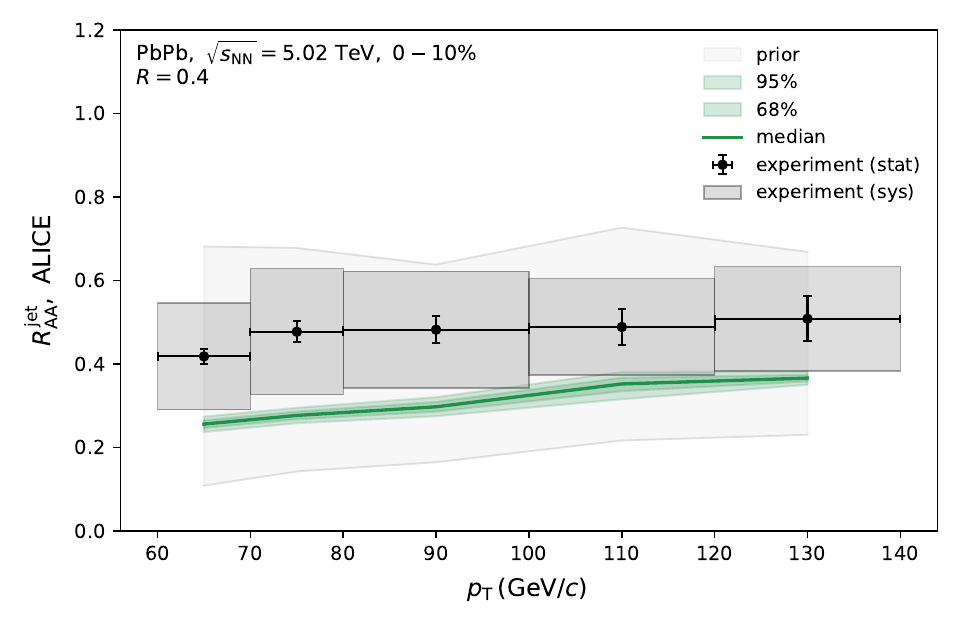}
\end{minipage}
\hfill
\begin{minipage}{0.48\columnwidth}
\centering
\includegraphics[width=\linewidth]{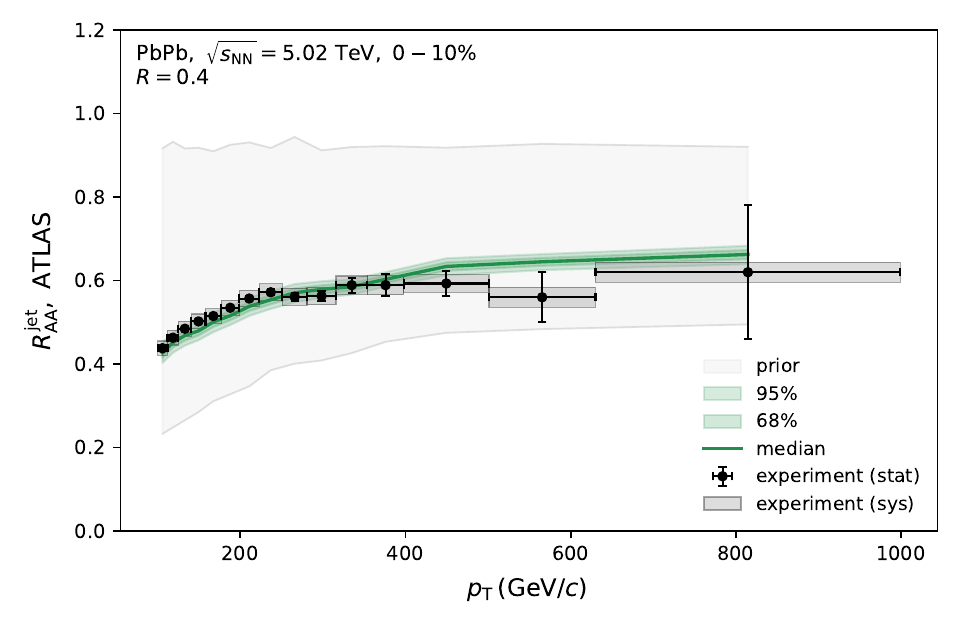}
\end{minipage}
\caption{Representative posterior predictions for the fitted observables. The first row shows charged-hadron $R_{\mathrm{AA}}$ at $\sqrt{s_{\mathrm{NN}}}=5.02$ TeV for ALICE and CMS in the $0$--$5\%$ centrality class. The second row shows charged-hadron $R_{\mathrm{AA}}$ at $\sqrt{s_{\mathrm{NN}}}=2.76$ TeV for ALICE and ATLAS in the $0$--$5\%$ centrality class. The third row shows inclusive-jet $R_{\mathrm{AA}}$ at $\sqrt{s_{\mathrm{NN}}}=5.02$ TeV for ALICE and ATLAS in the $0$--$10\%$ centrality class.}
\label{fig:experiment-observables-posterior-main}
\end{figure}

\subsection{Controlled subset decomposition}
\label{sec:subset-decomposition}
A successful global fit does not by itself show whether the all-observables baseline is subset-stable or an effective compromise among different data classes. We therefore decompose the calibration along three controlled axes. The centrality split tests geometry and path-length dependence; the beam-energy split tests transferability across different temperature histories and partonic spectra; and the observable-class split tests whether hadron-biased and inclusive-jet measurements prefer the same effective quenching scale. In all three comparisons, the model, emulator, priors, and reference point used for posterior comparison are kept fixed; only the calibration subset is changed. This makes the comparisons parameter-space stress tests of the common baseline rather than independent model variants.

The following posterior comparisons evaluate $\hat{q}/\mathrm{T}^3$ for a quark at a common reference point, $\mathrm{E}_{\mathrm{ref}}=100~\mathrm{GeV}$ and $\mathrm{T}_{\mathrm{ref}}=0.2~\mathrm{GeV}$. These reference values are used only as common coordinates for comparing posterior shifts. They should not be interpreted as an event-averaged medium state.

\begin{figure}[!htbp]
\centering
\begin{minipage}{\columnwidth}
\centering
\includegraphics[width=\linewidth]{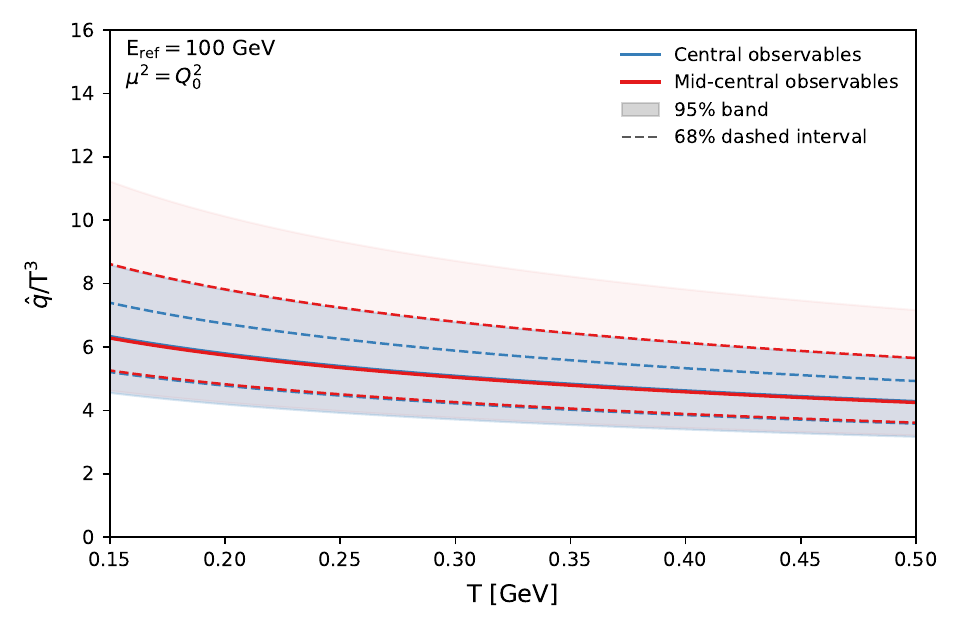}
\end{minipage}
\par\smallskip
\begin{minipage}{\columnwidth}
\centering
\includegraphics[width=\linewidth]{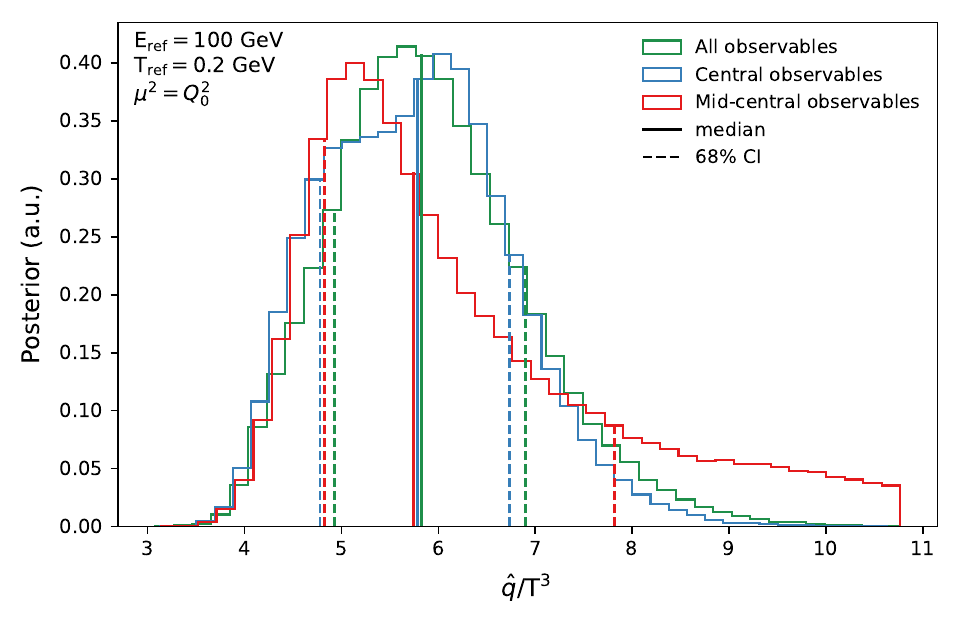}
\end{minipage}
\caption{Posterior estimates of $\hat{q}/\mathrm{T}^3$ for the central and mid-central comparisons. Top: temperature-dependent posterior with $\mathrm{E}_{\mathrm{ref}}=100~\mathrm{GeV}$, shown only for the central and mid-central calibrations for visual clarity. Bottom: posterior at the common reference point, $\mathrm{E}_{\mathrm{ref}}=100~\mathrm{GeV}$ and $\mathrm{T}_{\mathrm{ref}}=0.2~\mathrm{GeV}$, for the all-observable, central, and mid-central calibrations.}
\label{fig:qhat-over-t3-centralmidcentral}
\end{figure}

The centrality split gives the most stable parameter-space comparison among the subset decompositions, with central and mid-central posteriors strongly overlapping in Fig.~\ref{fig:qhat-over-t3-centralmidcentral}. Their median trends remain close over the displayed temperature range. The central subset gives the narrower constraint, while the mid-central subset keeps a broader high-$\hat{q}$ tail. This behavior is consistent with the stronger quenching leverage of central events and the larger residual degeneracy between path length and transport strength in mid-central events. Importantly, the comparison does not select two separated transport-coefficient branches. Within the current uncertainties, the centrality split is therefore consistent with an approximately common $\hat{q}/\mathrm{T}^3$ across these geometry classes, although this parameter-space overlap still needs to be tested in observable space.

\begin{figure}[!htbp]
\centering
\begin{minipage}{\columnwidth}
\centering
\includegraphics[width=\linewidth]{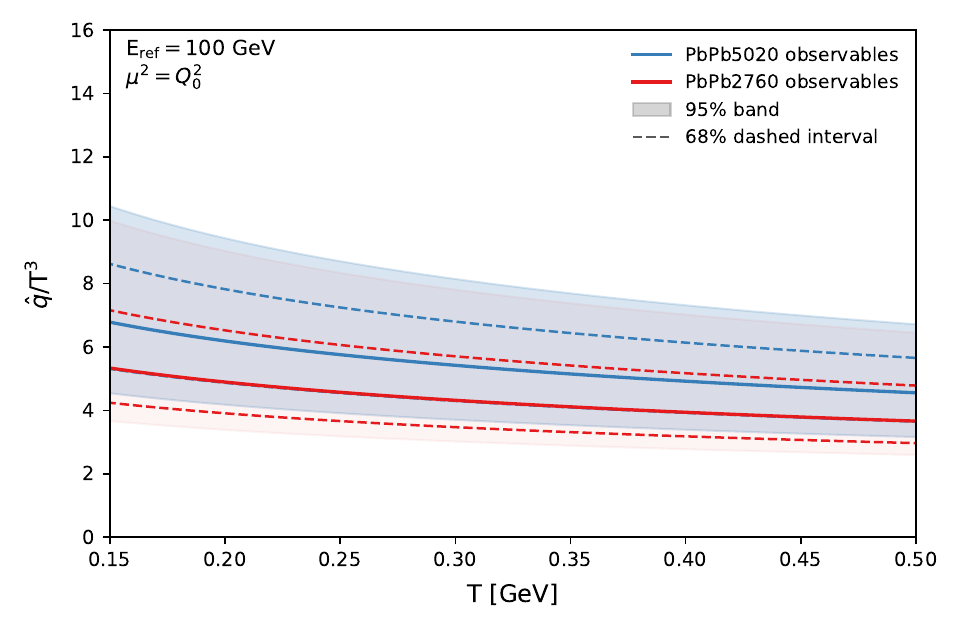}
\end{minipage}
\par\smallskip
\begin{minipage}{\columnwidth}
\centering
\includegraphics[width=\linewidth]{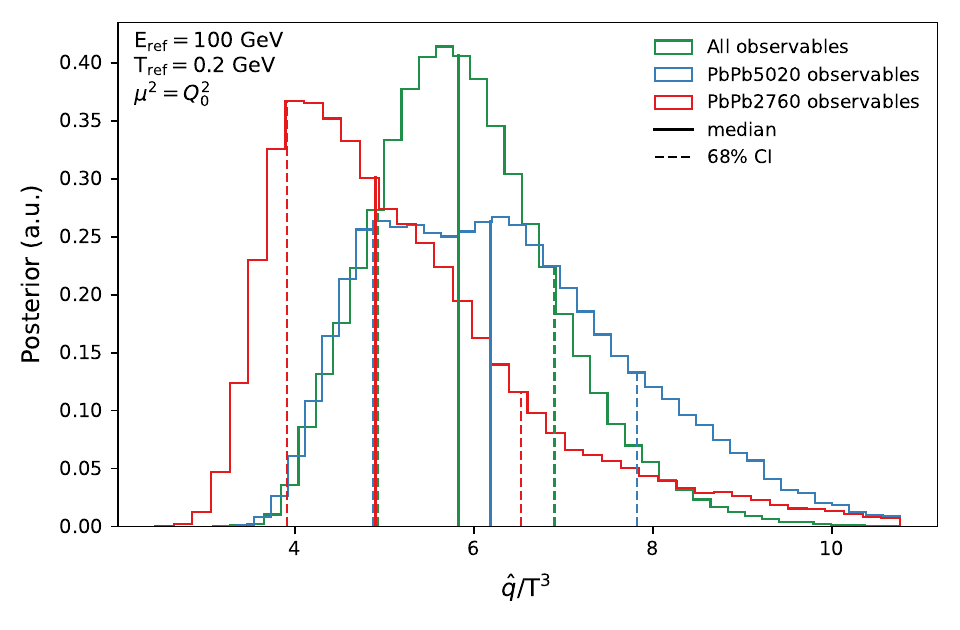}
\end{minipage}
\caption{Posterior estimates of $\hat{q}/\mathrm{T}^3$ for the beam-energy comparisons. Top: temperature-dependent posterior with $\mathrm{E}_{\mathrm{ref}}=100~\mathrm{GeV}$, shown only for the PbPb 5.02 TeV and PbPb 2.76 TeV calibrations for visual clarity. Bottom: posterior at the common reference point, $\mathrm{E}_{\mathrm{ref}}=100~\mathrm{GeV}$ and $\mathrm{T}_{\mathrm{ref}}=0.2~\mathrm{GeV}$, for the all-observable, PbPb 5.02 TeV, and PbPb 2.76 TeV calibrations.}
\label{fig:qhat-over-t3-pbpb5020pbpb2760}
\end{figure}

The beam-energy split produces a more visible transport-coefficient displacement than the centrality split, as shown in Fig.~\ref{fig:qhat-over-t3-pbpb5020pbpb2760}. The PbPb $5.02$ TeV subset favors the higher-$\hat{q}$ side, whereas the PbPb $2.76$ TeV subset favors lower values. The credible regions still overlap, so this is not a direct inconsistency. However, the shift indicates that using a single effective transport coefficient across beam energies is more restrictive than using one across centrality classes. Since this comparison changes both the temperature history and the hard-probe spectral environment, it is a natural place for residual energy- or scale-dependence in the present parametrization to appear.

\begin{figure}[!htbp]
\centering
\begin{minipage}{\columnwidth}
\centering
\includegraphics[width=\linewidth]{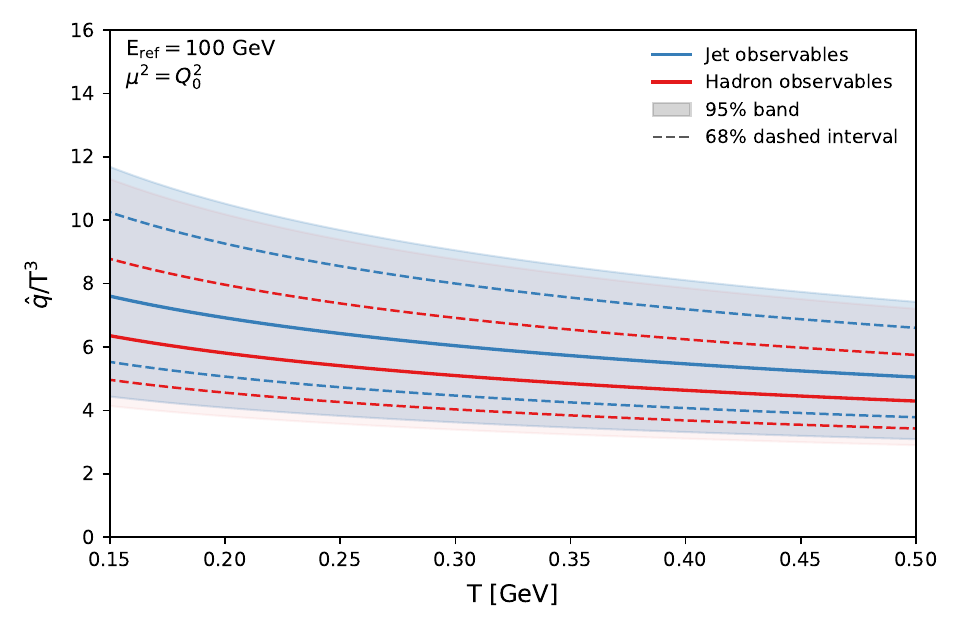}
\end{minipage}
\par\smallskip
\begin{minipage}{\columnwidth}
\centering
\includegraphics[width=\linewidth]{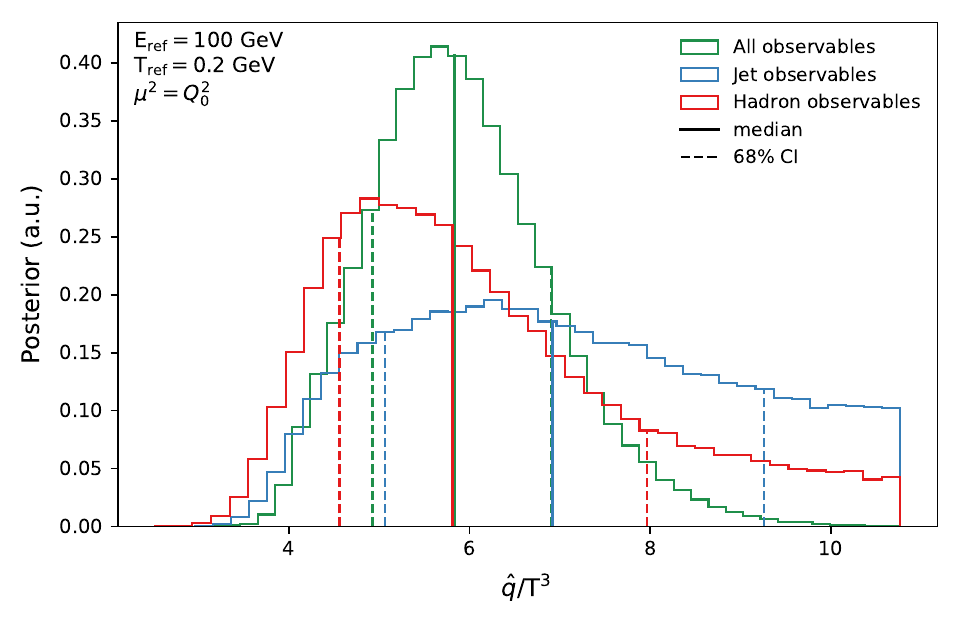}
\end{minipage}
\caption{Posterior estimates of $\hat{q}/\mathrm{T}^3$ for the observable-class comparison. Top: temperature-dependent posterior with $\mathrm{E}_{\mathrm{ref}}=100~\mathrm{GeV}$ for jet-only and hadron-only calibrations. Bottom: posterior at the common reference point, $\mathrm{E}_{\mathrm{ref}}=100~\mathrm{GeV}$ and $\mathrm{T}_{\mathrm{ref}}=0.2~\mathrm{GeV}$, for the all-observable, jet-only, and hadron-only calibrations.}
\label{fig:qhat-over-t3-jet-hadron}
\end{figure}

The observable-class split reveals a residual hadron--jet direction within a common posterior support, as shown in Fig.~\ref{fig:qhat-over-t3-jet-hadron}. The two observable classes overlap substantially, but they do not pull the posterior in exactly the same way. The jet-only calibration favors a somewhat larger $\hat{q}/\mathrm{T}^3$ and has broader high-$\hat{q}$ support, whereas the hadron-only calibration is more concentrated toward lower values. The all-observables posterior lies in the common support of the two subsets rather than forming an independent branch, so we treat this split as a residual direction within a common solution rather than as evidence for two incompatible solutions.

This observable-class separation is physically important because charged hadrons and reconstructed jets do not sample the same part of the in-medium shower. High-$p_{\mathrm{T}}$ charged hadrons emphasize events with a hard leading fragment and stronger surface bias, while inclusive jets retain sensitivity to the wider shower energy distribution, including out-of-cone transport and partial in-cone recovery from medium response~\cite{CasalderreySolana:2019hadjet,Du:2022jettomography,He:2019inclusivejet}. This provides a physical reason why a common suppression scale can still leave a hadron--jet direction in parameter space.

Table~\ref{tab:subset-decomposition-summary} summarizes the same three decompositions at the common reference point used in the bottom panels of Figs.~\ref{fig:qhat-over-t3-centralmidcentral}--\ref{fig:qhat-over-t3-jet-hadron}.
The signed normalized shift is defined as $\Delta\sigma_{68}=(m_2-m_1)/\sqrt{\sigma_{68,1}^2+\sigma_{68,2}^2}$, where $m_i$ is the posterior median and $\sigma_{68,i}=(q_{84,i}-q_{16,i})/2$.
Thus $|\Delta\sigma_{68}|$ measures the median separation in units of the combined $68\%$ posterior width, while $\mathrm{OVL}$ denotes the histogram overlap of the two one-dimensional posteriors.

The common-reference metrics reported in Table~\ref{tab:subset-decomposition-summary} make the visual hierarchy more explicit. The centrality split is nearly centered on the same value, with $|\Delta\sigma_{68}|\simeq 0.02$ and a large posterior overlap. The beam-energy split gives the largest displacement, with the PbPb $5.02$ TeV subset shifted upward relative to the PbPb $2.76$ TeV subset. The hadron--jet split is intermediate in this common-reference summary: the overlap remains sizable, but the jet-only calibration prefers a higher median and a broader high-$\hat{q}$ tail. These are parameter-space statements only. They identify centrality, beam energy, and observable class as the relevant stress-test axes, but they do not by themselves determine whether the subset shifts lead to observable prediction loss. That question is the purpose of the cross-prediction tests that follow.

\begin{center}
\captionof{table}{Quantitative summary of the controlled subset decomposition at $\mathrm{E}_{\mathrm{ref}}=100~\mathrm{GeV}$ and $\mathrm{T}_{\mathrm{ref}}=0.2~\mathrm{GeV}$ for the quark $\hat{q}/\mathrm{T}^3$ posterior. The posterior entries are reported as the median $[q_{16},q_{84}]$. The sign of $\Delta\sigma_{68}$ corresponds to the second subset minus the first subset.}
\label{tab:subset-decomposition-summary}
\scriptsize
\setlength{\tabcolsep}{3pt}
\begin{ruledtabular}
\begin{tabular}{llcc}
Axis & Subset posteriors & $\Delta\sigma_{68}$ & $\mathrm{OVL}$ \\
\hline
Centrality
& \begin{tabular}[t]{@{}l@{}}central: $5.78\,[4.78,6.73]$\\mid-central: $5.74\,[4.83,7.82]$\end{tabular}
& $-0.02$
& $0.81$ \\
Beam energy
& \begin{tabular}[t]{@{}l@{}}PbPb $5.02$ TeV: $6.19\,[4.87,7.82]$\\PbPb $2.76$ TeV: $4.89\,[3.91,6.52]$\end{tabular}
& $-0.66$
& $0.66$ \\
Observable class
& \begin{tabular}[t]{@{}l@{}}jet-only: $6.92\,[5.07,9.26]$\\hadron-only: $5.81\,[4.56,7.96]$\end{tabular}
& $-0.41$
& $0.77$ \\
\end{tabular}
\end{ruledtabular}
\end{center}

\section{DISCUSSION}
\label{sec:discussion}
\subsection{Cross-prediction tests}
Even when subset posteriors overlap, predictive transferability is not guaranteed. We therefore test directly whether the subset-posterior differences found above lead to observable-space prediction loss.

The test follows the same three axes used in Sec.~\ref{sec:subset-decomposition}, but it is organized as six directed transfers: central $\to$ mid-central, mid-central $\to$ central, PbPb $5.02\to2.76$ TeV, PbPb $2.76\to5.02$ TeV, hadron $\to$ jet, and jet $\to$ hadron. The posterior decomposition gives a simple hierarchy: centrality-selected posteriors almost overlap, beam-energy-separated posteriors show the largest shift, and the observable-class split moves the jet-only posterior toward larger $\hat{q}/T^3$ than the hadron-only posterior. Cross-prediction then asks whether these parameter-space shifts remain small in observable space or become visible transfer degradation when each source posterior is propagated without refitting.

To test transferability directly, we propagate each source-subset posterior, without refitting, to the observables of a target subset. We then compare the transferred prediction with the posterior predictive distribution from a dedicated fit to that target subset. For compact bin-by-bin comparison, we use two diagnostics derived from the cross-prediction summary tables. The first is the overlap fraction between the transferred and target-fit predictive distributions, denoted OVL. The second is the separation of their medians in units of the combined $68\%$ width, denoted $\Delta\sigma_{68}$.
For each matched bin $i$, let $p_{\mathrm{tr}}^{(i)}(y)$ and $p_{\mathrm{fit}}^{(i)}(y)$ denote the transferred and target-fit posterior predictive densities, with medians $m_{\mathrm{tr}}^{(i)}$ and $m_{\mathrm{fit}}^{(i)}$.
Defining the half-widths of the central $68\%$ intervals by $\sigma_{68,\mathrm{tr}}^{(i)} \equiv \tfrac{1}{2}\!\left(q_{84,\mathrm{tr}}^{(i)} - q_{16,\mathrm{tr}}^{(i)}\right)$ and $\sigma_{68,\mathrm{fit}}^{(i)} \equiv \tfrac{1}{2}\!\left(q_{84,\mathrm{fit}}^{(i)} - q_{16,\mathrm{fit}}^{(i)}\right)$, we use
\begin{align}
\mathrm{OVL}^{(i)} &\equiv \int dy\, \min\!\left[p_{\mathrm{tr}}^{(i)}(y),\, p_{\mathrm{fit}}^{(i)}(y)\right], \\
\Delta\sigma_{68}^{(i)} &\equiv \frac{\left|m_{\mathrm{tr}}^{(i)} - m_{\mathrm{fit}}^{(i)}\right|}{\sqrt{\left(\sigma_{68,\mathrm{tr}}^{(i)}\right)^2 + \left(\sigma_{68,\mathrm{fit}}^{(i)}\right)^2}} .
\end{align}
In practice, $\mathrm{OVL}^{(i)}$ is estimated numerically from histogram densities constructed from the posterior predictive samples. Thus $\mathrm{OVL}=1$ for identical predictive distributions and approaches zero as the two distributions separate. Likewise, $\Delta\sigma_{68}\lesssim 1$ indicates that the medians remain within roughly one quadrature-combined $68\%$ scale.

The observable-level cross-prediction summaries first report these bin-level quantities for each target observable. Table~\ref{tab:cross-prediction-summary} then combines the observables belonging to each directed transfer with the number of matched bins as weights. In the table, $N_{\mathrm{bin}}$ is the total number of matched target bins entering that transfer. The columns labeled $\mathrm{OVL}$ and $\Delta\sigma_{68}$ are the corresponding $N_{\mathrm{bin}}$-weighted averages of the observable-level median overlap and median separation. The columns $f_{\Delta<1}$, $f_{\mathrm{OVL}>0.5}$, and $f_{68>0}$ are the weighted fractions of matched bins satisfying $\Delta\sigma_{68}<1$, $\mathrm{OVL}>0.5$, and nonzero overlap of the transferred and target-fit $68\%$ intervals, respectively. Transferability is stronger when the overlap remains sizable, when $\Delta\sigma_{68}\lesssim 1$, and when the supporting fractions remain large.

\begin{widetext}
\begin{center}
\captionof{table}{Bin-weighted cross-prediction summary. $N_{\mathrm{bin}}$ is the total number of matched target bins. $\mathrm{OVL}$ and $\Delta\sigma_{68}$ are $N_{\mathrm{bin}}$-weighted averages of the observable-level median predictive overlap and median separation. The last three columns give weighted fractions satisfying $\Delta\sigma_{68}<1$, $\mathrm{OVL}>0.5$, and nonzero $68\%$-interval overlap.}
\label{tab:cross-prediction-summary}
\scriptsize
\setlength{\tabcolsep}{5pt}
\begin{ruledtabular}
\begin{tabular*}{0.96\textwidth}{@{\extracolsep{\fill}}llrrrrrr}
Axis & Source $\to$ target & $N_{\mathrm{bin}}$ & $\mathrm{OVL}$ & $\Delta\sigma_{68}$ & $f_{\Delta<1}$ & $f_{\mathrm{OVL}>0.5}$ & $f_{68>0}$ \\
\hline
Centrality & central $\to$ mid-central & 64 & 0.34 & 1.34 & 20\% & 16\% & 48\% \\
Centrality & mid-central $\to$ central & 86 & 0.45 & 0.85 & 79\% & 28\% & 100\% \\
Beam energy & PbPb $5.02\to2.76$ TeV & 70 & 0.42 & 1.14 & 36\% & 29\% & 74\% \\
Beam energy & PbPb $2.76\to5.02$ TeV & 80 & 0.45 & 1.04 & 54\% & 38\% & 79\% \\
Observable class & hadron $\to$ jet & 59 & 0.63 & 0.52 & 98\% & 86\% & 100\% \\
Observable class & jet $\to$ hadron & 91 & 0.65 & 0.45 & 100\% & 100\% & 100\% \\
\end{tabular*}
\end{ruledtabular}
\end{center}
\end{widetext}

\begin{figure}[!htbp]
\centering
\begin{minipage}{\columnwidth}
\centering
\includegraphics[width=\linewidth]{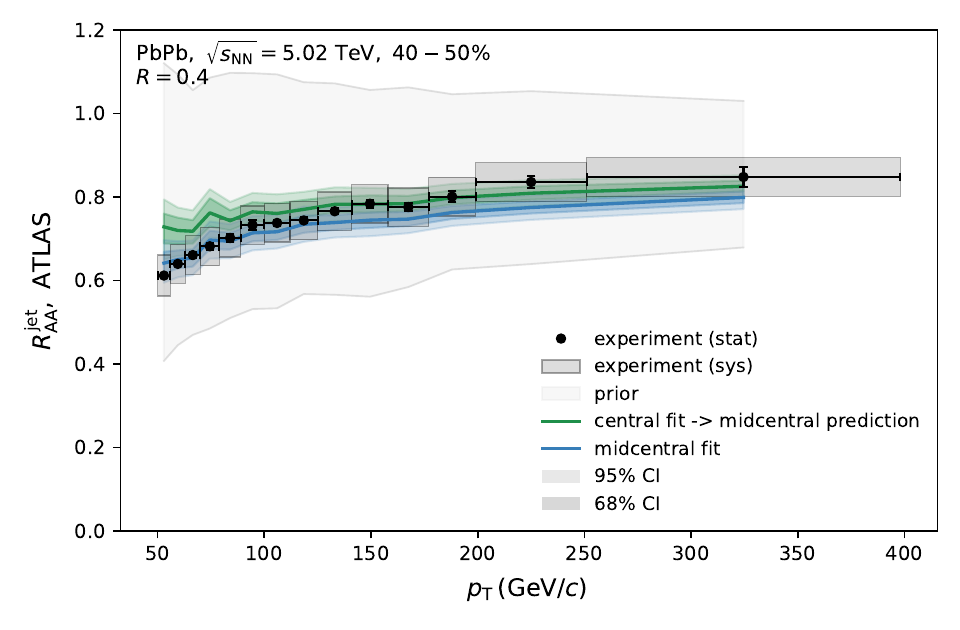}
\end{minipage}
\par\smallskip
\begin{minipage}{\columnwidth}
\centering
\includegraphics[width=\linewidth]{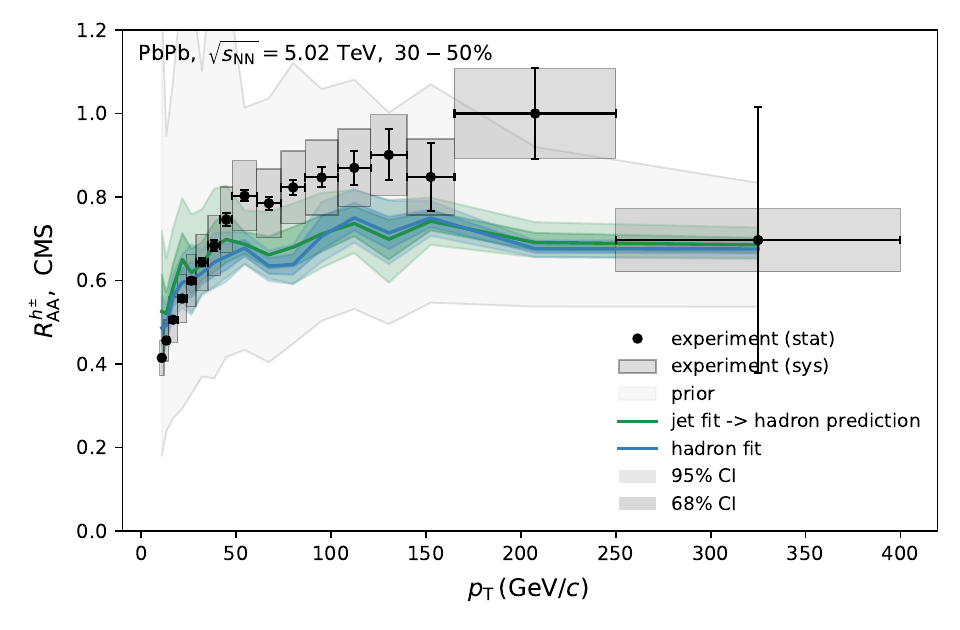}
\end{minipage}
\caption{Representative cross-prediction examples. Top: central $\to$ mid-central transfer for ATLAS inclusive-jet $R_{\mathrm{AA}}$ in PbPb $5.02$ TeV, $40$--$50\%$. Bottom: jet $\to$ hadron transfer for CMS charged-hadron $R_{\mathrm{AA}}$ in PbPb $5.02$ TeV, $30$--$50\%$. The transfer-level hierarchy is summarized in Table~\ref{tab:cross-prediction-summary}.}
\label{fig:crosspred-representative-best-worst}
\end{figure}

The cross-prediction examples span a weak transfer direction and a stable one, as illustrated in Fig.~\ref{fig:crosspred-representative-best-worst} and summarized quantitatively in Table~\ref{tab:cross-prediction-summary}. Table~\ref{tab:cross-prediction-summary} shows first that centrality transfer is not symmetric, even though the central and mid-central $\hat{q}/T^3$ posteriors nearly overlap in Table~\ref{tab:subset-decomposition-summary}. The central $\to$ mid-central transfer is the weakest of the six directions by the average separation, with $\mathrm{OVL}\approx0.34$, $\Delta\sigma_{68}\approx1.34$, and only about $48\%$ of matched bins retaining nonzero $68\%$-interval overlap. The reverse mid-central $\to$ central transfer is more permissive, with $\mathrm{OVL}\approx0.45$, $\Delta\sigma_{68}\approx0.85$, and nonzero $68\%$-interval overlap in all matched bins. This directionality explains why the posterior overlap in Table~\ref{tab:subset-decomposition-summary} is not sufficient: the mid-central calibration retains broader high-$\hat{q}$ support and can still cover much of the stronger-quenching central target, whereas the tighter central calibration transfers less flexibly to the less-quenched mid-central data.

The beam-energy transfers behave differently. They are less directional than the centrality test, but both directions remain displaced: PbPb $5.02\to2.76$ TeV gives $\mathrm{OVL}\approx0.42$ and $\Delta\sigma_{68}\approx1.14$, while PbPb $2.76\to5.02$ TeV gives $\mathrm{OVL}\approx0.45$ and $\Delta\sigma_{68}\approx1.04$. This pattern is consistent with the larger beam-energy posterior shift in Table~\ref{tab:subset-decomposition-summary}. In these cases, the issue is not a one-way loss of flexibility, but a persistent offset between the predictive distributions obtained from the two beam-energy calibrations.

The direct observable-class transfers are the most stable of the six tests. Although the jet-only posterior is shifted toward larger $\hat{q}/T^3$, the hadron $\to$ jet and jet $\to$ hadron predictions retain sizable overlap with the corresponding target fits for the inclusive $R_{\mathrm{AA}}$ observables used here. The weighted values are $\mathrm{OVL}\approx0.63$ and $\Delta\sigma_{68}\approx0.52$ for hadron $\to$ jet, and $\mathrm{OVL}\approx0.65$ and $\Delta\sigma_{68}\approx0.45$ for jet $\to$ hadron; all or nearly all matched bins satisfy $\Delta\sigma_{68}<1$ and retain nonzero $68\%$-interval overlap. Thus, the observable-class split identifies a residual parameter-space direction, but within the present inclusive observable basis, it does not produce the same level of observable-space degradation as the centrality and beam-energy transfers.

Taken together, the cross-prediction tests show why posterior overlap alone is not a sufficient universality criterion. The centrality comparison is the cleanest example: the posteriors overlap strongly at the common reference point, but the central $\to$ mid-central transfer still degrades. The beam-energy comparison gives the complementary case: the posterior shift is larger, and the predictive distributions remain offset in both transfer directions. By contrast, the direct hadron--jet transfer remains comparatively stable in the aggregate. Within the present inclusive $R_{\mathrm{AA}}$ basis, the most visible predictive limitations therefore appear in transfer across collision conditions rather than in the direct exchange between the inclusive hadron and jet observable classes.

\subsection{Sensitivity to different observables}
The cross-prediction tests identify where transferred posterior predictions degrade. As a complementary local diagnostic, we quantify how each observable responds to each model parameter around the posterior reference point.
The reference point is taken as the MAP sample from the MCMC chain, and each parameter is varied by a relative step of $\delta=0.1$ while keeping all other parameters fixed.
The MAP vector used for this local expansion is listed in Table~\ref{tab:sensitivity-map-point}.

\begin{table}[!htbp]
\caption{All-observables MAP parameter vector used as the baseline point for the local sensitivity map.}
\label{tab:sensitivity-map-point}
\begin{ruledtabular}
\scriptsize
\begin{tabular*}{\columnwidth}{@{\extracolsep{\fill}}lcccccc}
Parameter & $Q_0$ & $\tau_0$ & $A$ & $B$ & $C$ & $\alpha_s$ \\
\hline
MAP value & 1.314 & 0.104 & 10.638 & 140.353 & 0.282 & 0.342 \\
\end{tabular*}
\end{ruledtabular}
\end{table}

Following the implementation in the plotting script, the perturbed parameter vector for parameter $i$ is
\begin{equation}
\theta_j^{(i)}=
\begin{cases}
\theta_j^{(0)}(1+\delta), & j=i, \\
\theta_j^{(0)}, & j\neq i ,
\end{cases}
\end{equation}
where $\theta^{(0)}$ is the baseline (MAP) point.
For each parameter--observable-bin pair, the sensitivity index is then defined as
\begin{equation}
S_i^{(b)}
=
\frac{\left|y_{\mathrm{pert},i}^{(b)}-y_{\mathrm{base}}^{(b)}\right|}
{\delta\,\left|y_{\mathrm{base}}^{(b)}\right|},
\end{equation}
with $y_{\mathrm{base}}^{(b)}=y^{(b)}(\theta^{(0)})$ and $y_{\mathrm{pert},i}^{(b)}=y^{(b)}(\theta^{(i)})$.
This corresponds directly to the definition
$\left| \mathrm{pert}-\mathrm{base} \right| / \left(\delta\,\left|\mathrm{base}\right|\right)$.
For Fig.~\ref{fig:sensitivity-map-mean}, these bin-wise values are averaged within each observable block to obtain a compact mean-sensitivity summary.

Using these $10\%$ local variations around the MAP point, the resulting sensitivity map shows that the current observable set is most sensitive to $Q_0$ and next most sensitive to $\alpha_s$.
The influence of the $\hat{q}$-shape parameter $B$ is only intermediate, while the sensitivities to $\tau_0$, $A$, and $C$ remain uniformly small.
The largest local responses are concentrated in the central charged-hadron $R_{\mathrm{AA}}$ bins, whereas the jet $R_{\mathrm{AA}}$ observables are comparatively less sensitive.
This pattern explains why the posterior contraction in Fig.~\ref{fig:experiment-params-posterior} is concentrated mainly in $Q_0$ and $\alpha_s$, while the remaining $\hat{q}$-shape parameters retain broad support.
Because the construction is a local finite-difference probe around the MAP point, it should be interpreted as a diagnostic of nearby response rather than as a replacement for the full posterior analysis.

\begin{widetext}
\begin{center}
\centering
\includegraphics[width=\textwidth,height=0.45\textheight,keepaspectratio]{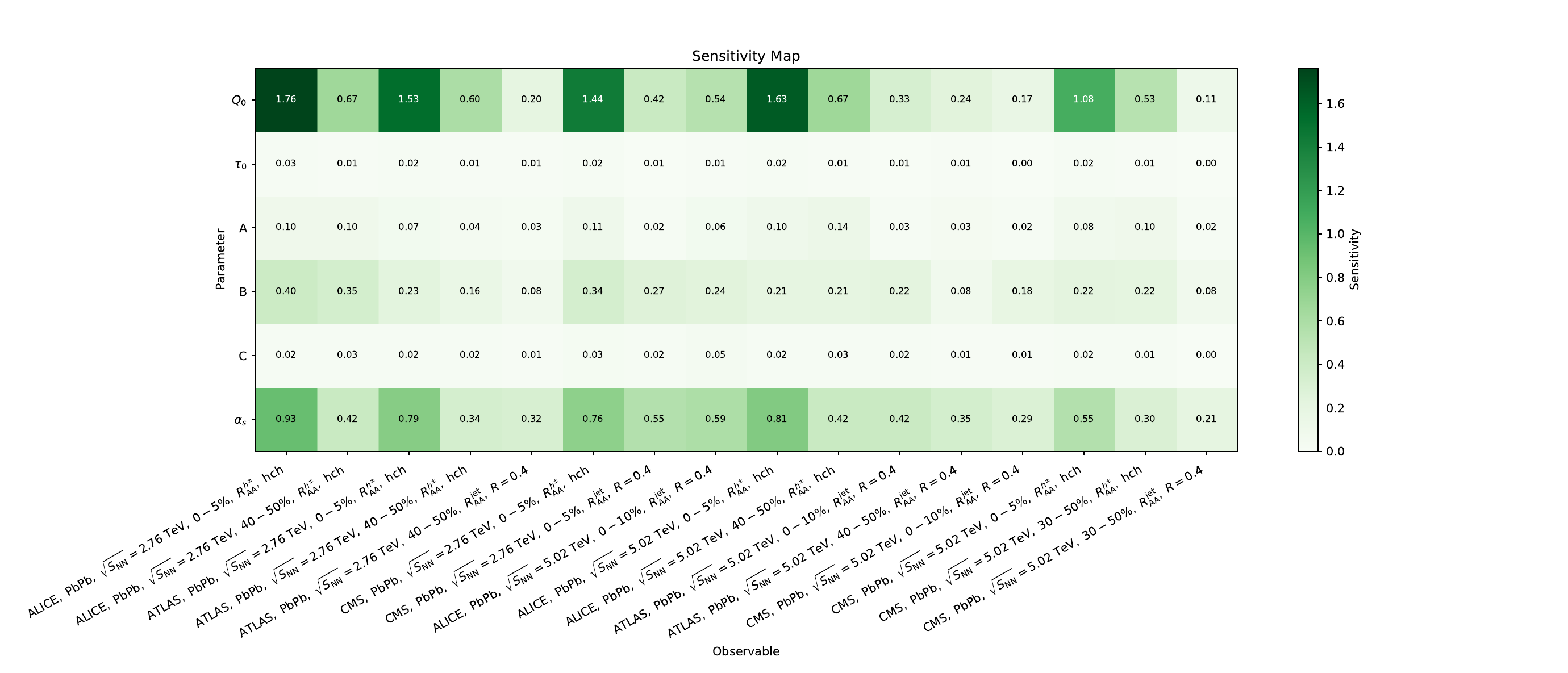}
\captionof{figure}{Mean sensitivity map for the calibrated observables, evaluated around the MAP point with a relative parameter variation $\delta=0.1$. The shown values correspond to bin-averaged sensitivities within each observable block.}
\label{fig:sensitivity-map-mean}
\end{center}
\end{widetext}

\subsection{Leading-hadron-selected jet as a suggested extension}
\label{sec:leading-hadron-jet}

To introduce a scale-dependent jet observable with minimal additional reconstruction complexity, we evaluate full-jet $R_{\mathrm{AA}}$ in PbPb $0$--$10\%$ at $\sqrt{s_{\mathrm{NN}}}=5.02$~TeV with a leading-hadron requirement applied to the PbPb jet sample.
Jets are reconstructed with anti-$k_T$ and $R=0.4$, with $|\eta_{\mathrm{jet}}|<0.5$, and the leading charged-hadron threshold is scanned as
$p_{\mathrm{T,leading}} > 3,\,5,\,8,\,12,\,15,\,20,\,25,$ and $30$~GeV/$c$.
This choice is physically motivated because jet fragmentation proceeds in a strongly ordered structure in transverse momentum and angle, so varying $p_{\mathrm{T,leading}}$ provides a controlled handle on the internal fragmentation scale.
Since medium-induced modifications are expected to be most pronounced for softer and wider-angle fragments in heavy-ion collisions, this selection provides information qualitatively similar to jet-shape measurements while avoiding an explicit jet-shape analysis.
For all selections, the denominator is held fixed at the default $pp$ jet spectrum, so each curve probes how imposing a harder in-jet leading-scale in PbPb modifies the effective suppression pattern relative to the same baseline.

The leading-hadron study in Sec.~\ref{sec:leading-hadron-jet} is not refitted to data. It uses a fixed default type-6 parameter vector from the PbPb $5.02$ TeV, $0$--$10\%$ leading-hadron configuration, summarized in Table~\ref{tab:leading-hadron-params}.

\begin{table}[!htbp]
\caption{Fixed type-6 energy-loss parameters used for the leading-hadron-selected jet calculation in Sec.~\ref{sec:leading-hadron-jet}.}
\label{tab:leading-hadron-params}
\begin{ruledtabular}
\scriptsize
\begin{tabular*}{\columnwidth}{@{\extracolsep{\fill}}lcccccc}
Parameter & $Q_0$ & $\tau_0$ & $A$ & $B$ & $C$ & $\alpha_s$ \\
\hline
Value & 2.0 & 0.5 & 10 & 100 & 0.2 & 0.4 \\
\end{tabular*}
\end{ruledtabular}
\end{table}

\begin{center}
\begin{minipage}{\columnwidth}
\centering
\includegraphics[width=\columnwidth]{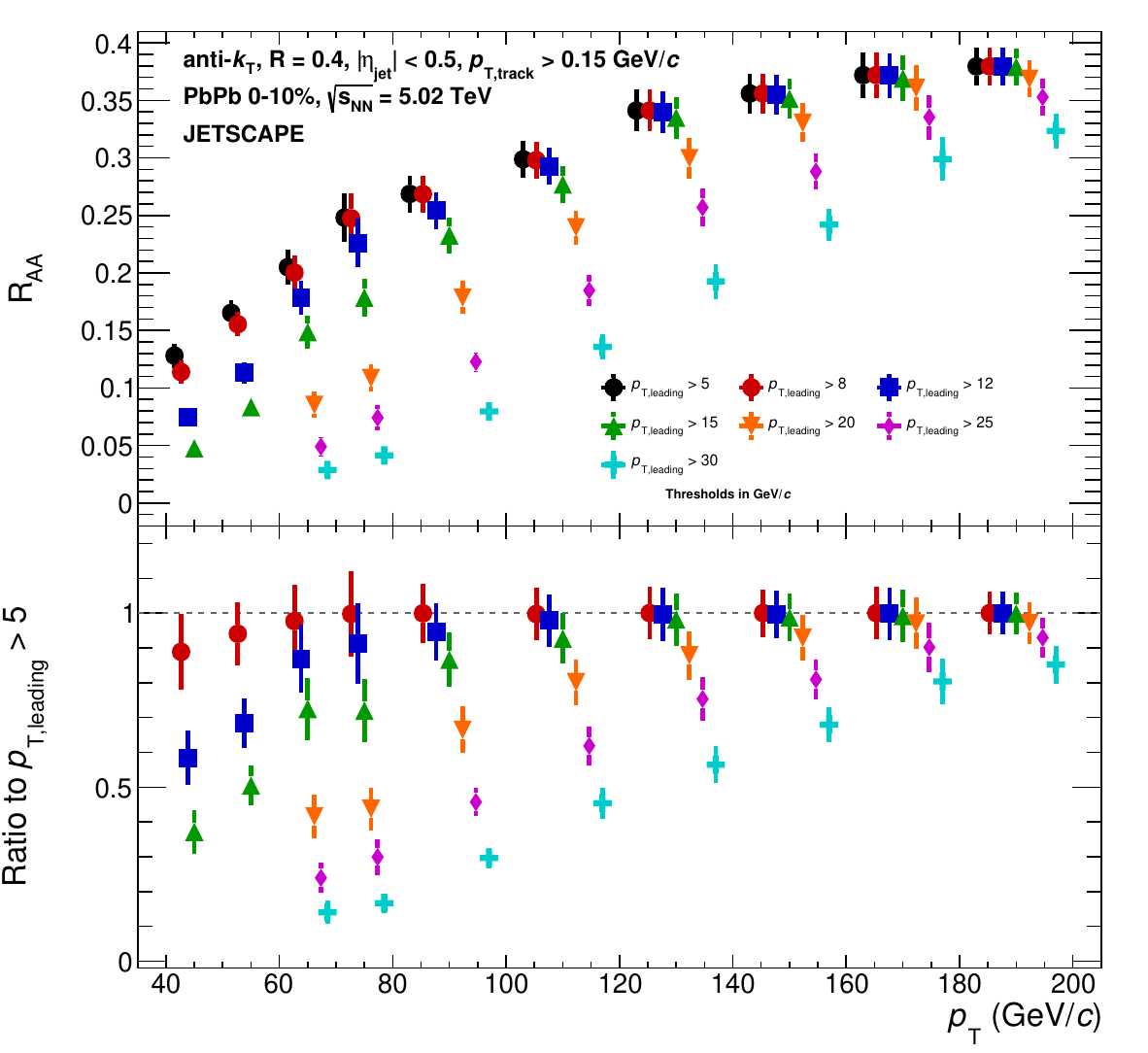}
\captionof{figure}{Leading-hadron-selected full-jet $R_{\mathrm{AA}}$ in PbPb $0$--$10\%$ at $\sqrt{s_{\mathrm{NN}}}=5.02$~TeV for anti-$k_T$ $R=0.4$ jets with $|\eta_{\mathrm{jet}}|<0.5$. The upper panel shows the selected $R_{\mathrm{AA}}$ curves and the lower panel shows each selection divided by the default model result. Increasing the leading-hadron threshold produces a systematic suppression relative to the default case, with stronger separation at lower jet $p_T$.}
\label{fig:leading-hadron-jet-raa-5020}
\end{minipage}
\end{center}

Figure~\ref{fig:leading-hadron-jet-raa-5020} shows the resulting observable.
The upper panel presents the selected $R_{\mathrm{AA}}$ curves, while the lower panel shows the ratio of each selected curve to the default model result.
A clear hierarchy is observed: tighter leading-hadron thresholds generally shift the ratio below unity in the lower panel, indicating stronger suppression than the default selection.
The separation is largest in the lower-$p_T$ part of the plotted range and becomes less pronounced toward higher jet $p_T$, where several selections approach the default behavior.
This trend is consistent with the interpretation that leading-hadron selection acts as a controllable internal jet scale, providing additional differential sensitivity beyond inclusive-jet $R_{\mathrm{AA}}$ alone.
Because it continuously varies the degree of leading-fragment bias inside a reconstructed-jet observable, it is also a natural candidate for future Bayesian calibration studies aimed at resolving the hadron--jet tension identified in the main text.

\section{CONCLUSION}
\label{sec:conclusion}
We calibrated a six-parameter JETSCAPE effective energy-loss model to charged-hadron and inclusive-jet $R_{\mathrm{AA}}$ data in PbPb collisions at $\sqrt{s_{\mathrm{NN}}}=5.02$ and $2.76$ TeV, and evaluated model performance across centrality, beam energy, and observable class.
This global Bayesian agreement alone, however, does not establish effective jet-transport universality: extracting $\hat{q}$ from inclusive suppression data entangles jet--medium interaction strength with geometry, path length, temperature history, hard-parton spectra, and observable bias.

Accordingly, we define universality operationally: subset $\hat{q}/T^3$ posteriors should be mutually compatible, and a posterior inferred from one subset should remain predictively transferable to complementary observables without refitting. In these comparisons, central and mid-central calibrations are consistent within current uncertainties. Beam-energy and observable-class splits also show overlapping credible regions, but with visible shifts and different posterior shapes, suggesting residual tensions that are partially averaged out in the all-observables posterior.

The observable-space transfer study shows why posterior compatibility alone is not sufficient. When each subset posterior is propagated to complementary observables without refitting, the six directed transfers are not symmetric. The central $\to$ mid-central transfer is weaker than the reverse direction, and beam-energy transfer remains displaced in both directions. By contrast, direct transfer between charged-hadron and inclusive-jet data is comparatively stable for the inclusive $R_{\mathrm{AA}}$ observables used here. Thus, in the present inclusive $R_{\mathrm{AA}}$ dataset, the main prediction loss arises when transferring across centrality and beam energy, whereas charged-hadron $\leftrightarrow$ inclusive-jet transfers remain comparatively stable.

Physically, these results suggest that the present calibration mainly fixes the overall quenching scale while leaving some scale-dependent details unresolved. The centrality comparison does not require separate transport scales for central and mid-central collisions, but the weaker central $\to$ mid-central transfer indicates residual path-length and geometry dependence. The beam-energy comparison is more restrictive because changing $\sqrt{s_{\mathrm{NN}}}$ changes both the temperature history and hard-parton spectra. The observable-class comparison is less limiting in direct transfer, but charged hadrons and inclusive jets still probe different parts of the shower: high-$p_T$ charged-hadron measurements are more sensitive to hard leading fragments and surface-biased parent showers, whereas inclusive jets retain more sensitivity to redistributed shower energy. The sensitivity analysis supports this picture by showing that several shape directions of the six-parameter transport coefficient remain weakly constrained.

The natural next step is to add observables that interpolate more directly between hadron-biased and jet-inclusive constraints. The leading-hadron-selected jet observable presented in Sec.~\ref{sec:leading-hadron-jet} is one candidate because it introduces a tunable leading-fragment bias within the reconstructed-jet framework. The complementary sensitivity results presented in Sec.~\ref{sec:discussion} clarify why this type of observable can add leverage beyond inclusive $R_{\mathrm{AA}}$ alone. Combined with a denser design and more uniformly matched hadron and jet measurements across beam energies, such observables can assess whether the residual hadron--jet and beam-energy tensions can be absorbed into a more flexible common transport description.

\begin{acknowledgments}
We thank the JETSCAPE Collaboration for developing and maintaining the JETSCAPE source code used in this work.
This work made use of computational resources provided by the KISTI Analysis Facility (KIAF), operated by the Global Scientific Data Hub Center at the Korea Institute of Science and Technology Information (KISTI).
This work was supported by grants from the National Research Foundation of Korea (NRF), funded by the Ministry of Science and ICT (MSIT), Republic of Korea, under Grant Nos.~RS-2026-25489046, RS-2008-NR007226, and RS-2023-00279977. DJK is supported by the Research Council of Finland, the Center of Excellence in Quark Matter (project no 346328).
\end{acknowledgments}

\clearpage
\appendix
\section{Production Details}
\label{app:production-details}

This appendix collects the technical production settings needed for reproducibility. The main-text argument relies on these details only insofar as the same production setup is applied consistently across all calibrations and subset comparisons.

\subsection{Hard-process stitching and smoothing}
Because the charged-hadron and inclusive-jet observables used in this work extend to relatively high transverse momentum, the event generation is performed in $\hat{p}_{\mathrm{T}}$-binned samples rather than in a single fully inclusive sample. This improves the statistical coverage of the rare high-$p_T$ region while keeping the total computational cost manageable. The weighted recombination of these bins is defined in Eq.~\eqref{eq:raa}.

At the same time, the handoff between neighboring nominal $\hat{p}_{\mathrm{T}}$ intervals can produce isolated bin-to-bin spikes when the Monte Carlo statistics fluctuate strongly in a limited part of the spectrum. Before emulator training, we therefore preprocess the stitched model outputs with a dedicated smoothing procedure. The preprocessing framework distinguishes between filtering and smoothing. Filtering would remove entire design points whose statistical quality is systematically poor across many observable bins, but it is disabled in the present analysis to retain all 50 design points are retained. Instead, we apply only local smoothing at the observable-bin level. For each design point and observable, candidate outlier bins are identified in two passes using an RMS-based threshold with $n_{\mathrm{rms}}=2.5$: first from unusually large relative statistical errors, and then from anomalously large first-difference jumps in the central values. Only isolated bins or runs of at most two consecutive bins are eligible for replacement; longer runs are flagged as unresolved and left unchanged. Eligible central values and statistical uncertainties are replaced by linear interpolation in $p_T$ from neighboring retained bins. Across the 16 selected calibration observables, this procedure modifies 85 of 19,250 central-value cells and the corresponding 85 statistical-uncertainty cells ($0.44\%$ each), while 45 flagged cells remain unresolved and are not altered. The smoothing, therefore, acts as a conservative local correction to nonphysical stitching artifacts rather than a broad reshaping of the spectra.

\subsection{Background generation settings}
The event-by-event initial entropy-density profiles used for the medium background are generated with the TRENTo model~\cite{Moreland:2014oya,Bernhard:2016tnd}. In the present analysis, the TRENTo events are generated in advance and stored as HDF5 files, which are then used as the initial-condition ensemble for the later medium-evolution stage. For both PbPb systems, we generate $100{,}000$ minimum-bias events in order to ensure stable centrality selection and sufficient event-by-event coverage of the geometric fluctuations. The chosen setup is effectively boost-invariant in rapidity, as indicated by $\eta_{\max}=0.0$, and therefore provides a two-dimensional transverse initial condition for the background used in this work.

Table~\ref{tab:trento-settings} summarizes the TRENTo sampling setup used for the two collision energies. The normalization is adjusted separately for PbPb at $\sqrt{s_{\mathrm{NN}}}=5.02$ and $2.76$ TeV, while the reduced-thickness parameter, fluctuation strength, nucleon width, minimum nucleon separation, skew settings, and grid definition are kept identical between the two systems. This choice isolates the beam-energy dependence primarily through the overall entropy normalization while preserving the same geometric model assumptions across the two collision energies.

\begin{table}[!htbp]
\caption{TRENTo initial-condition generation settings used to produce the HDF5 event files for PbPb at $\sqrt{s_{\mathrm{NN}}}=5.02$ and $2.76$ TeV.}
\label{tab:trento-settings}
\centering
\scriptsize
\resizebox{\columnwidth}{!}{%
\begin{tabular}{lcc}
\hline\hline
Item & PbPb5020 & PbPb2760 \\
\hline
Number of events & 100000 & 100000 \\
Normalization & 21.044 & 14.373 \\
Reduced thickness & 0.0056 & 0.0056 \\
Fluctuation & 1.0468 & 1.0468 \\
Nucleon width & 0.86 & 0.86 \\
Nucleon minimum distance & 1.2367 & 1.2367 \\
Mean / std / skew & \shortstack[l]{1.0 / 3.0 / 0.0\\ (skew-type 1)} & \shortstack[l]{1.0 / 3.0 / 0.0\\ (skew-type 1)} \\
Jacobian & 0.8 & 0.8 \\
Grid setup & \shortstack[l]{xy-max 15, xy-step 0.3,\\ eta-max 0.0, eta-step 0.5} & \shortstack[l]{xy-max 15, xy-step 0.3,\\ eta-max 0.0, eta-step 0.5} \\
\hline\hline
\end{tabular}
}
\end{table}

The TRENTo profiles then pass through a short freestream stage before being evolved with MUSIC hydrodynamics~\cite{Song:2007ux}. The fixed freestream+MUSIC background settings used to generate the temperature and flow fields are summarized in Table~\ref{tab:hydro-background-settings}. As emphasized in the main text, the hydrodynamic start time $\tau_{\mathrm{hydro}}$ is part of the fixed background production setup and is distinct from the calibrated medium-onset parameter $\tau_0$ of the energy-loss model.

\begin{table}[!htbp]
\caption{Fixed freestream+MUSIC background settings used to evolve the TRENTo \texttt{initial.hdf5} profiles.}
\label{tab:hydro-background-settings}
\begin{ruledtabular}
\scriptsize
\begin{tabular}{lc}
Parameter & Value \\
\hline
Hydrodynamic start time $\tau_{\mathrm{hydro}}$ [fm/$c$] & 0.71 \\
$\eta/s(T)$ parametrization mode & 2 \\
Minimum $\eta/s$ & 0.093 \\
$\eta/s$ slope & 0.8024 \\
$\eta/s$ curvature & 0.1568 \\
$\zeta/s(T)$ parametrization mode & 3 \\
Maximum $\zeta/s$ & 0.01844 \\
$T_{\mathrm{peak}}^{\zeta/s}$ [GeV] & 0.1889 \\
Width of $\zeta/s$ [GeV] & 0.04252 \\
Asymmetry parameter $\lambda_{\zeta/s}$ & 0 \\
Freeze-out temperature [GeV] & 0.1595 \\
\end{tabular}
\end{ruledtabular}
\end{table}

\section{Inference and Design Details}
\label{app:inference-details}

This appendix records the emulator-construction choices and design diagnostics that support the main-text inference but are not required for the main paper's logical flow.

In the present analysis, no separate $\Sigma_{\mathrm{model}}(\theta)$ term is added to the likelihood. The model calculations at the design points are obtained from finite event samples, so residual Monte Carlo fluctuations are already present in the emulator training targets. In practice, this effect is handled at the GP-training stage byconstructing the emulator, including an additive white-noise term in the kernel. Consequently, the effective uncertainty associated with finite model statistics is absorbed into $\Sigma_{\mathrm{emu}}(\theta)$ rather than introduced as an additional covariance term.

In the present emulator construction, the emulator covariance is decomposed as
\[
\Sigma_{\mathrm{emu}}(\theta)
=
\Sigma_{\mathrm{GP}}(\theta)
+
\Sigma_{\mathrm{PCA}}^{\mathrm{trunc}},
\]
where $\Sigma_{\mathrm{GP}}(\theta)$ is the predictive covariance of the retained GP-emulated principal components projected back to observable space, and $\Sigma_{\mathrm{PCA}}^{\mathrm{trunc}}$ accounts for the residual variance associated with the discarded PCA components. Thus, $\Sigma_{\mathrm{emu}}(\theta)$ contains the GP predictive uncertainty, including the effective residual training-noise contribution learned through the white-noise term, together with the uncertainty induced by reducing the output space to a finite number of principal components.

The GP emulator is built in a PCA-reduced output space. The calibration outputs are partitioned into hadron and jet groups, corresponding here to charged-hadron and inclusive-jet $R_{\mathrm{AA}}$, and separate GP emulators are trained for the two groups before being reassembled into a common observable ordering. Before GP training, the six input parameters are standardized over the design set to zero mean and unit variance, and the observable outputs are likewise standardized over the design points before dimensional reduction. In addition, we apply output weights within each observable block so that different $p_T$ regions contribute more evenly to the reduced representation.

The training outputs are constructed after the analysis $p_T$ selections used in the emulator configuration: for charged-hadron $R_{\mathrm{AA}}$, only bins with $p_T > 10~\mathrm{GeV}/c$ are retained, whereas for inclusive-jet $R_{\mathrm{AA}}$ no additional emulator-specific $p_T$ cut is imposed beyond the experimental jet binning. Following Ref.~\cite{Fan:2023metric}, we retain five principal components in each group, with the residual variance of the discarded components carried by $\Sigma_{\mathrm{PCA}}^{\mathrm{trunc}}$.

Each retained component is modeled with a Gaussian process using a squared-exponential (RBF) kernel with an additive white-noise term:
\begin{equation}
k(\theta,\theta')
=
\sigma_f^2
\exp\!\left[
-\frac{1}{2}
\sum_i \frac{(\theta_i-\theta_i')^2}{\ell_i^2}
\right]
+
\sigma_n^2\,\delta_{\theta\theta'}.
\label{eq:kernel}
\end{equation}
Again, following Ref.~\cite{Fan:2023metric}, the kernel hyperparameters are determined by maximizing the GP log-marginal likelihood with five optimizer restarts. This repeated optimization reduces sensitivity to local optima in the nonconvex hyperparameter fit and helps stabilize the learned length scales and noise level across both the hadron and jet emulators. In the present analysis, the RBF form imposes a smooth response over parameter space, while the additive white-noise term provides regularization in the GP fit.

Using this surrogate, the posterior density in Eq.~\eqref{eq:posterior} can be evaluated repeatedly at negligible cost compared with a full event simulation. We therefore sample the posterior with Markov chain Monte Carlo (MCMC), using the \texttt{emcee} affine-invariant ensemble sampler with 64 walkers, 30,000 steps, and a burn-in of 5,000 steps. To reduce initialization bias, the walkers are seeded through a pilot prior-screening stage rather than being drawn purely at random from the prior volume. During sampling, we use a mixed move set consisting of StretchMove, DEMove, and DESnookerMove in order to improve exploration of correlated and elongated posterior directions. The resulting chain is then used to construct marginal posteriors and posterior predictive distributions for the observables and for the derived $\hat{q}/T^3$ quantities.

\subsection{Simulation statistics and design coverage}
A design point denotes a unique parameter vector ($Q_0$, $\tau_0$, $A$, $B$, $C$, $\alpha_s$) at which a full event simulation is performed. For each design point, we generate 10,000 hard-scattering events for each centrality class at each collision energy. These are propagated through an ensemble of 100 distinct QCD medium background events, so each medium event is reused for multiple hard-scattering events (about 100 on average).

Across the retained calibration bins, the relative statistical uncertainty of the model outputs used in the calibration has a mean of 7.84\% and a maximum of 35.72\%, while the corresponding experimental relative uncertainty has a mean of 11.00\% and a maximum of 54.29\%. This indicates that finite-statistics fluctuations in the simulation outputs are not negligible and motivates the white-noise treatment adopted in the GP emulator training. The full production cost corresponds to about 1,700 core-hours per design point for a single centrality bin at a single collision energy. Due to this computational cost, we limit the calibration set to 50 design points.

\begin{figure}[!htbp]
\centering
\includegraphics[width=\columnwidth]{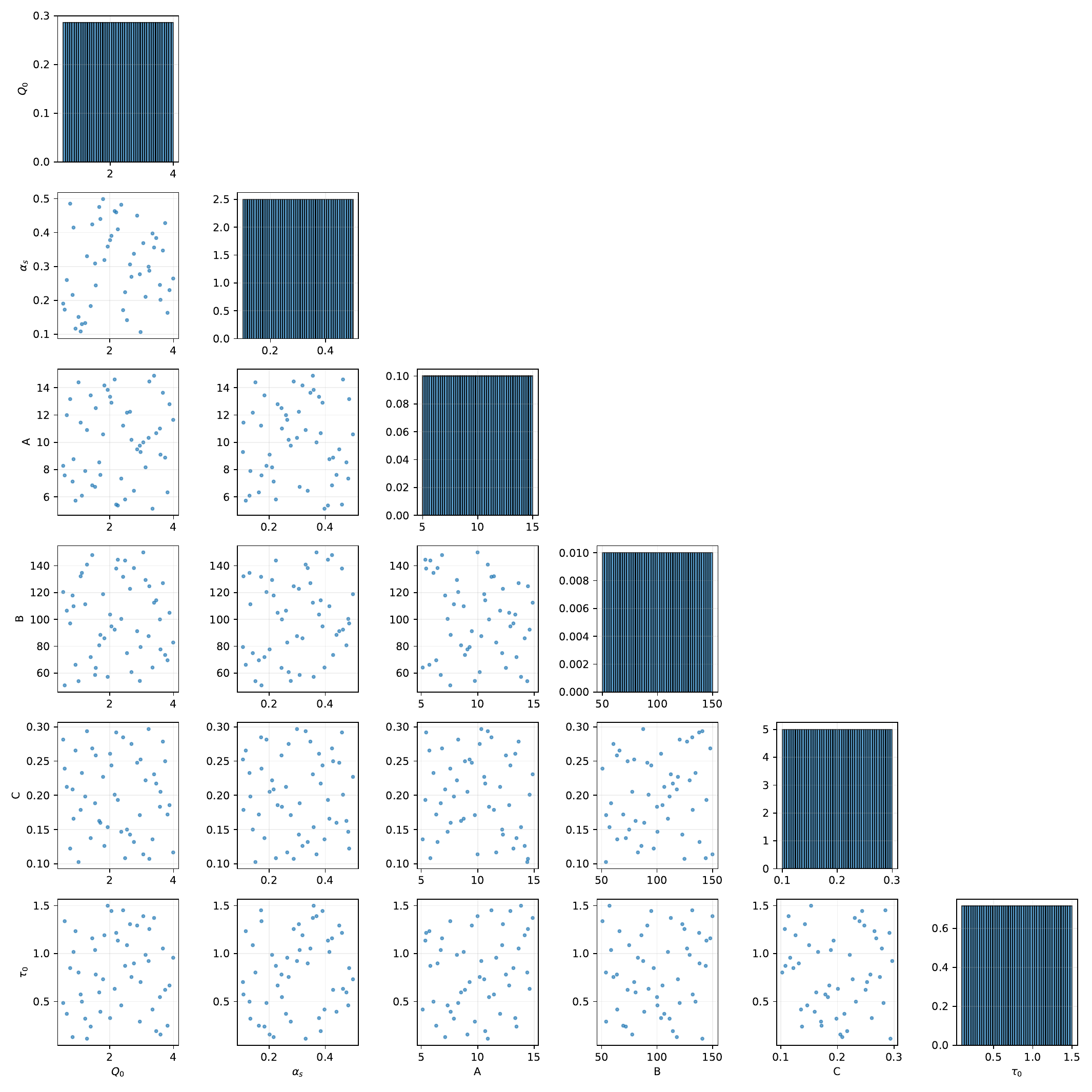}
\caption{50 design points sampled within the prior parameter ranges.}
\label{fig:lhs-pairplot}
\end{figure}

\begin{figure}[!htbp]
\centering
\includegraphics[width=\columnwidth]{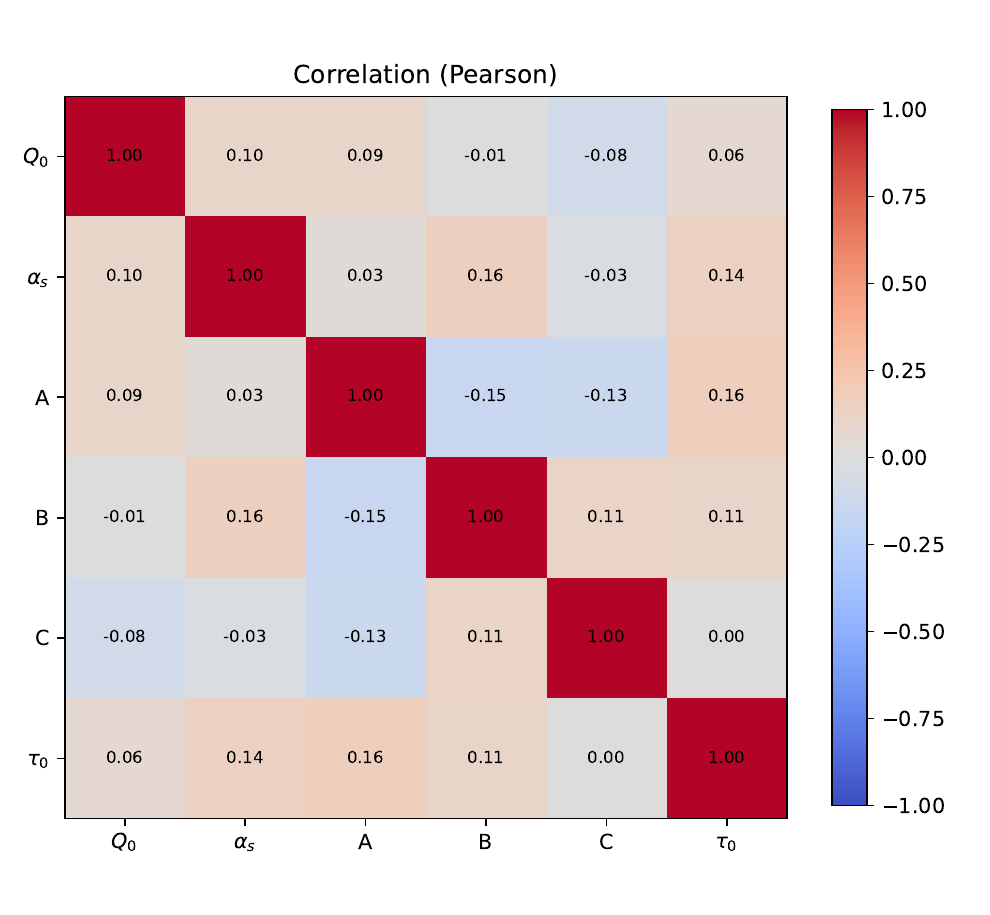}
\caption{Correlation matrix among calibration parameters.}
\label{fig:lhs-corr}
\end{figure}

These design points are generated using LHS (Latin hypercube sampling)~\cite{McKay:1979lhs,TangHypercube,MORRIS1995381} from independent uniform priors to ensure uniform coverage of the multidimensional parameter space. Fig.~\ref{fig:lhs-pairplot} visualizes the resulting space-filling design within the prior ranges. Fig.~\ref{fig:lhs-corr} shows the Pearson correlation coefficients among the calibration parameters for the LHS design. The off-diagonal elements are close to zero, consistent with the assumption of independent priors and indicating that the sampling procedure does not introduce artificial correlations among parameters.

\section{Validation and Closure Details}
\label{app:validation-details}

This appendix expands the compact validation statements given in the main text.

We validate the GP emulator with two complementary tests. The validation uses the same hadron/jet grouping, preprocessing, and five principal components of the PCA representation as the production inference. In the grouped full-train emulator bundle, the retained components account for 99.22\% of the jet variance and 88.84\% of the hadron variance, corresponding to 92.92\% overall when weighted by the retained output bins. These fractions should be interpreted as the variance attributed to the retained PCs. The lower hadron value reflects a broader variance spectrum rather than discarded structure being ignored: the omitted hadron components are propagated through the PCA-truncation covariance in the likelihood, and the adequacy of the resulting reduced representation is assessed empirically by the validation and closure tests below.

Following Ref.~\cite{Fan:2023metric}, Method 1 performs a partial leave-one-out check: five design points are randomly selected, and for each selected point the emulator is trained on the remaining 49 points and then used to predict the held-out calculation. Representative Method 1 panels and the corresponding Method 2 panels for the same representative observables highlighted in Fig.~\ref{fig:prior-panels-main} are shown in the main text, while the remaining panels are collected in Appendices~\ref{app:method1-panels} and \ref{app:method2-panels}. These comparisons provide qualitative evidence that the GP emulator reproduces the model calculations well for both charged-hadron and inclusive-jet $R_{\mathrm{AA}}$ across the selected systems and centrality intervals.

Method 2 evaluates the full emulator performance by comparing its predictions with model calculations across all design points. The fit-overlay panel in Fig.~\ref{fig:closure-method2-main} shows no localized mismatch, and the points follow the $x=y$ trend. Taken together, the Method 1 and Method 2 diagnostics indicate that the present 50-point design is adequate for stable emulator-based inference at the precision level targeted here, although it does not provide exact local parameter resolution throughout the prior volume.

Even if the GP emulator predicts observables accurately, this alone does not guarantee that the inferred posterior is well constrained. To test parameter recovery explicitly, we perform a leave-one-design-point-out closure test using the same Bayesian inference setup as in the main analysis. In each trial, one design point is removed from the set of 50 and treated as the pseudo-truth. The GP emulator is retrained on the remaining 49 points, pseudo-data are generated from the full model at the held-out point, and MCMC sampling is repeated with the same likelihood and covariance treatment.

Because the true parameter vector is known by construction, the posterior can be checked directly against the truth values. This setup, therefore, tests whether localized pathological behavior exists in specific parameter regions, such as systematic peak shifts or artificially sharp posteriors from overfitting. We repeat this procedure for 9 randomly selected held-out design points, and the corresponding runs satisfy the MCMC chain-quality criteria summarized in Appendix~\ref{app:mcmc-diagnostics}.

The recovery is qualitative rather than exact: across these hold-out cases, the true parameter values are generally captured by the posterior support, but the posterior maxima do not coincide perfectly with the truth. This level of mismatch is consistent with the limited resolution of the current 50 design points, which constrains how accurately the emulator can represent local parameter dependence across the full prior volume.

\section{Additional Prior-Distribution Panels}
\label{app:prior-panels}

The main text shows representative prior-distribution panels in order to keep the discussion compact.
The remaining prior-distribution panels are collected here for completeness.

\begin{figure}[!tbp]
\centering
\begin{minipage}{0.48\columnwidth}
\centering
\includegraphics[width=\linewidth]{figures/202604/project_4/02_prior_distributions/prior_distributions/prior_chargedALICE_HadronAnalysis_Hadron_hRaa_PbPb5020_cent_0_5.pdf}
\end{minipage}
\hfill
\begin{minipage}{0.48\columnwidth}
\centering
\includegraphics[width=\linewidth]{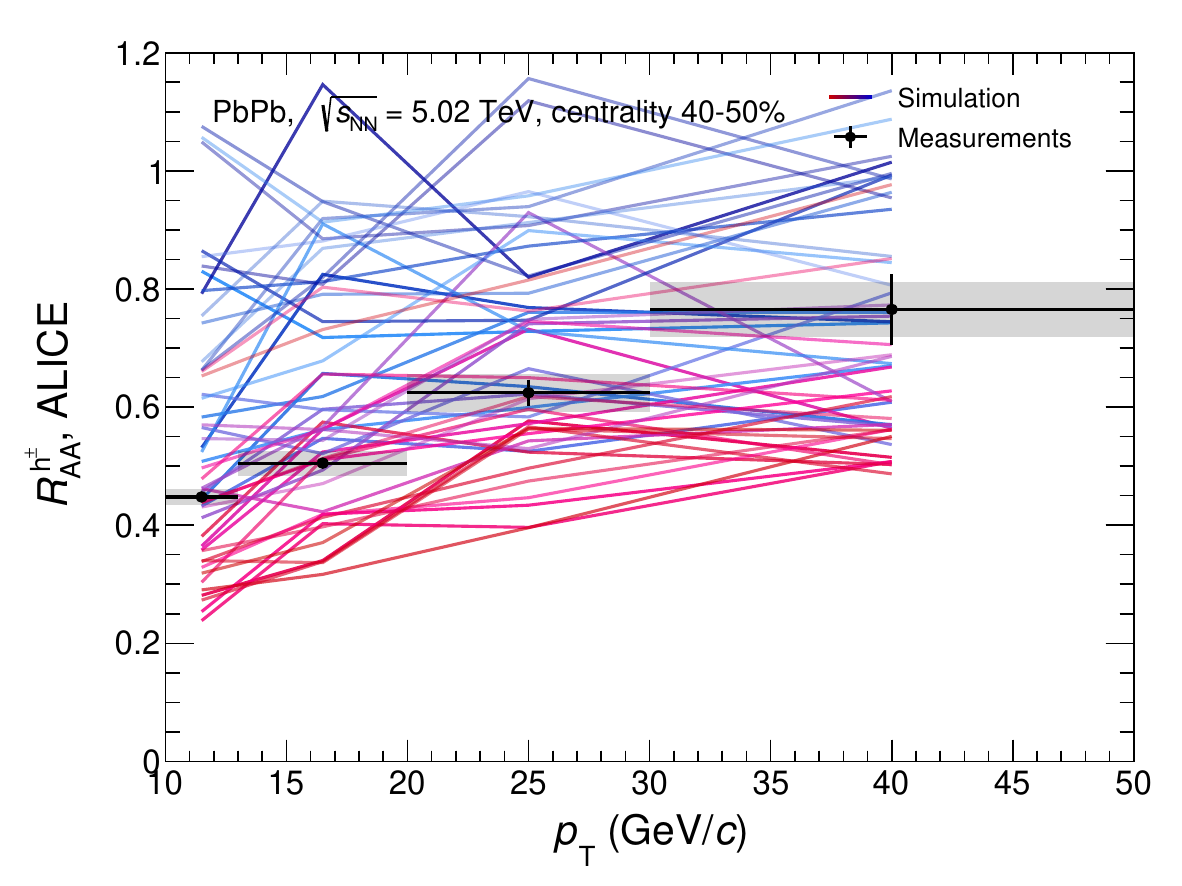}
\end{minipage}

\vspace{0.5em}

\begin{minipage}{0.48\columnwidth}
\centering
\includegraphics[width=\linewidth]{figures/202604/project_4/02_prior_distributions/prior_distributions/prior_chargedALICE_HadronAnalysis_Hadron_hRaa_PbPb2760_cent_0_5.pdf}
\end{minipage}
\hfill
\begin{minipage}{0.48\columnwidth}
\centering
\includegraphics[width=\linewidth]{figures/202604/project_4/02_prior_distributions/prior_distributions/prior_chargedATLAS_HadronAnalysis_Hadron_hRaa_PbPb2760_cent_0_5.pdf}
\end{minipage}

\vspace{0.5em}

\begin{minipage}{0.48\columnwidth}
\centering
\includegraphics[width=\linewidth]{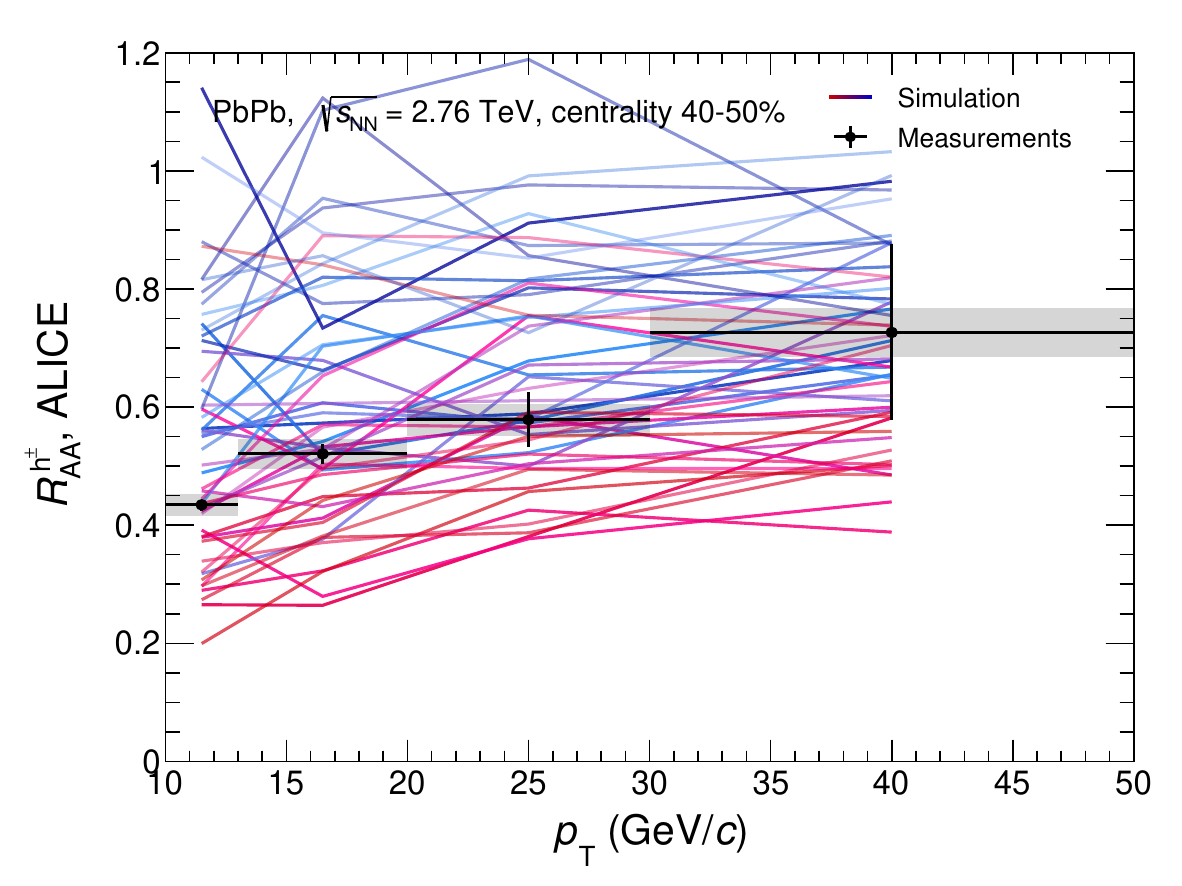}
\end{minipage}
\hfill
\begin{minipage}{0.48\columnwidth}
\centering
\includegraphics[width=\linewidth]{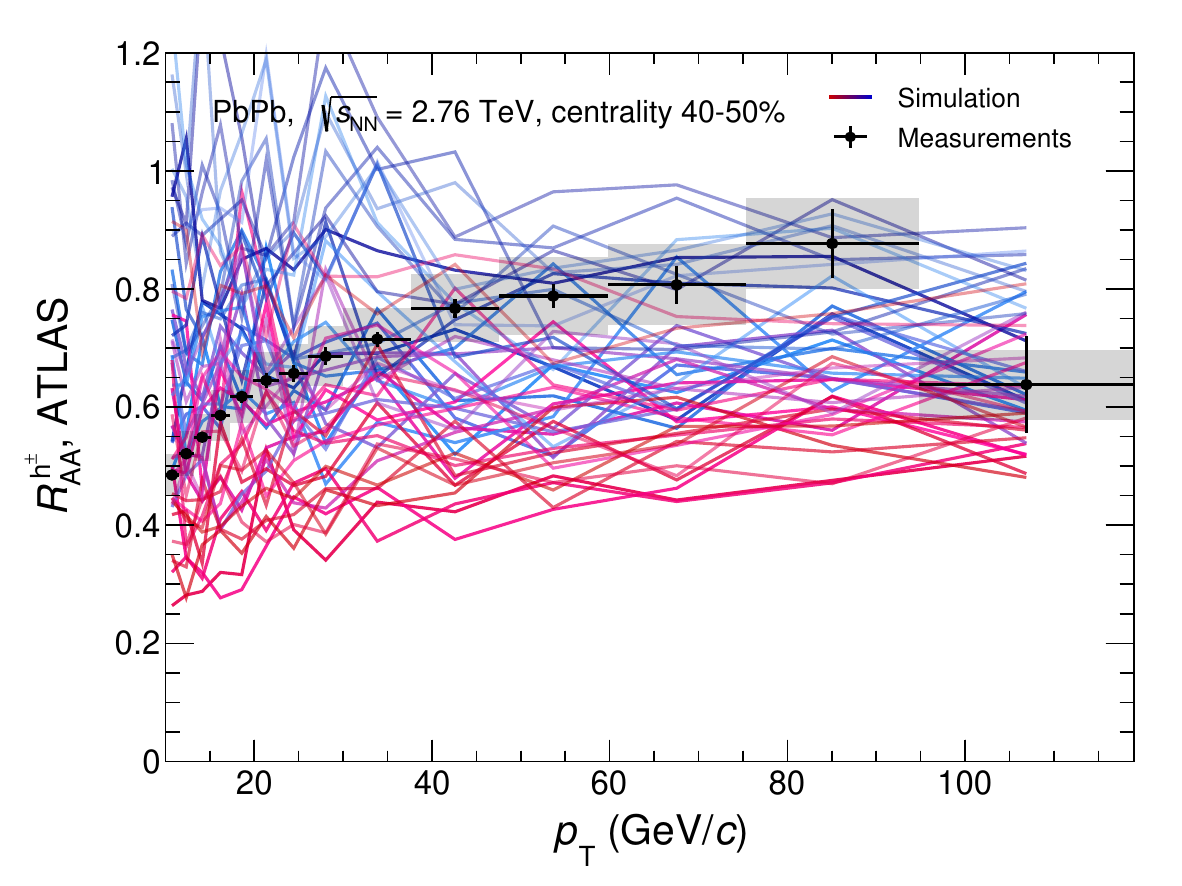}
\end{minipage}
\caption{Additional ALICE and ATLAS prior distributions for charged-hadron $R_{\mathrm{AA}}$.}
\label{fig:prior-hadron-panels-extra}
\end{figure}

\begin{figure}[!tbp]
\centering
\begin{minipage}{0.48\columnwidth}
\centering
\includegraphics[width=\linewidth]{figures/202604/project_4/02_prior_distributions/prior_distributions/prior_ALICE_JetAnalysis_Jet_hRaa_PbPb5020_cent_0_10.pdf}
\end{minipage}
\hfill
\begin{minipage}{0.48\columnwidth}
\centering
\includegraphics[width=\linewidth]{figures/202604/project_4/02_prior_distributions/prior_distributions/prior_ATLAS_JetAnalysis_Jet_hRaa_PbPb5020_cent_0_10.pdf}
\end{minipage}

\vspace{0.5em}

\begin{minipage}{0.48\columnwidth}
\centering
\includegraphics[width=\linewidth]{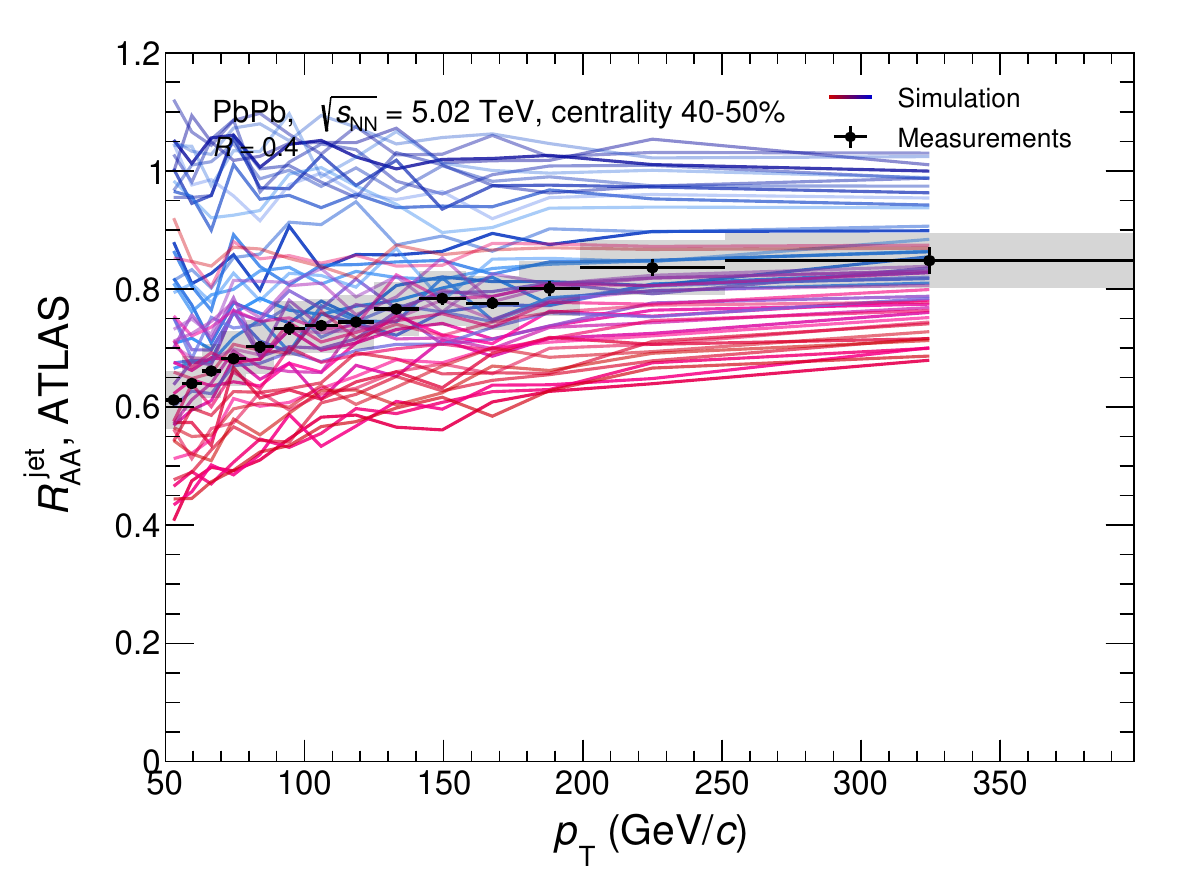}
\end{minipage}
\hfill
\begin{minipage}{0.48\columnwidth}
\centering
\includegraphics[width=\linewidth]{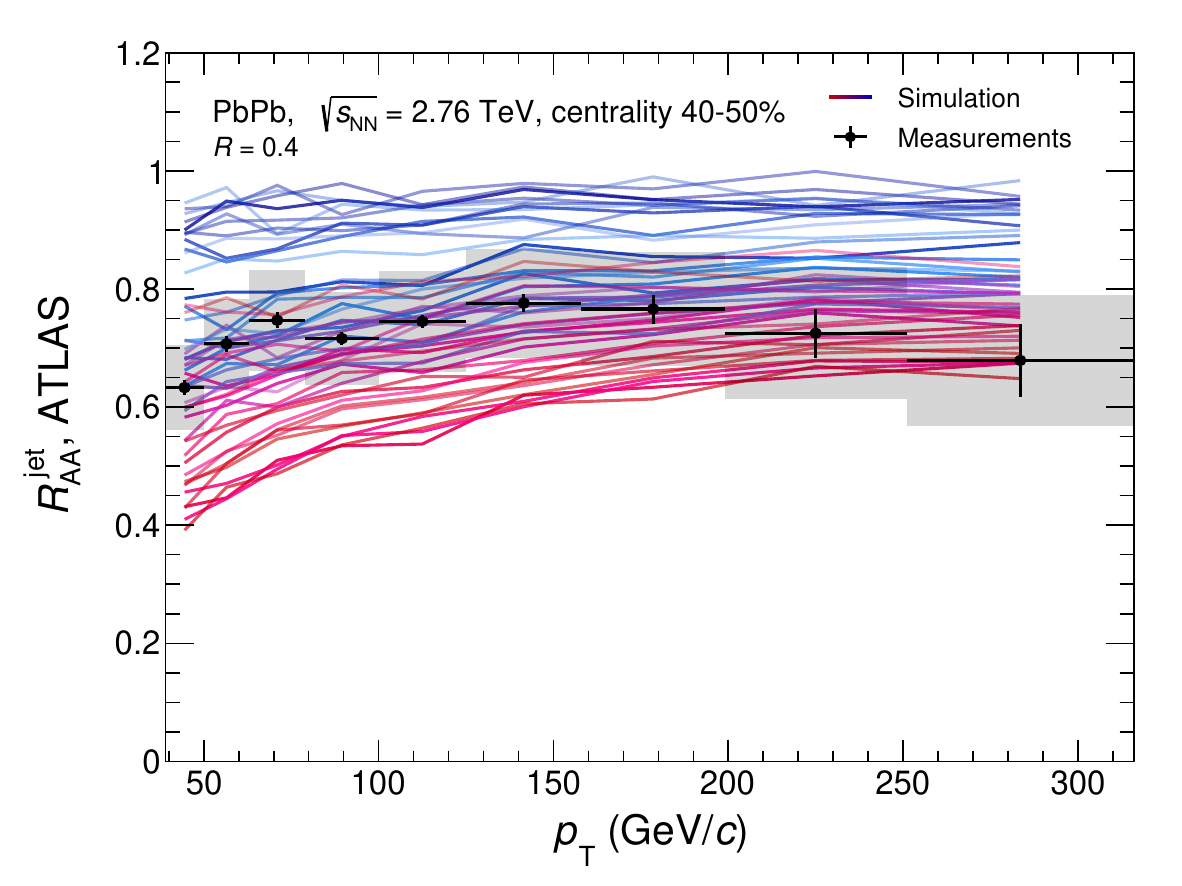}
\end{minipage}
\caption{Additional ALICE and ATLAS prior distributions for inclusive-jet $R_{\mathrm{AA}}$.}
\label{fig:prior-jet-panels-extra}
\end{figure}

\section{Additional Posterior-Predictive Panels}
\label{app:posterior-panels}

The main text shows representative posterior-predictive panels in order to keep the discussion compact.
The remaining posterior-predictive panels are collected here for completeness.

\begin{figure}[!tbp]
\centering
\begin{minipage}{0.48\columnwidth}
\centering
\includegraphics[width=\linewidth]{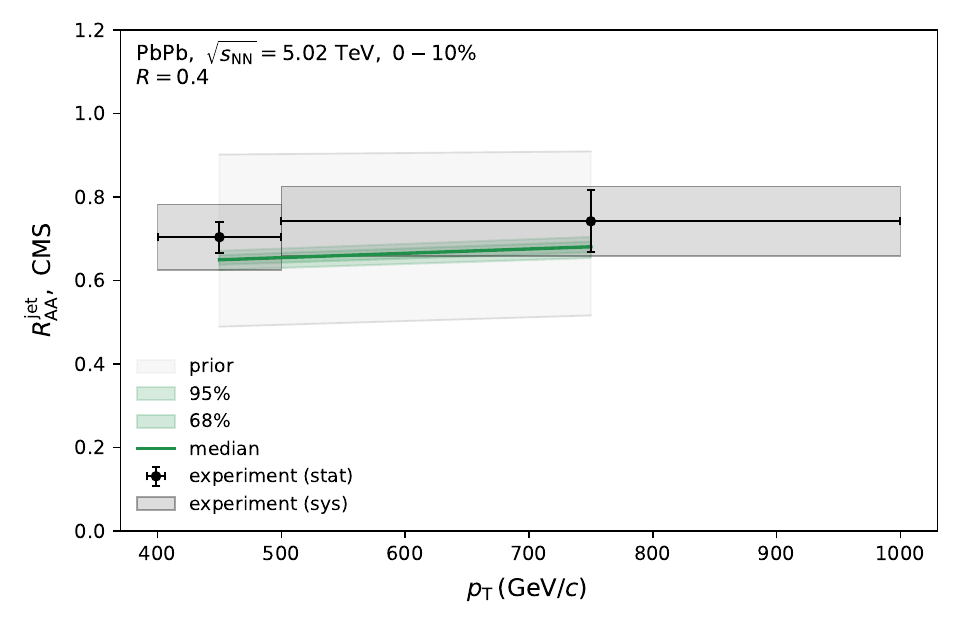}
\end{minipage}
\hfill
\begin{minipage}{0.48\columnwidth}
\centering
\includegraphics[width=\linewidth]{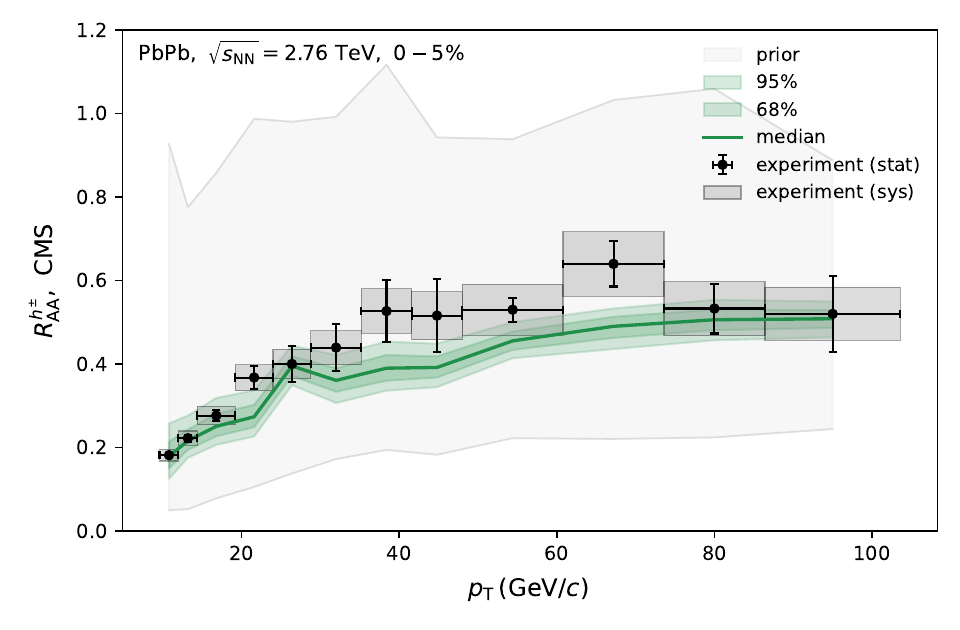}
\end{minipage}

\vspace{0.5em}

\makebox[\columnwidth][c]{%
\begin{minipage}{0.48\columnwidth}
\centering
\includegraphics[width=\linewidth]{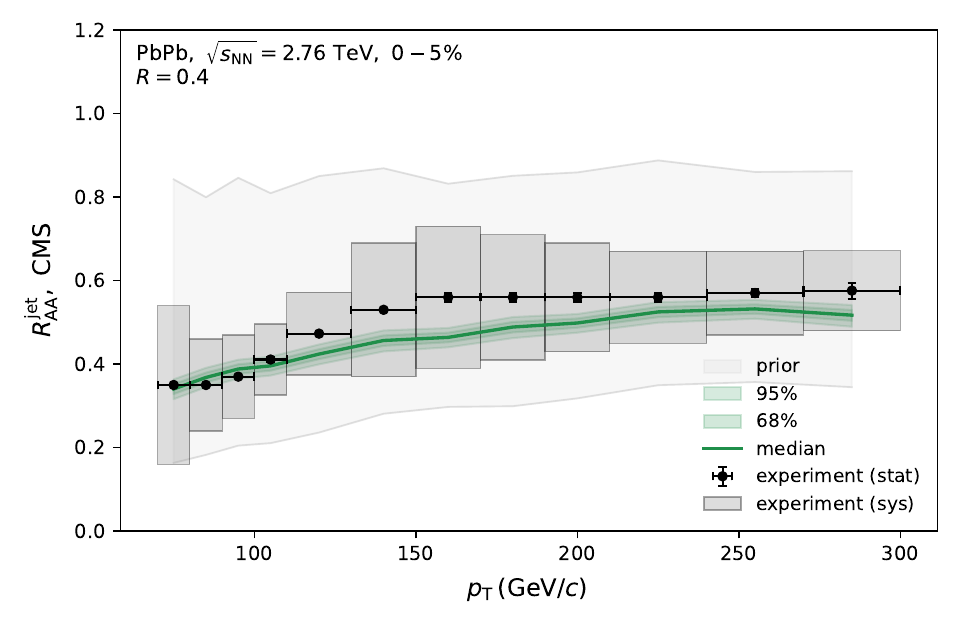}
\end{minipage}%
}
\caption{Additional posterior-predictive central panels not shown in the main text. These panels complete the PbPb $\sqrt{s_{\mathrm{NN}}}=5.02$ and $2.76$ TeV central selections beyond the representative subset shown in the main text.}
\label{fig:posterior-observables-central-extra}
\end{figure}

\begin{figure}[!tbp]
\centering
\begin{minipage}{0.48\columnwidth}
\centering
\includegraphics[width=\linewidth]{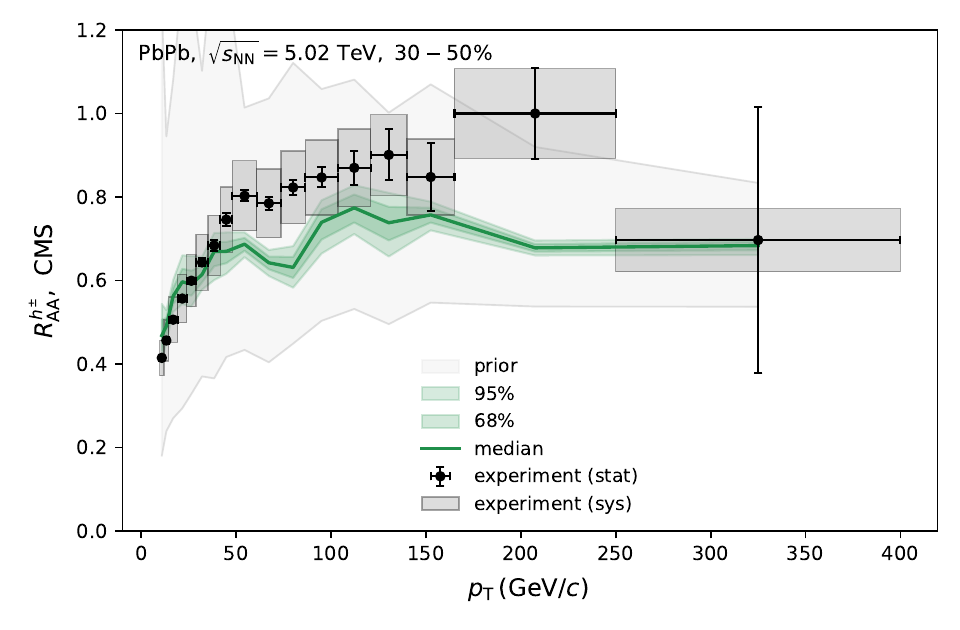}
\end{minipage}
\hfill
\begin{minipage}{0.48\columnwidth}
\centering
\includegraphics[width=\linewidth]{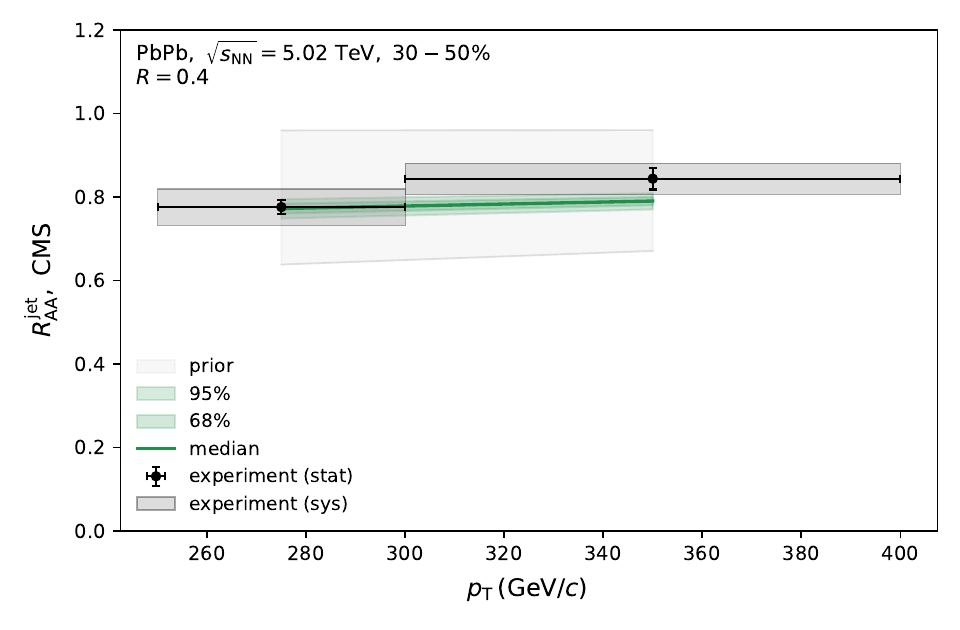}
\end{minipage}

\vspace{0.5em}

\begin{minipage}{0.48\columnwidth}
\centering
\includegraphics[width=\linewidth]{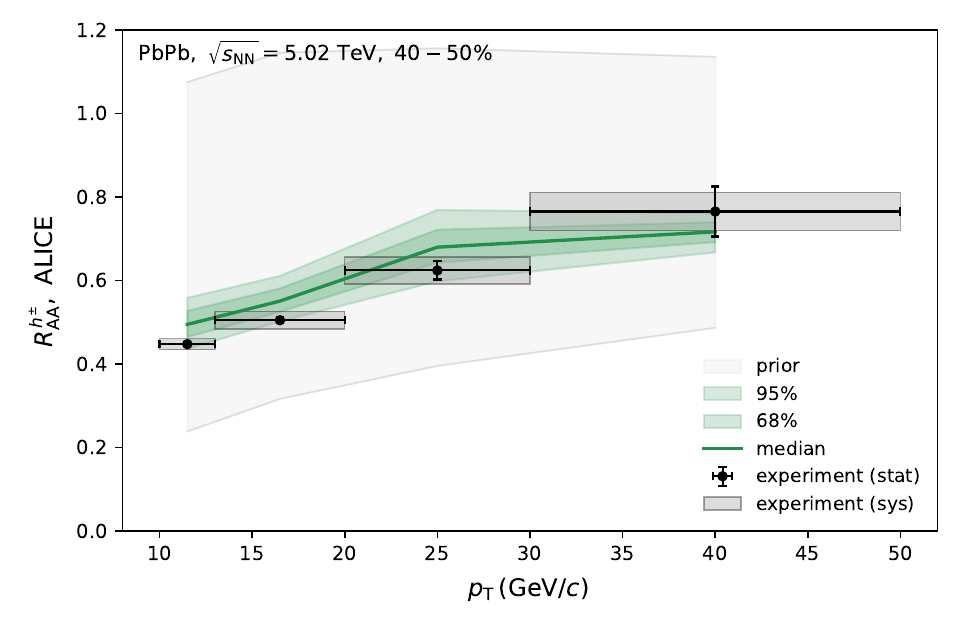}
\end{minipage}
\hfill
\begin{minipage}{0.48\columnwidth}
\centering
\includegraphics[width=\linewidth]{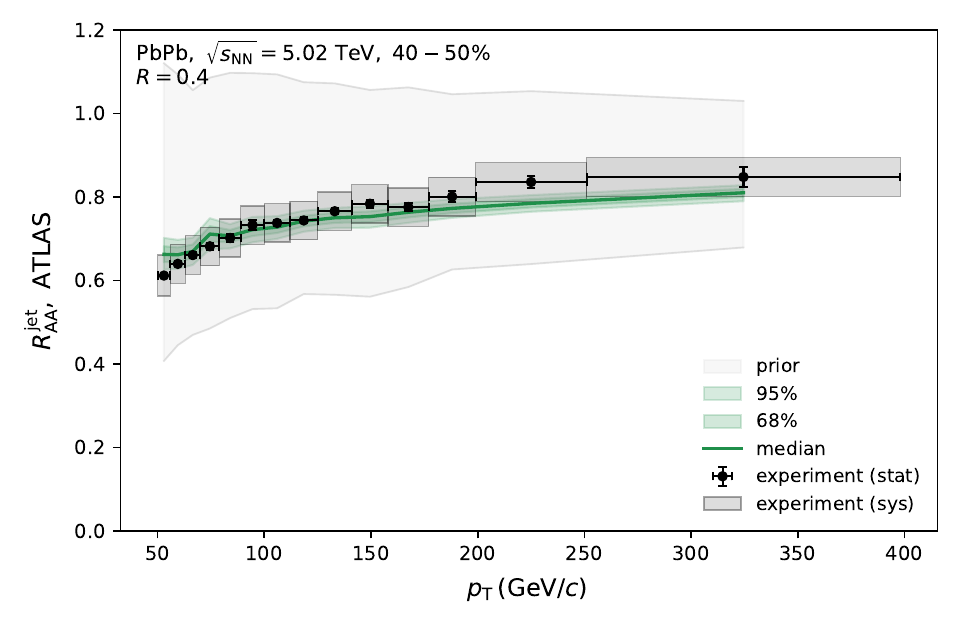}
\end{minipage}
\caption{Additional posterior-predictive mid-central panels for PbPb at $\sqrt{s_{\mathrm{NN}}}=5.02$ TeV.}
\label{fig:posterior-observables-5020-midcentral-extra}
\end{figure}

\begin{figure}[!tbp]
\centering
\begin{minipage}{0.48\columnwidth}
\centering
\includegraphics[width=\linewidth]{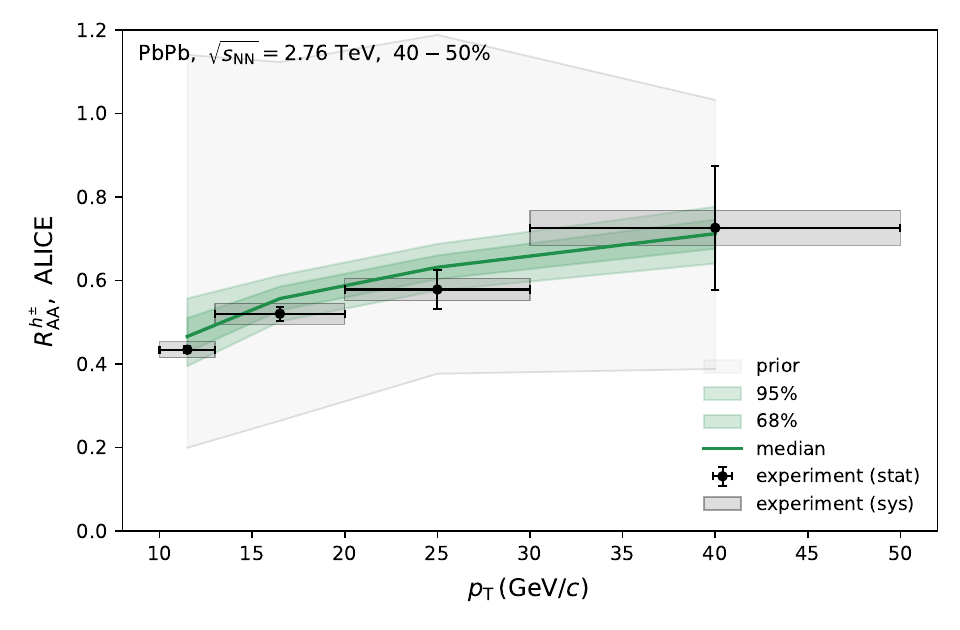}
\end{minipage}
\hfill
\begin{minipage}{0.48\columnwidth}
\centering
\includegraphics[width=\linewidth]{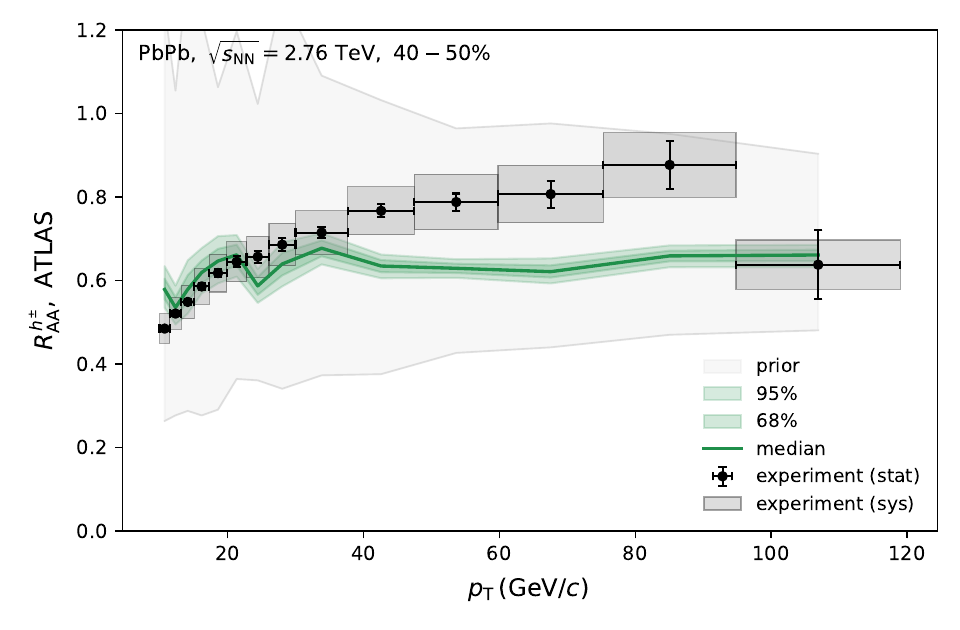}
\end{minipage}

\vspace{0.5em}

\makebox[\columnwidth][c]{%
\begin{minipage}{0.48\columnwidth}
\centering
\includegraphics[width=\linewidth]{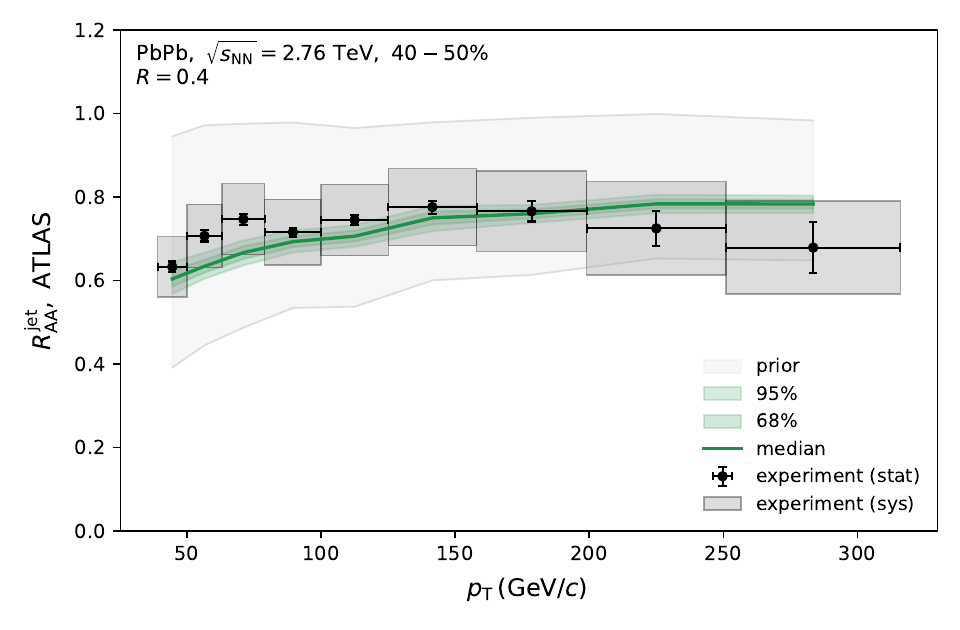}
\end{minipage}%
}
\caption{Additional posterior-predictive mid-central panels for PbPb at $\sqrt{s_{\mathrm{NN}}}=2.76$ TeV.}
\label{fig:posterior-observables-2760-midcentral-extra}
\end{figure}

\section{Additional Method 1 Validation Panels}
\label{app:method1-panels}

The main text shows the Method 1 validation panels for the same representative observables as the prior figure.
The remaining Method 1 panels are collected here for completeness.

\begin{figure}[!tbp]
\centering
\begin{minipage}{0.48\columnwidth}
\centering
\includegraphics[width=\linewidth]{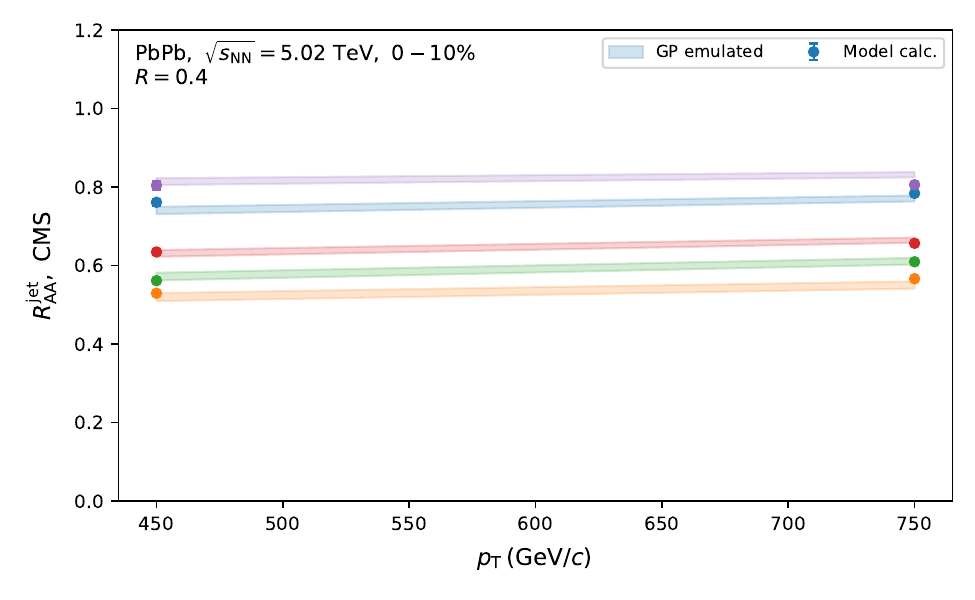}
\end{minipage}
\hfill
\begin{minipage}{0.48\columnwidth}
\centering
\includegraphics[width=\linewidth]{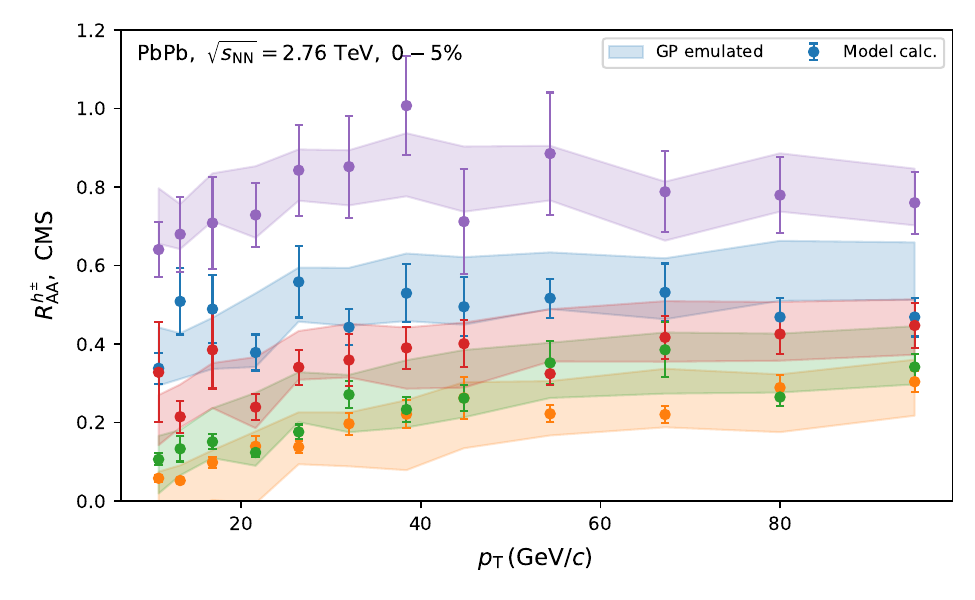}
\end{minipage}

\vspace{0.5em}

\makebox[\columnwidth][c]{%
\begin{minipage}{0.48\columnwidth}
\centering
\includegraphics[width=\linewidth]{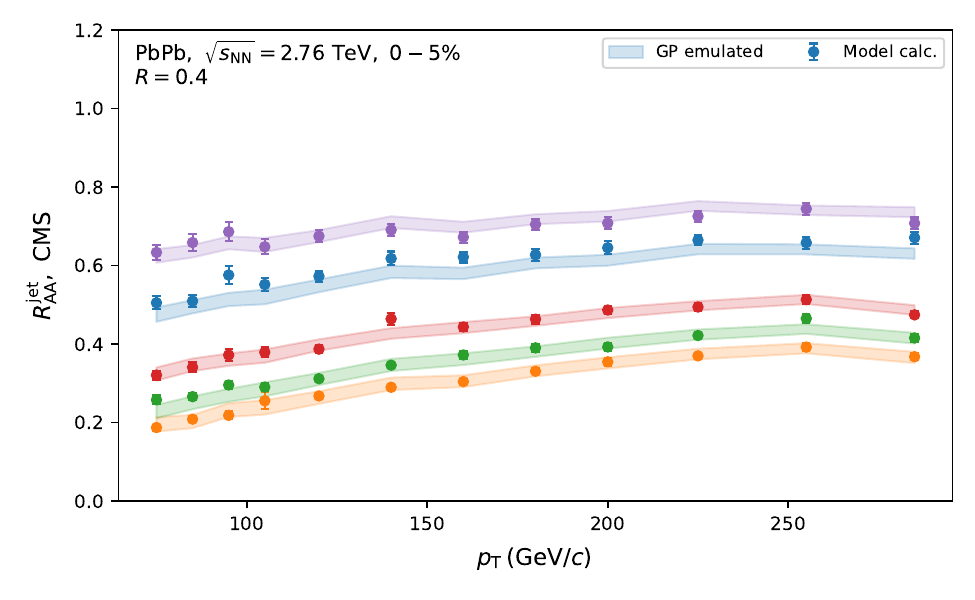}
\end{minipage}%
}
\caption{Additional Method 1 central panels not shown in the main text. These panels complete the PbPb $\sqrt{s_{\mathrm{NN}}}=5.02$ and $2.76$ TeV central selections beyond the representative subset shown in the main text.}
\label{fig:closure-method1-central-extra}
\end{figure}

\begin{figure}[!tbp]
\centering
\begin{minipage}{0.48\columnwidth}
\centering
\includegraphics[width=\linewidth]{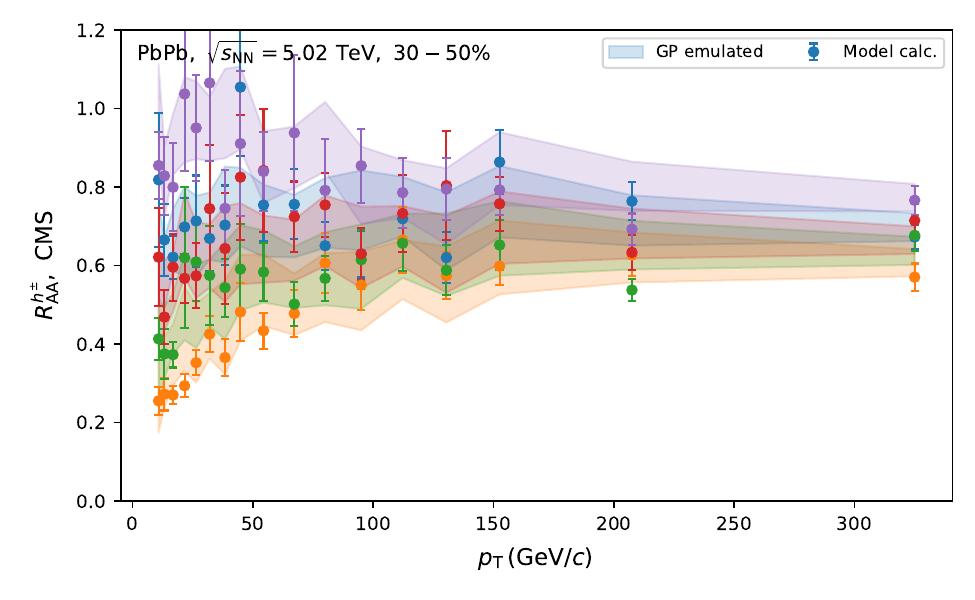}
\end{minipage}
\hfill
\begin{minipage}{0.48\columnwidth}
\centering
\includegraphics[width=\linewidth]{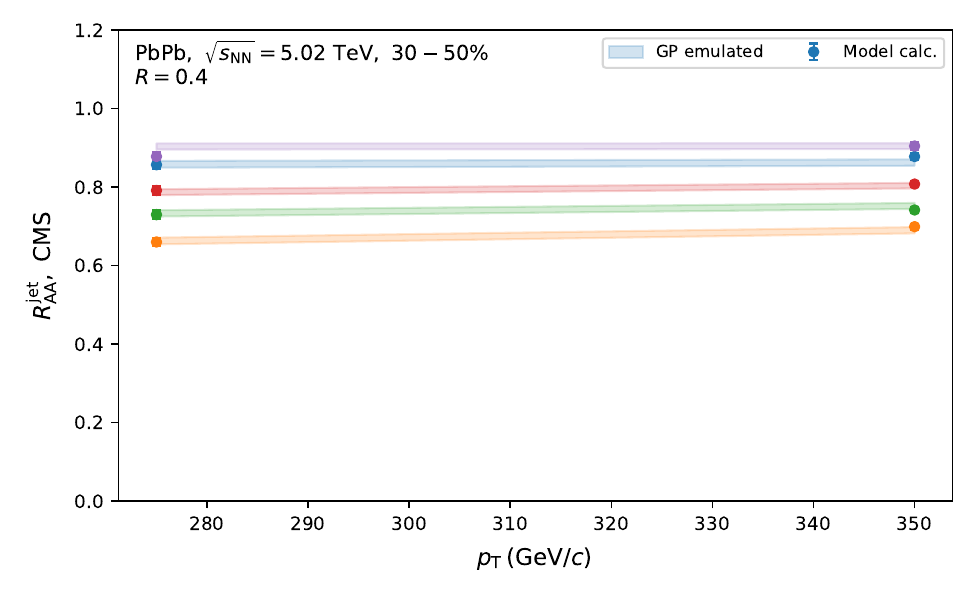}
\end{minipage}

\vspace{0.5em}

\begin{minipage}{0.48\columnwidth}
\centering
\includegraphics[width=\linewidth]{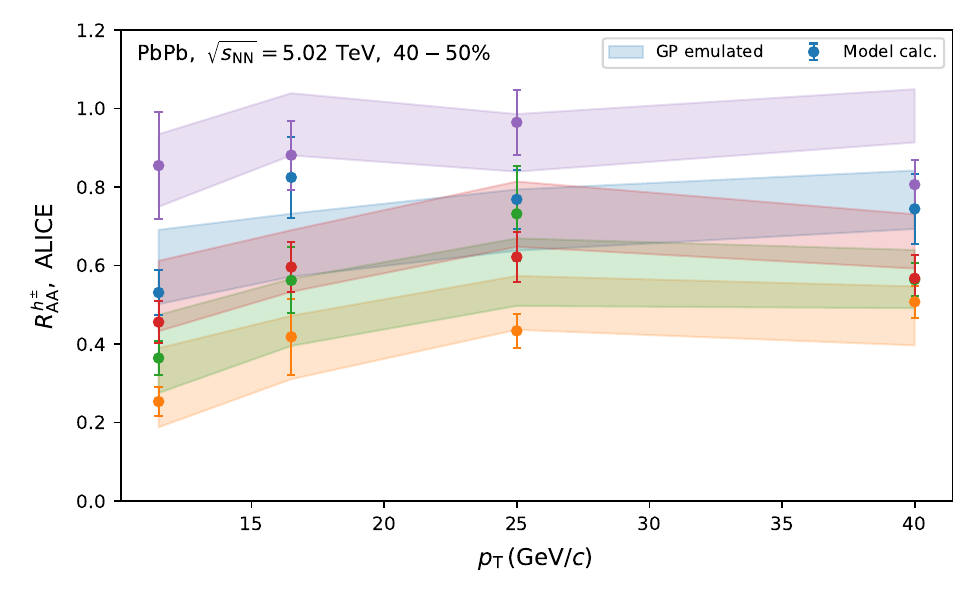}
\end{minipage}
\hfill
\begin{minipage}{0.48\columnwidth}
\centering
\includegraphics[width=\linewidth]{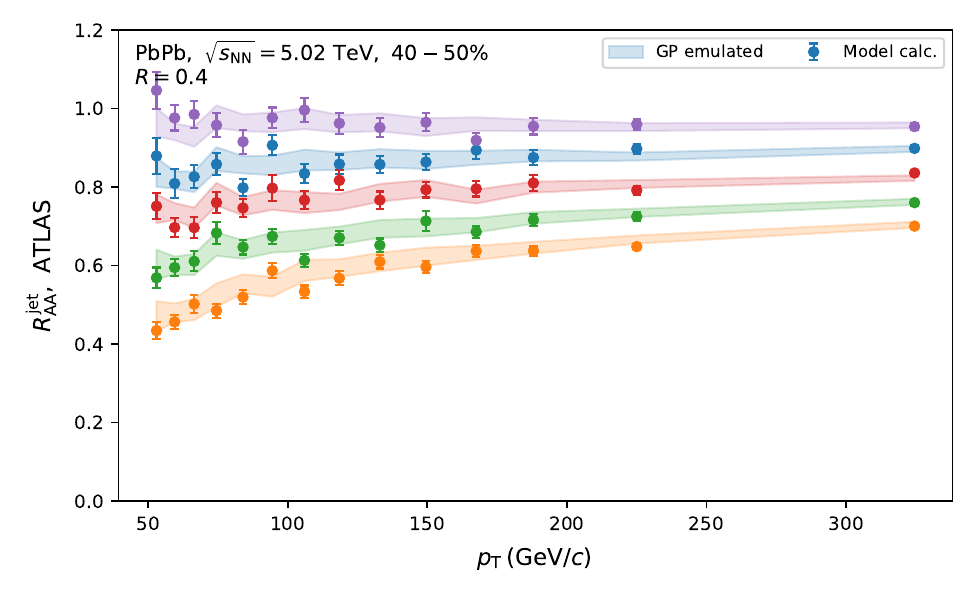}
\end{minipage}
\caption{Additional Method 1 mid-central panels for PbPb at $\sqrt{s_{\mathrm{NN}}}=5.02$ TeV.}
\label{fig:closure-method1-5020-midcentral-extra}
\end{figure}

\begin{figure}[!tbp]
\centering
\begin{minipage}{0.48\columnwidth}
\centering
\includegraphics[width=\linewidth]{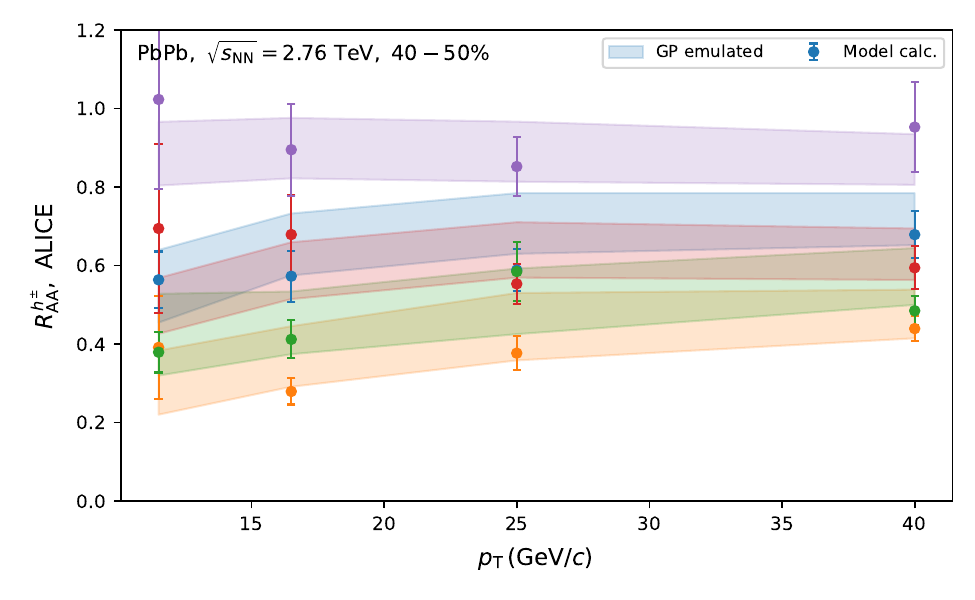}
\end{minipage}
\hfill
\begin{minipage}{0.48\columnwidth}
\centering
\includegraphics[width=\linewidth]{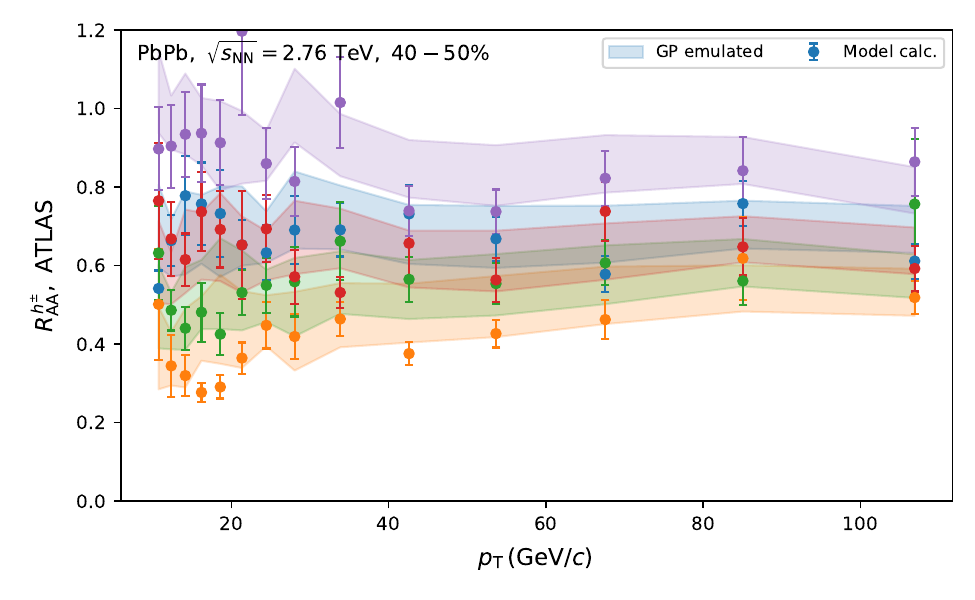}
\end{minipage}

\vspace{0.5em}

\makebox[\columnwidth][c]{%
\begin{minipage}{0.48\columnwidth}
\centering
\includegraphics[width=\linewidth]{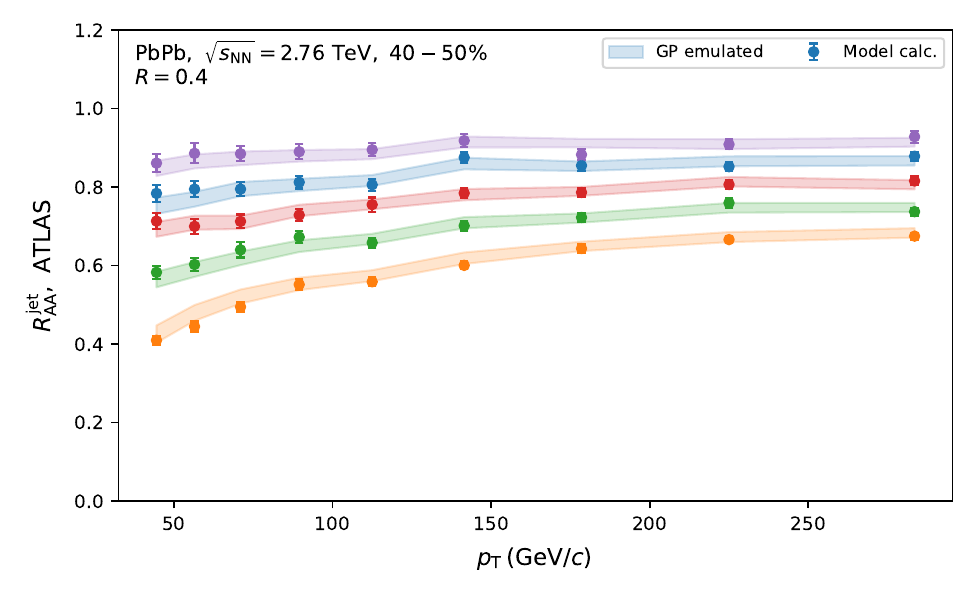}
\end{minipage}%
}
\caption{Additional Method 1 mid-central panels for PbPb at $\sqrt{s_{\mathrm{NN}}}=2.76$ TeV.}
\label{fig:closure-method1-2760-midcentral-extra}
\end{figure}

\section{Additional Method 2 Validation Panels}
\label{app:method2-panels}

The main text shows representative Method 2 panels alongside their Method 1 counterparts.
The remaining Method 2 panels are collected here for completeness.

\begin{figure}[!tbp]
\centering
\begin{minipage}{0.48\columnwidth}
\centering
\includegraphics[width=\linewidth]{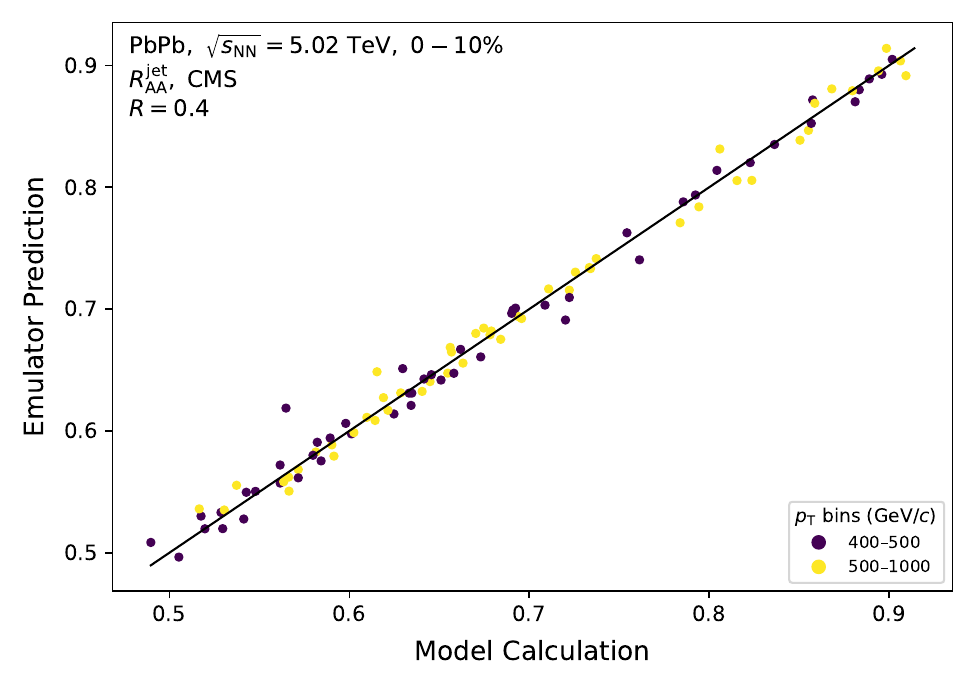}
\end{minipage}
\hfill
\begin{minipage}{0.48\columnwidth}
\centering
\includegraphics[width=\linewidth]{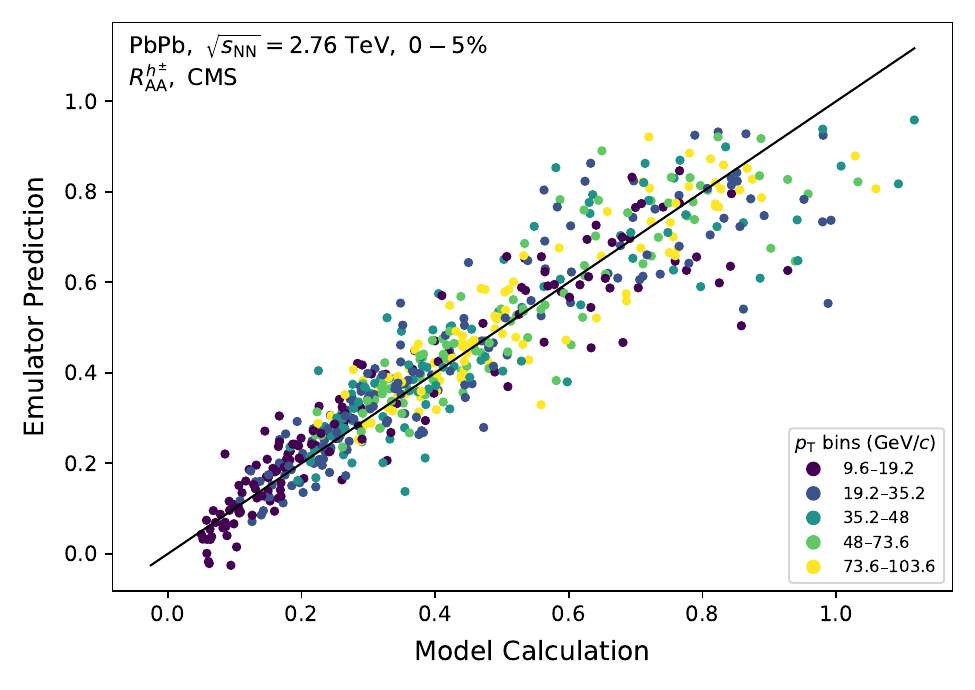}
\end{minipage}

\vspace{0.5em}

\makebox[\columnwidth][c]{%
\begin{minipage}{0.48\columnwidth}
\centering
\includegraphics[width=\linewidth]{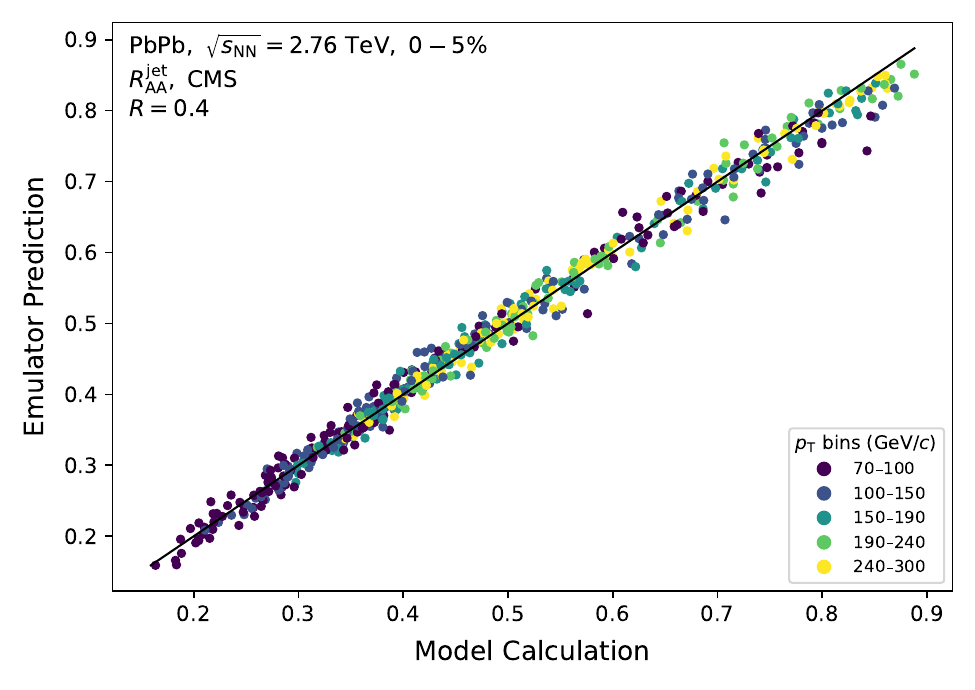}
\end{minipage}%
}
\caption{Additional Method 2 central panels not shown in the main text. These panels complete the PbPb $\sqrt{s_{\mathrm{NN}}}=5.02$ and $2.76$ TeV central selections beyond the representative subset shown in the main text.}
\label{fig:closure-method2-central-extra}
\end{figure}

\begin{figure}[!tbp]
\centering
\begin{minipage}{0.48\columnwidth}
\centering
\includegraphics[width=\linewidth]{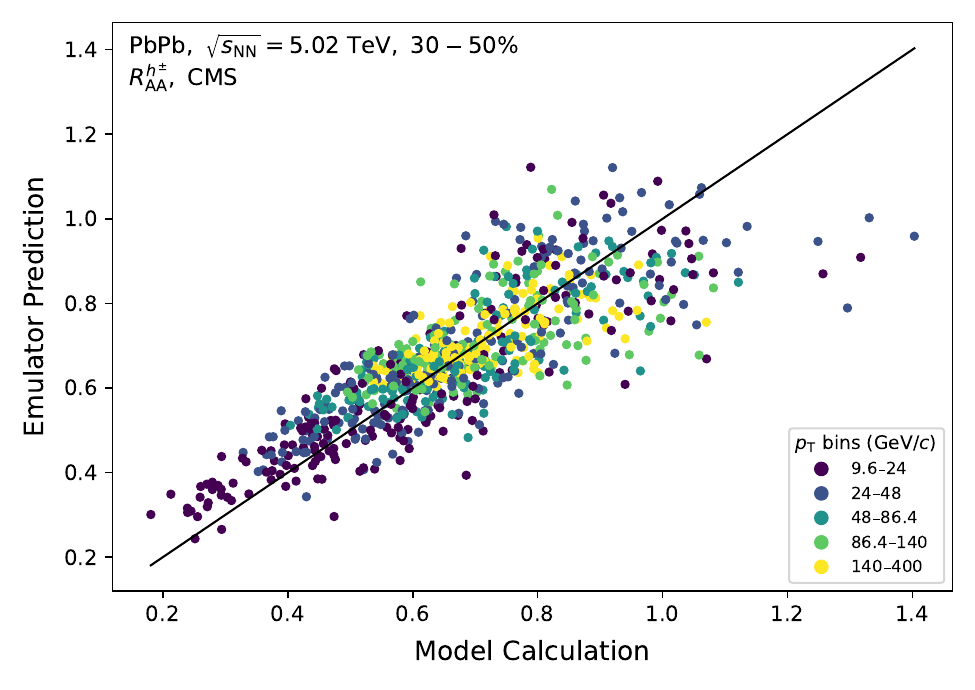}
\end{minipage}
\hfill
\begin{minipage}{0.48\columnwidth}
\centering
\includegraphics[width=\linewidth]{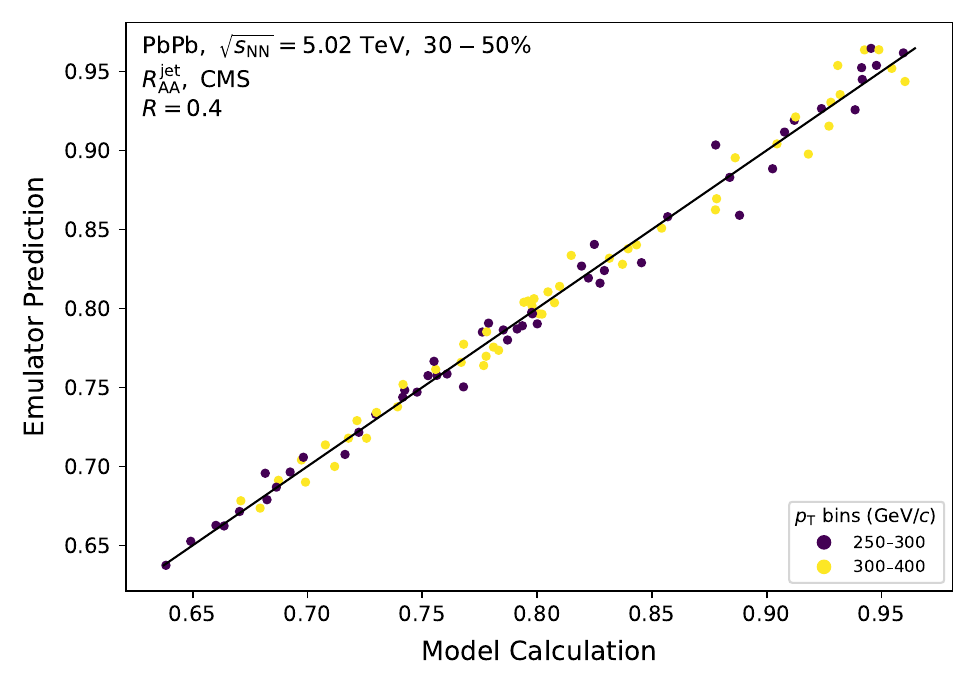}
\end{minipage}

\vspace{0.5em}

\begin{minipage}{0.48\columnwidth}
\centering
\includegraphics[width=\linewidth]{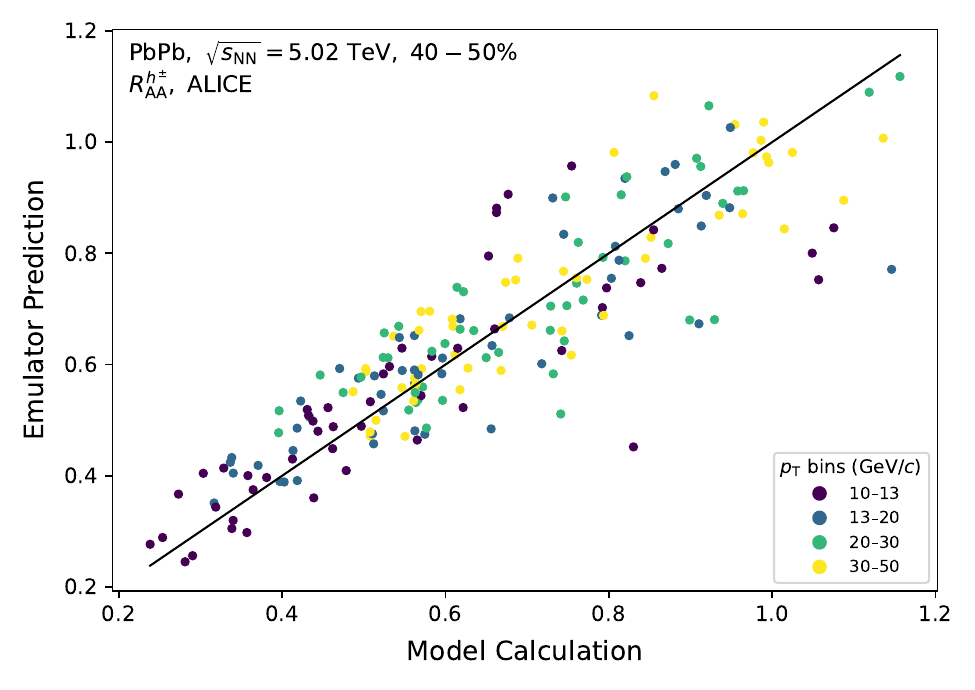}
\end{minipage}
\hfill
\begin{minipage}{0.48\columnwidth}
\centering
\includegraphics[width=\linewidth]{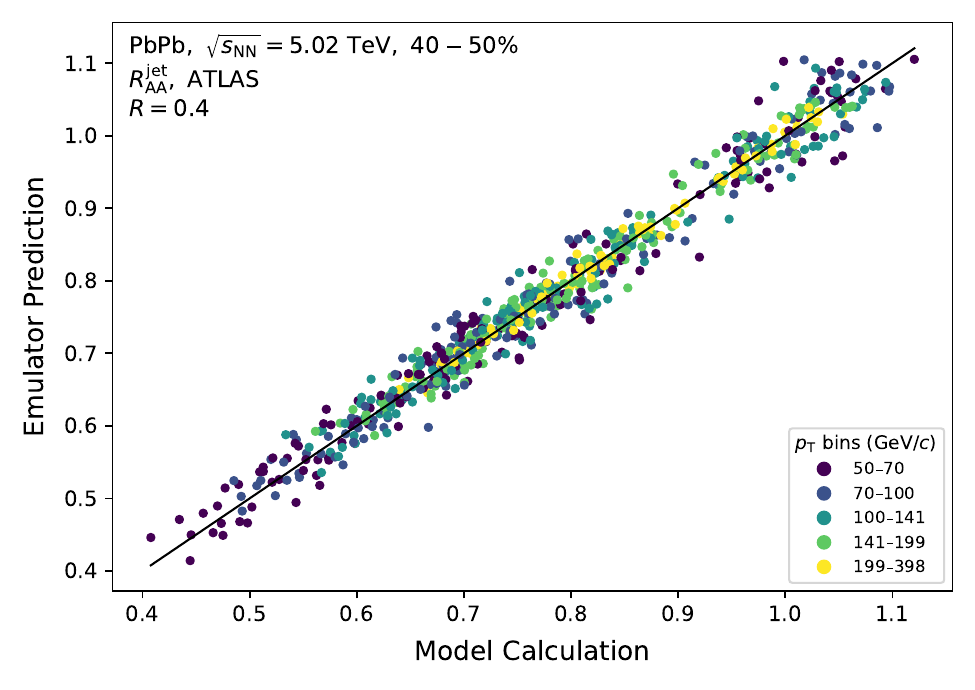}
\end{minipage}
\caption{Additional Method 2 mid-central panels for PbPb at $\sqrt{s_{\mathrm{NN}}}=5.02$ TeV.}
\label{fig:closure-method2-5020-midcentral-extra}
\end{figure}

\begin{figure}[!tbp]
\centering
\begin{minipage}{0.48\columnwidth}
\centering
\includegraphics[width=\linewidth]{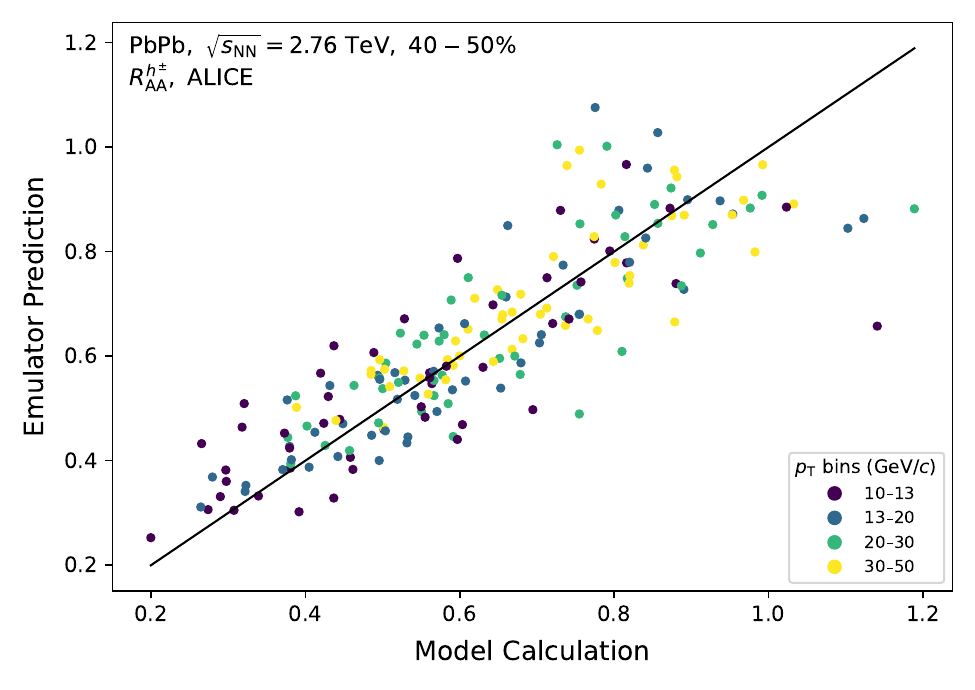}
\end{minipage}
\hfill
\begin{minipage}{0.48\columnwidth}
\centering
\includegraphics[width=\linewidth]{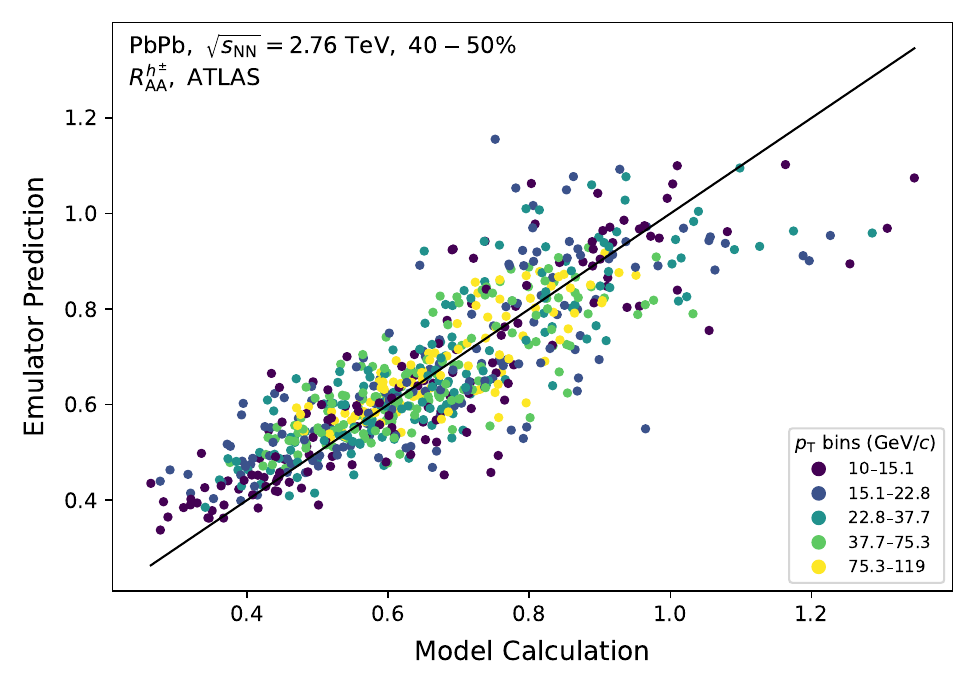}
\end{minipage}

\vspace{0.5em}

\makebox[\columnwidth][c]{%
\begin{minipage}{0.48\columnwidth}
\centering
\includegraphics[width=\linewidth]{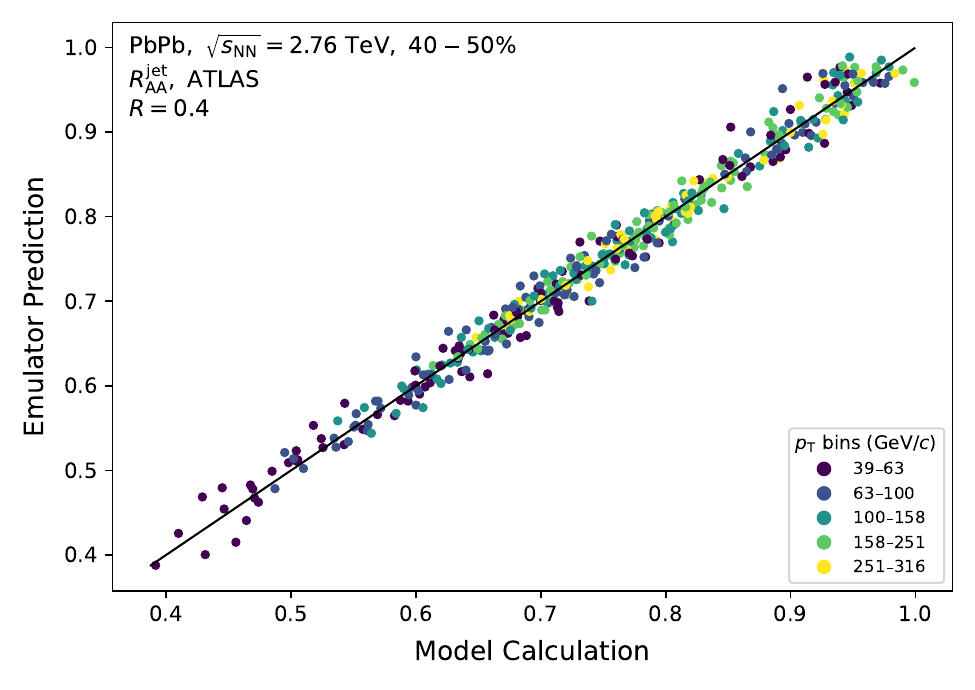}
\end{minipage}%
}
\caption{Additional Method 2 mid-central panels for PbPb at $\sqrt{s_{\mathrm{NN}}}=2.76$ TeV.}
\label{fig:closure-method2-2760-midcentral-extra}
\end{figure}

\section{MCMC Mixing Diagnostics}
\label{app:mcmc-diagnostics}

To assess the reliability of the posterior sampling, we monitor several standard MCMC quality measures for each run.
These diagnostics include the mean walker acceptance fraction, the split-$\hat{R}$ statistic, the integrated autocorrelation time, and the corresponding effective sample size (ESS).
In the implementation used here, the chain is considered adequately mixed when the mean acceptance fraction lies in the range 0.2--0.5, the maximum split-$\hat{R}$ remains below 1.05, and the minimum ESS across parameters exceeds 400.
We additionally monitor whether an appreciable fraction of posterior samples accumulates within 2\% of a prior boundary; if more than 5\% of the samples for a parameter lie in that near-boundary region, the run is flagged as potentially prior-limited.
These checks are used as diagnostic criteria for validating the production chains reported in the main text.

\clearpage
\nocite{*}
\bibliographystyle{apsrev4-2}
\bibliography{main/paper}

\end{document}